\documentclass[12pt, draftclsnofoot, onecolumn]{IEEEtran}

\usepackage{amsmath,graphicx}
\usepackage{amssymb}
\usepackage{subfigure}
\usepackage{graphicx}
\usepackage{graphics}
\usepackage{epstopdf}
\usepackage{amsmath}
\usepackage{multirow}
\usepackage{bm}
\usepackage{enumerate}
\usepackage{mathdots}
\usepackage{todonotes}
\usepackage{algorithmicx}
\usepackage{algpseudocode}
\usepackage{algorithm}
\usepackage{verbatim}
\usepackage{balance}
\usepackage{cite}
\usepackage{url}
\usepackage{booktabs}
\usepackage{tabularx}
\usepackage{acronym}
\usepackage{capt-of}

\usepackage{color}
\definecolor{purple(x11)}{rgb}{0.63, 0.36, 0.94}
\definecolor{cadmiumgreen}{rgb}{0.0, 0.42, 0.24}

\newacro{RSS}{received signal strength}
\newacro{OFDM}{orthogonal frequency division multiplexing}
\newacro{V2X}{vehicle-to-everything}
\newacro{A2X}{air-to-everything}
\newacro{SINR}{signal-to-interference plus noise ratio}
\newacro{ML}{maximum likelihood}
\newacro{LLF}{log-likelihood function}
\newacro{S-OMP}{simultaneous - orthogonal matching pursuit}
\newacro{WLS}{weighted least squares}
\newacro{GLM}{general linear model}
\newacro{BLUE}{best linear unbiased estimator}
\newacro{SW-OMP}{simultaneous weighted - orthogonal matching pursuit}
\newacro{SS-SW-OMP+Th}{subcarrier selection - simultaneous weighted - orthogonal matching pursuit + thresholding}
\newacro{QuaDRiGa}{quasi deterministic radio channel generator}
\newacro{AM}{alternating maximization}
\newacro{CSI}{channel state information}
\newacro{TTI}{transmission time interval}
\newacro{AoA}{angle-of-arrival}
\newacro{AoD}{angle-of-departure}
\newacro{LSP}{large-scale parameters}
\newacro{SSP}{small-scale parameters}
\newacro{CFO}{carrier frequency offset}
\newacro{TO}{timing offset}
\newacro{DS}{delay spread}
\newacro{AS}{angular spread}
\newacro{mmWave}{millimeter wave}
\newacro{LOS}{line-of-sight}
\newacro{NLOS}{non line-of-sight}
\newacro{CRLB}{Cram\'{e}r-Rao lower bound}
\newacro{NR}{New Radio}
\newacro{SU}{single-user}
\newacro{RF}{Radio-Frequency}
\newacro{MU}{multi-user}
\newacro{UMi}{urban micro cell}
\newacro{QAM}{quadrature amplitude modulation}
\newacro{NMSE}{normalized mean square error}
\newacro{MIMO}{multiple-input multiple-output}
\newacro{MMSE}{minimum mean square error}
\newacro{LMMSE}{linear minimum mean square error}
\newacro{MVUE}{minimum variance unbiased estimator}
\newacro{KF}{Kalman filter}
\newacro{PF}{particle filter}
\newacro{MPF}{marginalized particle filter}
\newacro{GMPF}{generalized marginalized particle filter}
\newacro{MT}{mobile terminal}
\newacro{ADC}{analog-to-digital converter}
\newacro{DAC}{digital-to-analog converter}
\newacro{SNR}{signal-to-noise ratio}
\newacro{LS}{least squares}
\newacro{SSAMP}{structured sparsity-adaptive matching pursuit}
\newacro{DGMP}{distributed grid matching pursuit}
\newacro{LKF}{linearized Kalman filter}
\newacro{KF}{Kalman filter}
\newacro{EKF}{extended Kalman filter}
\newacro{SISO}{single-input single-output}
\newacro{PN}{phase noise}
\newacro{HCRLB}{hybrid Cram\'{e}r-Rao lower bound}
\newacro{FIM}{Fisher information matrix}
\newacro{OMP}{orthogonal matching pursuit}
\newacro{ZP}{zero prefix}
\newacro{ISI}{inter symbol interference}
\newacro{BER}{bit error rate}
\newacro{W-OMP}{weighted - orthogonal matching pursuit}
\newacro{FET}{field-effect transistor}
\newacro{VCO}{voltage-controlled oscillator}
\newacro{QPSK}{quadrature phase-shift keying}
\newacro{SC-FDE}{single carier - frequency domain equalization}
\newacro{BS}{base station}
\newacro{UE}{user equipment}
\newacro{CPE}{common phase error}
\newacro{ICI}{inter carrier interference}
\newacro{SIC}{successive interference cancellation}
\newacro{GLM}{general linear model}
\newacro{MP}{matching pursuit}
\newacro{ULA}{uniform linear array}
\newacro{UPA}{uniform planar array}
\newacro{EM}{expectation-maximization}
\newacro{ZC}{Zadoff-Chu}
\newacro{CS}{compressed sensing}
\newacro{AWGN}{additive white Gaussian noise}
\newacro{GLRT}{generalized likelihood ratio test}
\newacro{RIP}{restricted isometry property}
\newacro{BPDN}{basis pursuit de-noising}
\newacro{MAP}{maximum a posteriori}
\newacro{PDF}{probability density function}
\newacro{SOCP}{second-order cone program}
\newacro{LASSO}{least absolute shrinkage and selection operator}
\newacro{CoSaMP}{compressive sampling matching pursuit}
\newacro{MSE}{mean squared error}
\newacro{HIM}{hybrid information matrix}
\newacro{GLM}{general linear model}
\newacro{QMMSE}{quadratic minimum mean square error estimator}
\newacro{RTS}{Rauch-Tung-Striebel}
\newacro{SVD}{singular value decomposition}
\newacro{GHP}{greedy hybrid precoding}
\newacro{PC}{per-antenna constrained}
\newacro{SSP-OMP}{spatially sparse precoding - orthogonal matching pursuit}
\newacro{LF}{likelihood function}
\newacro{PSD}{power spectral density}
\newacro{CP}{cyclic prefix}
\newacro{DFT}{Discrete Fourier Transform}

\newcommand{\vect}{\mathop{\mathrm{vec}}}
\newcommand{\var}{\mathop{\mathrm{var}}}
\newcommand{\covar}{\mathop{\mathrm{covar}}}
\newcommand{\Real}{\mathop{\mathrm{Re}}}
\newcommand{\Imag}{\mathop{\mathrm{Im}}}
\newcommand{\trace}{\mathop{\mathrm{trace}}}
\newcommand{\diag}{\mathop{\mathrm{diag}}}

\newcommand{\SNR}{\mathop{\mathrm{SNR}}}

\usepackage{pifont}

\newcommand{\Nr}{N_{\mathrm{r}}}
\newcommand{\Nt}{N_{\mathrm{t}}}
\newcommand{\Ns}{N_{\mathrm{s}}}

\newcommand{\Ts}{T_{\mathrm{s}}}

\newcommand{\rhoL}{\rho_{\mathrm{L}}}
\newcommand{\Gr}{G_{\mathrm{r}}}
\newcommand{\Gt}{G_{\mathrm{t}}}
\newcommand{\Lt}{L_{\mathrm{t}}}
\newcommand{\Lr}{L_{\mathrm{r}}}
\newcommand{\Lc}{L_{\mathrm{c}}}
\newcommand{\Ntr}{N_{\mathrm{tr}}}

\newcommand{\ar}{{\mathbf{a}}_{\mathrm{R}}}
\newcommand{\at}{{\mathbf{a}}_{\mathrm{T}}}

\newcommand{\jj}{{\mathrm{j}}}

\newcommand{\be}{\begin{eqnarray}}
\newcommand{\ee}{\end{eqnarray}}

\def\ba{{\mathbf{a}}}

\def\bc{{\mathbf{c}}}
\def\bd{{\mathbf{d}}}
\def\bee{{\mathbf{e}}}
\def\bff{{\mathbf{f}}}
\def\bg{{\mathbf{g}}}
\def\bh{{\mathbf{h}}}
\def\bi{{\mathbf{i}}}

\def\br{{\mathbf{r}}}

\def\bv{{\mathbf{v}}}

\def\bx{{\mathbf{x}}}
\def\by{{\mathbf{y}}}

\def\b0{{\mathbf{0}}}

\def\bC{{\mathbf{C}}}

\def\bF{{\mathbf{F}}}

\def\bH{{\mathbf{H}}}
\def\bI{{\mathbf{I}}}
\def\bJ{{\mathbf{J}}}

\def\bM{{\mathbf{M}}}

\def\bP{{\mathbf{P}}}

\def\bW{{\mathbf{W}}}

\def\bsfA{\bm{\mathsf{A}}}

\def\bsfC{\bm{\mathsf{C}}}
\def\bsfD{\bm{\mathsf{D}}}

\def\bsfF{\bm{\mathsf{F}}}
\def\bsfG{\bm{\mathsf{G}}}
\def\bsfH{\bm{\mathsf{H}}}

\def\bsfS{\bm{\mathsf{S}}}

\def\bsfW{\bm{\mathsf{W}}}

\def\sfg{{\mathsf{g}}}

\def\sfj{{\mathsf{j}}}

\def\sfs{{\mathsf{s}}}

\def\sf0{{\mathsf{0}}}

\def\bsfg{{\bm{\mathsf{g}}}}

\def\bsfq{{\bm{\mathsf{q}}}}

\def\bsfs{{\bm{\mathsf{s}}}}

\def\bsfv{{\bm{\mathsf{v}}}}

\def\bsf0{{\bm{\mathsf{0}}}}

\begin{document}
\title{Broadband Synchronization and Compressive Channel Estimation for Hybrid mmWave MIMO Systems}
\author{Javier Rodr\'{i}guez-Fern\'{a}ndez ({\em Student Member IEEE}) \\
The University of Texas at Austin \\ Email: javi.rf@utexas.edu
}

\maketitle

\begin{abstract}

Synchronization is a fundamental procedure in cellular systems whereby an \ac{UE} acquires the time and frequency information required to decode the data transmitted by a \ac{BS}. Due to the necessity of using large antenna arrays to obtain the beamforming gain required to compensate for small antenna aperture, synchronization must be performed either jointly with beam training as in 5G \ac{NR}, or at the low \ac{SNR} regime if the high-dimensional \ac{mmWave} \ac{MIMO} channel is to be estimated. To circumvent this problem, this work proposes the first synchronization framework for \ac{mmWave} \ac{MIMO} that is robust to both \ac{TO}, \ac{CFO}, and \ac{PN} synchronization errors and, unlike prior work, implicitly considers the use of multiple RF chains at both transmitter and receiver. I provide a theoretical analysis of the estimation problem and derive the \ac{HCRLB} for the estimation of both the \ac{CFO}, \ac{PN}, and equivalent beamformed channels seen by the different receive RF chains. I also propose two novel algorithms to estimate the different unknown parameters, which rely on approximating the \ac{MMSE} estimator for the \ac{PN} and the \ac{ML} estimators for both the \ac{CFO} and the equivalent beamformed channels. Thereafter, I propose to use the estimates for the equivalent beamformed channels to perform compressive estimation of the high-dimensional frequency-selective \ac{mmWave} \ac{MIMO} channel and thus undergo data transmission. For performance evaluation, I consider the QuaDRiGa channel simulator, which implements the 5G \ac{NR} channel model, and show that both compressive channel estimation without prior synchronization is possible, and the proposed approaches outperform current solutions for joint beam training and synchronization currently considered in 5G \ac{NR}.

\end{abstract}


\section{Introduction}

Time-frequency synchronization is one of the most important design aspects in cellular systems. In \ac{mmWave} systems, however, acquiring synchronization information is significantly more challenging than for traditional sub-$6$ GHz \ac{MIMO} systems. Due to the necessity of using large antenna arrays to obtain the beamforming gain required to compensate for small antenna aperture, time and frequency synchronization must be performed either jointly with beam training as in 5G \ac{NR}, or at the low \ac{SNR} regime if the high-dimensional \ac{mmWave} \ac{MIMO} channel is to be estimated. Unfortunately, at such high operating frequency bands, synchronization, channel estimation, and data transmission are impacted by \ac{PN} impairments, which consist of random fluctuations in the phase of the carrier generated by local oscillators.

\subsection{Prior work and Motivation}

In the context of \ac{mmWave} \ac{MIMO} systems, synchronization parameters need to be properly estimated and compensated for before \ac{CSI} can be acquired. This sets new challenges as synchronization acquisition must be performed at the low \ac{SNR} regime, before transmit and receive communication beams can be aligned for data transmission. 

The problem of \ac{CS}-based joint beam training and synchronization is studied in \cite{CS_Initial_Beamforming,Cabric:CompressiveIA:JSTSP:2019}. In \cite{CS_Initial_Beamforming}, the problem of beam training under \ac{PN} errors and unknown \ac{CFO} was studied for narrowband \ac{mmWave} \ac{MIMO} systems using analog architectures. Therein, an \ac{EKF}-based solution is proposed to track the joint phase of the unknown \ac{PN} and beamformed narrowband channel, the phase of the received signal is compensated, and then \ac{MP} is used to estimate the dominant \ac{AoD} and \ac{AoA}, thereby discovering a single communication path. In \cite{Cabric:CompressiveIA:JSTSP:2019}, a compressive initial access approach based on omnidirectional pseudorandom analog beamforming is proposed as an alternative to the directional initial access procedure used during beam management in 5G \ac{NR}, and the effects of imperfect \ac{TO} and \ac{CFO} are studied therein. The main limitations of the algorithm proposed in \cite{Cabric:CompressiveIA:JSTSP:2019} are: i) the algorithm is tailored to \ac{LOS} channel models, thereby implicitly ignoring the presence of spatio-temporal clusters in the propagation environment; ii) the proposed signal model assumes the presence of phase measurement errors only due to \ac{CFO}, thereby ignoring the \ac{PN} impairment. Prior work on joint broadband channel estimation and synchronization for \ac{mmWave} \ac{MIMO} is limited, since much (if not most) of the prior work on channel estimation assumes perfect synchronization at the receiver side \cite{VenAlkGon:Channel-Estimation-for-Hybrid:17}, \cite{ChEstHybridPrecmmWave}, \cite{Marzi2016:ChanEstTracking}, \cite{GaoDaiWan:Channel-estimation-mmWave-massiveMIMO:16}, \cite{GaoHuDai:Channel-estimation-mmWave-massiveMIMO-FS:16}. 

In the context of broadband channel models, prior work is limited to \cite{Zhang_Iterative_Channel_Phase_Noise_ICC_2015,CFO_Channel_Est,Swift-Link,MessagePassingCFOMyers}. In \cite{Zhang_Iterative_Channel_Phase_Noise_ICC_2015}, the problem of joint channel and \ac{PN} estimation for a \ac{SISO} system is considered, which is unrealistic at \ac{mmWave}, and the proposed algorithms are only evaluated in very high \ac{SNR} regime. In \cite{CFO_Channel_Est}, analog-only architectures with a single RF chain are assumed, and an autocorrelation-based iterative algorithm is proposed to jointly estimate the \ac{CFO} and the \ac{mmWave} channel. Prior work in \cite{CFO_Channel_Est} assumes that analog beamformers and combiners can be instantaneously reconfigured for two consecutive transmitted time-domain samples, which is unrealistic since phase-shifters need an adjustment time for phase reconfiguration \cite{IEEE:11ad}. Further, the algorithm proposed in \cite{CFO_Channel_Est} has only been evaluated for \ac{mmWave} channels having a very small number of non-clustered multipath components, which is not realistic at \ac{mmWave} \cite{StatisticalChannelModel:TWC:2017}. In addition, owing to the nature of the autocorrelation function, the proposed algorithm does not perform both well when the \ac{CFO} is considerably large and the \ac{SNR} is low. In \cite{Swift-Link}, a \ac{CFO}-robust beam alignment technique is developed to find the beam pairs maximizing the received \ac{SNR}. The main limitation of \cite{Swift-Link} is that the algorithm proposed therein can only be applied to analog \ac{MIMO} architectures, and its \ac{CFO} correction capability is limited by both the number of delay taps in the \ac{mmWave} \ac{MIMO} channel, as well as the length of the training sequence, thereby making the algorithm impractical for practical \ac{mmWave} deployments with more significant \ac{CFO}. In \cite{MessagePassingCFOMyers}, the joint \ac{CFO} and broadband channel estimation problem is formulated as a sparse bilinear optimization problem, which is solved using the parametric bilinear generalized approximate message passing (PBiGAMP) algorithm in \cite{PBiGAMP}. The main limitation of \cite{MessagePassingCFOMyers} is that the proposed estimation strategy is tailored to all-digital \ac{MIMO} architectures with low-resolution \ac{ADC} converters, thereby not being directly applicable to hybrid \ac{MIMO} architectures. In \cite{JointCFOChEstMyers1bit}, a similar strategy to the one in \cite{MessagePassingCFOMyers} is followed, in which the joint \ac{CFO} and channel estimation problem is studied for all-digital \ac{MIMO} architectures. The problem is formulated as a quantized sparse bilinear optimization problem, which is solved using sparse lifting to increase the dimension of the \ac{CFO} and channel estimation problem \cite{BiconvexCS}, and then applying the generalized approximate message passing (GAMP) algorithm in \cite{GAMP} to solve the lifted problem. 

\section{Contributions}

In this paper, I develop efficient and robust solutions to the problem of estimating the \ac{TO}, \ac{CFO}, \ac{PN}, and frequency-selective channel for hybrid \ac{mmWave} \ac{MIMO} systems. The proposed solutions can leverage the spatial design degrees of freedom brought by having several RF chains at both the transmitter and receiver to perform synchronization and compressive channel estimation, without relying on any prior channel knowledge. The contributions of this paper are summarized hereinafter:

\begin{itemize}
	\item Based on a protocol of forwarding several training frames from the transmitter to the receiver \cite{Ahmed:2014}, \cite{Cabric:CompressiveIA:JSTSP:2019}, \cite{VenAlkGon:Channel-Estimation-for-Hybrid:17}, \cite{RodGon:CFO:TWC:2019}, I formulate and find a solution to the problem of \ac{TO}, \ac{CFO}, \ac{PN} and frequency-selective \ac{mmWave} \ac{MIMO} channel estimation for systems employing hybrid architectures. Further, the focus is on analyzing the synchronization problem at the low SNR regime.

	\item I propose to forward several training frames using \ac{ZC}-based beamforming in combination with random subarray switching and antenna selection in order to both acquire synchronization and enable compressive channel estimation at the low \ac{SNR} regime. 	
	
	\item For every training frame, which comprises several \ac{OFDM} symbols, as in the 5G \ac{NR} wireless standard \cite{5G_standard}, I theoretically analyze the hybrid \ac{CRLB} for the problem of estimating the \ac{CFO}, \ac{PN}, and equivalent frequency-selective beamformed channel collecting the joint effect of the transmit hybrid precoders, frequency-selective \ac{mmWave} \ac{MIMO} channel, receive hybrid combiners, and equivalent transmit-receive pulse-shaping that bandlimits the complex baseband equivalent channel.

	\item I propose two novel iterative algorithms based on the \ac{EM} method, which aim at finding the \ac{ML} estimates for the \ac{CFO} and beamformed equivalent channels, as well as the \ac{LMMSE} estimates for the \ac{PN} samples that impair the receive signals. The first proposed algorithm exhibits very good performance, yet it exhibits high computational complexity. The second proposed algorithm, conversely, offers a trade-off between estimation performance and computational complexity, and exhibits a very small performance gap with respect to the first algorithm.
	
	\item Using estimates of the unknown parameters for every training frame, I formulate the problem of estimating the high-dimensional frequency-selective \ac{mmWave} \ac{MIMO} channel, and find a solution to this problem using a variation of the \ac{SW-OMP} algorithm in \cite{RodGonVenHea:TWC:2018}. 

\end{itemize}

I evaluate the performance of the proposed algorithms in terms of \ac{NMSE} and spectral efficiency. I use all-digital precoders and combiners to show the effectiveness of the proposed algorithms. Simulation results obtained from the estimated channel show that both the \ac{TO}, \ac{CFO}, and equivalent channels can be accurately estimated even in the presence of strong \ac{PN}, and when the \ac{MIMO} channel has several clusters with non-negligible \ac{AS}. Furthermore, I show that near-optimum spectral efficiency can be attained, without incurring in significant overhead and/or computational complexity. To the best of my knowledge, this is the first work that theoretically analyzes and provides solutions to the problem of joint synchronization and compressive channel estimation at \ac{mmWave} considering hybrid \ac{MIMO} architectures, and that is robust to both \ac{CFO}, \ac{PN}, and low \ac{SNR} regime.

\section{System model with synchronization impairments}
\label{sec:system_model_broadband}

\begin{figure}[t!]
\centering
\includegraphics[width=\textwidth]{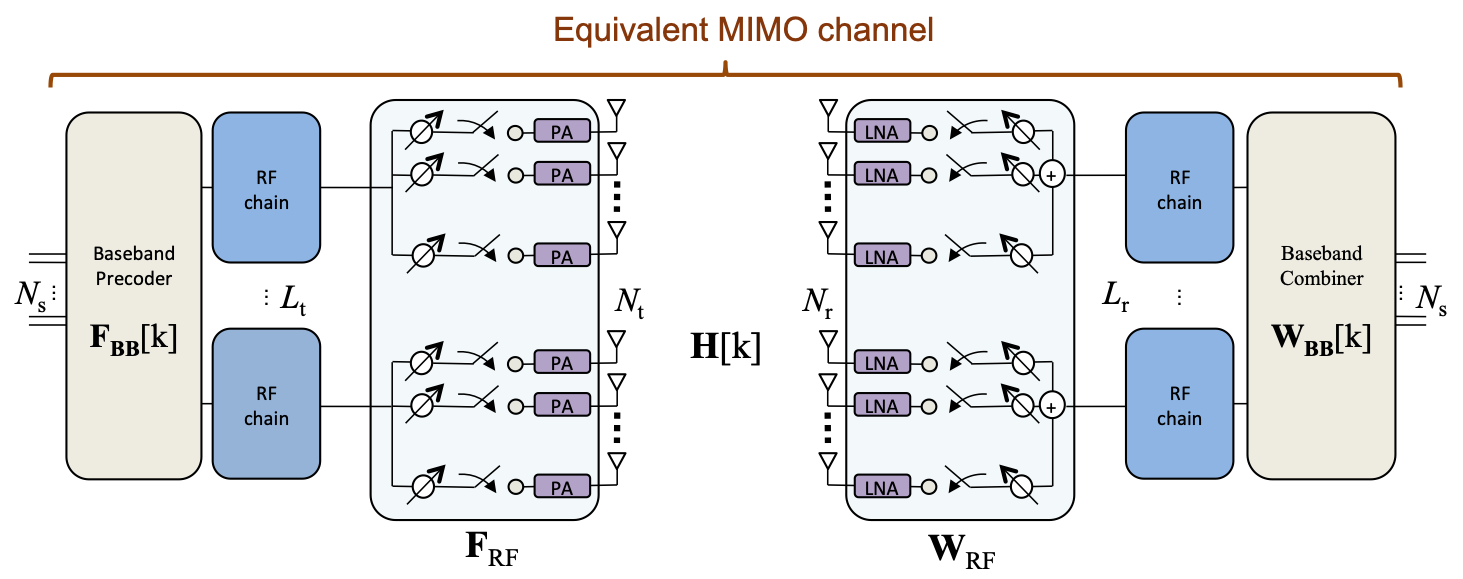} 
\caption{Illustration of the structure of a partially-connected hybrid \ac{MIMO} architecture, which includes analog and digital precoders and combiners. This figure has been taken from \cite{RodGon-Globecom-2019-arxiv}.}     
\label{fig:hybrid_architecture}        
\end{figure}

I consider a single-user \ac{mmWave} \ac{MIMO}-\ac{OFDM} communications link in which a transmitter equipped with $\Nt$ antennas sends $\Ns$ data streams to a receiver having $\Nr$ antennas. Both transmitter and receiver are assumed to use partially-connected hybrid \ac{MIMO} architectures \cite{Rial_switchOrShifter:Access2016}, as shown in Fig. \ref{fig:hybrid_architecture}, with $\Lt$ and $\Lr$ RF chains. A frequency-selective hybrid precoder is used at the transmitter, with $\bsfF[k]= \bsfF_\text{RF}\bsfF_\text{BB}[k] \in {\mathbb{C}}^{\Nt\times\Ns}$, where $\bsfF_\text{RF} \in \mathbb{C}^{\Nt \times \Lt}$ is the analog precoder and $\bsfF_\text{BB}[k] \in \mathbb{C}^{\Lt \times \Ns}$ is the digital one at subcarrier $k$, $0 \leq k \leq K-1$. The \ac{RF} precoder and combiner are implemented using a partially-connected network of phase-shifters and switches, as described in \cite{Rial_switchOrShifter:Access2016}. Likewise, the receiver applies a hybrid linear combiner $\bsfW[k] = \bsfW_\text{RF} \bsfW_\text{BB}[k] \in {\mathbb{C}}^{\Nr\times\Lr}$, where $\bsfW_\text{RF} \in \mathbb{C}^{\Nr \times \Lr}$ is the analog combiner, and $\bsfW_\text{BB}[k] \in \mathbb{C}^{\Lr \times \Ns}$ is the baseband combiner at the $k$-th subcarrier.

Without loss of generality, I assume that the transmitted signal comprises $\Ntr$ \ac{OFDM} symbols with a \ac{CP} of length $\Lc$, similarly to the 5G \ac{NR} wireless standard \cite{5G_standard}. Let us define the rectangular pulse-shape $w_N[n] = 1$ for $n \in [0,N-1]$, and $w_N[n] = 0$ otherwise. Then, the hybrid-precoded transmitted signal can be expressed as
\begin{equation}
\begin{split}
	\bx[n] &= \frac{1}{K} \bsfF_\text{RF}\sum_{k=0}^{K-1}\sum_{t = 0}^{\Ntr-1}\bsfF_\text{BB}[k]\bsfs_t[k] e^{\sfj \frac{2\pi k (n - \Lc - t(K+\Lc))}{K}} w_{K+\Lc}[n - t(K+\Lc)] , \\ & \quad n = 0,\ldots,(\Ntr-1)(K+\Lc)-1,
\end{split}
\label{equation:OFDM_t_signal}
\end{equation}

Then, let $n_0 \in \mathbb{K}_+$, $\Delta f \in \mathbb{R}$, $\theta[n] \in \mathbb{R}$ denote the unknown \ac{TO}, \ac{CFO} normalized to the sampling rate $f_\text{s} = 1/\Ts$, and $n$-th receive \ac{PN} sample. Also, let $\{\bW[\ell]\}_{\ell=0}^{K-1}$ denote the time-domain hybrid combiner, given by the IFFT of the frequency-selective hybrid combiner $\{\bsfW[k\}_{k=0}^{K-1}$. Then, the received signal at discrete time instant $n$ can be written as
\begin{equation}
\by[n]  = \left\{\bW^*[\ell]\right\}_{\ell = 0}^{K-1} *\left(\sum_{d=0}^{D-1} \bH[d] \bx[n-d-n_0] e^{\sfj (2\pi \Delta f n + \theta[n])} \right) + \bv[n], 
\label{equation:rx_signal}
\end{equation}
for $n = 0,\ldots,N+D+n_0-1$, with $N$ being the length of the time-domain transmitted signal $\bx[n]$, and $\bv[n] \sim {\cal CN}\left(\bm 0, \sigma^2 \sum_{\ell=0}^{K-1} \bW^*[\ell] \bW[\ell] \right)$ is the post-combining received noise, where $\sigma^2$ denotes the variance of the noise at any receive antenna. 

In this paper, I focus on the problem of estimating the unknown \ac{CFO} $\Delta f$, \ac{PN} samples $\theta[n]$, and frequency-selective \ac{mmWave} \ac{MIMO} channel $\{\bH[d]\}_{d=0}^{D-1}$. Given the high dimensionality of the channel matrices, I consider a training protocol in which the transmitter forwards $M$ training frames to the receiver \cite{RodGon:CFO:TWC:2019}, \cite{VenAlkGon:Channel-Estimation-for-Hybrid:17}, \cite{Ahmed:2014}, \cite{RodGonVenHea:TWC:2018}, which must estimate the different unknown synchronization parameters.
In view of this, for the $m$-th training frame, $1 \leq m \leq M$, I set $\bsfF^{(m)}[k] = \bsfF_\text{tr}^{(m)}$, and $\bsfW^{(m)}[k] = \bsfW_\text{tr}^{(m)}$, for every $0 \leq k \leq K-1$. Furthermore, I design the training symbols in \eqref{equation:OFDM_t_signal} as $\bsfs_t^{(m)}[k] = \bsfq^{(m)} \bsfs_t^{(m)}[k]$, where $\bsfq^{(m)}$ operates as an equivalent baseband precoder for this particular design of the training sequence \cite{RodGon:CFO:TWC:2019}. Therefore, for the transmission of the $m$-th training frame, comprising of $\Ntr$ \ac{OFDM} symbols, the received signal reads as
\begin{equation}
	\by^{(m)}[n] = \bsfW_\text{tr}^{(m)*} \sum_{d=0}^{D-1}\bH[d] \bsfF_\text{tr}^{(m)} \bsfq^{(m)} s[n-d-n_0] e^{\sfj (2\pi \Delta f^{(m)}n + \theta^{(m)}[n])} + \bv^{(m)}[n],
	\label{equation:rx_signal_frame}
\end{equation}
in which $\bv^{(m)}[n] \sim {\cal CN}\left(\bm 0, \sigma^2 \bsfW_\text{tr}^{(m)*} \bsfW_\text{tr}^{(m)}\right)$ is the received post-combining circularly-symmetric complex additive Gaussian noise. As shown in \cite{CAMSAP_2017_Offgrid}, \cite{RodGon:CFO:TWC:2019}, the \ac{ML} criterion establishes that the baseband combiner must whiten the received signal to estimate the different unknown parameters. For this purpose, let us consider the Cholesky decomposition of $\bsfC_\text{w}^{(m)} = \bsfW_\text{tr}^{(m)*}\bsfW_\text{tr}^{(m)}$ as $\bsfC_\text{w}^{(m)} = \bsfD_\text{w}^{(m)*}\bsfD_\text{w}^{(m)}$, with $\bsfD_\text{w}^{(m)} \in \mathbb{C}^{\Lr \times \Lr}$ an upper triangular matrix. Now, let us define a vector $\bg^{(m)}[d] \in \mathbb{C}^{\Lr \times 1}$, $\bg^{(m)}[d] = \bsfD_\text{w}^{(m)-*}\bsfW_\text{tr}^{(m)*} \bH[d] \bsfF_\text{tr}^{(m)}\bsfq^{(m)}$, containing the complex equivalent channel samples for a given training step $1 \leq m \leq M$. Accordingly, for the $m$-th transmitted frame, the received signal in \eqref{equation:rx_signal_frame} can be expressed as \cite{RodGon-Globecom-2019-arxiv}
\begin{equation}
\begin{split}
	\by^{(m)}[n] &= e^{\sfj (2\pi \Delta f^{(m)} n + \theta^{(m)}[n])} \sum_{d=0}^{D-1} \underbrace{\bg^{(m)}[d] s^{(m)}[n-d-n_0]}_{\bx^{(m)}[n,d,n_0]} \\ &+ \bv^{(m)}[n], \quad n = 0,\ldots,n_0 + (\Lc + K) \Ntr + D - 1
\end{split}
	\label{equation:rx_frame}
\end{equation}
with $\bv^{(m)}[n] \sim {\cal CN}(\bm 0, \sigma^2 \bI_{\Lr})$ being the post-whitened spatially white received noise vector, and $\bg^{(m)}[d] = [\alpha_1[d] e^{j\beta_1[d]},\ldots,\alpha_{\Lr}[d] e^{j\beta_{\Lr}[d]}]^T$ is the complex equivalent beamformed channel for the $m$-th training step and $d$-th delay tap. Therefore, the interest here lies on estimating the vector of parameters $\bm \xi^{(m)} \triangleq \left[\left\{\bg^{(m)T}[d]\right\}_{d=0}^{D-1},\Delta f^{(m)},\{\theta^{(m)}[n]\}_{n=0}^{N+n_0 + D - 1}, n_0, \sigma^2 \right]^T$ for every training frame. In the next section, I theoretically analyze this estimation problem and find the hybrid \ac{CRLB} for the different parameters in $\bm \xi^{(m)}$.

The \ac{PN} impairment is modeled according to the IEEE 802.11ad wireless standard, whose details are given in \cite{Draft_IEEE_802.11_PN}. The \ac{PSD} of the \ac{PN} is given in \cite{Blind_PN_ICC_2015_mmWave}, \cite{RodGon-Globecom-2019-arxiv}, and the reader is referred therein for more details regarding the mathematical model of this parameter.

The discrete-time \ac{MIMO} channel between the transmitter and the receiver is modeled as a set of $\Nr \times \Nt$ matrices denoted as $\bH[d]$, for a given delay tap $d = 0,\ldots,D-1$, with $D$ the delay tap length of the channel. Each of the matrices $\bH[d]$ is assumed to be a sum of the contributions of $C$ spatial clusters, each contributing with $R_c$ rays,  $c = 1,\ldots,C$. I use $\rhoL$ to denote the pathloss, $\alpha_{c,r} \in \mathbb{C}$ is the complex gain of the $r$-th ray within the $c$-th cluster, $\tau_{c,r} \in \mathbb{R}_{+}$ is the time delay of the $r$-th ray within the $c$-th cluster, $\phi_{c,r}, \theta_{c,r} \in [0,2\pi)$ are the \ac{AoA} and \ac{AoD}, $\ar(\cdot) \in {\mathbb{C}}^{\Nr\times1}$ and $\at(\cdot) \in {\mathbb{C}}^{\Nt\times1}$ denote the receive and transmit array steering vectors, $p_\text{RC}(\tau)$ is the equivalent transmit-receive baseband pulse shape including analog filtering effects evaluated at $\tau$ \cite{schniter_sparseway:2014}, and $\Ts$ is the sampling interval. Using this notation, the frequency-domain channel matrix at the $k$-th subcarrier is given by
\begin{equation}
\begin{split}
\bsfH[k] &= \sqrt{\frac{\Nr\Nt}{\rho_\text{L} \sum_{c=1}^{C}{R_c}}}\sum_{k=0}^{K-1}\sum_{c = 1}^{C}\sum_{r=1}^{R_c}\alpha_{c,r} p_\text{RC}(d\Ts - \tau_{c,r}) \times \\ &\times \ar\left(\phi_{c,r}\right)\at^*(\theta_{c,r})e^{-\sfj \frac{2\pi k}{K}}. 
\end{split}
\label{eqn:channel_model}
\end{equation}
Taking the inverse Fourier transform of \eqref{eqn:channel_model} allows obtaining the discrete-time \ac{MIMO} channel used in \eqref{equation:rx_signal_frame}. The channel matrix can be represented more compactly as
\begin{equation}
\begin{split}
	\bsfH[k] &= \bsfA_\text{R} \bsfG[k] \bsfA_\text{T}^*,
\end{split}
\label{eq:channel_compact_fd}
\end{equation}
where $\bsfG[k] \in \mathbb{C}^{\sum_{c=1}^{C}{R_c} \times \sum_{c=1}^{C}{R_c}}$ is a diagonal matrix containing the path gains and the equivalent pulse-shaping effect, and $\bsfA_\text{T} \in \mathbb{C}^{\Nt \times \sum_{c=1}^{C}R_c}$, $\bsfA_\text{R} \in \mathbb{C}^{\Nr \times \sum_{c=1}^{C}R_c}$ are the frequency-selective array response matrices evaluated on the \ac{AoD} and \ac{AoA}, respectively.
Finally, the matrix $\bsfH[k]$ in \eqref{eq:channel_compact_fd} can be approximated using the extended virtual channel model \cite{mmWavetutorial} as
\begin{equation}
	\bsfH[k] \approx \tilde{\bsfA}_\text{R} \bsfG^\text{v}[k] \tilde{\bsfA}_\text{T}^*,
	 \label{equation:channel_extended_fd}
\end{equation}
where $\bsfG^\text{v}[k] \in \mathbb{C}^{\Gr \times \Gt}$ is a sparse matrix containing the path gains of the quantized spatial frequencies in the non-zero elements, and the dictionary matrices $\tilde{\bsfA} _\text{T} \in \mathbb{C}^{\Nt \times \Gt}$, $\tilde{\bsfA}_\text{R} \in \mathbb{C}^{\Nr \times \Gr}$ contain the transmit and receive array response vectors evaluated on spatial grids of sizes $\Gt$ and $\Gr$, respectively.

\section{Theoretical analysis of the estimation problem}
\label{subsec:FIM_CRLB_broadband}

In this section, I theoretically analyze the problem of estimating the unknown parameters in $\bm \xi^{(m)}$. Let us assume, without loss of generality, that $\Ntr$ \ac{OFDM} symbols are transmitted for the $m$-th training frame, and that the number of available received time-domain samples of $\br^{(m)}[n]$ is given by $N = n_0 + (\Lc + K) \Ntr + D$. Assuming that the received time-domain noise samples in \eqref{equation:rx_frame} are independent and identically distributed, the received signal in \eqref{equation:rx_frame} has \ac{LLF} given by
\begin{equation}
\begin{split}
&\log{p\left(\{\br^{(m)}[n]\}_{n=0}^{N-1};\bm \xi^{(m)}\right)} = -N \log{(\pi \sigma^2)} - \frac{1}{\sigma^2} \sum_{n=0}^{N-1} \left\| \br^{(m)}[n] \right\|_2^2 \\ & +2\sum_{n=0}^{N-1} \Real\left\{ \br^{(m)*} e^{\sfj (2\pi \Delta f^{(m)} n + \theta^{(m)}[n])} \sum_{d=0}^{D-1} \bx^{(m)}[n,d,n_0]\right\} - \sum_{n=0}^{N-1} \left\| \sum_{d=0}^{D-1} \bx^{(m)}[n,d,n_0] \right\|_2^2.
\end{split}
\label{equation:LLF_rx_signal}
\end{equation}

To ensure robustness of the synchronization algorithm, I will focus on finding the \ac{ML} estimators for the different unknown parameters in $\bm \xi^{(m)}$. From \eqref{equation:LLF_rx_signal}, it is observed that maximizing ${\cal L}(\bm \xi^{(m)}) = \log{p\left(\{\br^{(m)}[n]\}_{n=0}^{N-1}; \bm \xi^{(m)}\right)}$ as a function of $n_0$ requires knowledge of the other parameters contained in $\bm \xi^{(m)}$, which suggests that the \ac{ML} estimator exhibits high computational complexity. To reduce computational complexity, I propose to exploit the good correlation properties of Golay sequences \cite{IEEE:11ad}, \cite{IEEE:11ay}, and append a $64$-point $G_{\text{a},64}$ sequence at the beginning of the training frame, which has been shown to offer excellent performance in the absence of \ac{PN} \cite{Asilomar_2018}. Thereby, a practical \ac{TO} estimator can be devised by maximizing the third term in \eqref{equation:LLF_rx_signal}, which is given by \cite{Asilomar_2018,Cabric:CompressiveIA:JSTSP:2019}
\begin{equation}
\begin{split}
\hat{n}_0 &= \underset{n_0}{\arg\,\max} \sum_{i=1}^{\Lr} \sum_{n=0}^{63}\left| r_i^{(m)*}[n] s^{(m)}[n-d-n_0] \right|,
\end{split}
\label{equation:TO_estimator}
\end{equation}
which explicitly exploits the information coming from having $\Lr \geq 1$ at the receiver side. Assuming that the \ac{TO} has been estimated perfectly using \eqref{equation:TO_estimator}, this parameter can be compensated by advancing the receive signal by $\hat{n}_0$ as $\br^{(m)}[n] = \by^{(m)}[n + n_0]$, $n = 0,\ldots, (\Lc + K)\Ntr + D - 1$. Now, let the initial sample of the $t$-th \ac{OFDM} symbol after \ac{CP} removal be defined as $k_0[t] \triangleq \Lc + n_0 + t(K+\Lc)$, let $\phi^{(m)}[n_0,t] \triangleq e^{\sfj 2\pi \Delta f^{(m)} k_0[t]}$ be the common phase change at the $t$-th \ac{OFDM} symbol due to \ac{TO}, let $\bm \Omega_t \left(\Delta f^{(m)}\right) \triangleq \phi^{(m)}[n_0,t]\bigoplus_{n=0}^{K-1} e^{\sfj 2\pi \Delta f^{(m)}n}$ be the \ac{CFO} matrix impairing the $t$-th \ac{OFDM} symbol, and let $\bsfS_t^{(m)} \triangleq \bigoplus_{k=0}^{K-1} \sfs_t^{(m)}[k]$ and $\bsfS^{(m)} \triangleq \left[\begin{array}{ccc}
	\bsfS_0^{(m)T} & \ldots & \bsfS_{\Ntr}^{(m)T} \\ \end{array}\right]^T$ be matrices containing the $t$-th \ac{OFDM} training symbol and the $\Ntr$ \ac{OFDM} training symbols, respectively. Also, let $\bsfg_i^{(m)} \triangleq \left[\begin{array}{ccc} \sfg_i^{(m)}[0] & \ldots & \sfg_i^{(m)}[K-1] \\ \end{array}\right]$ be the frequency-response of the equivalent beamformed channel seen by the $i$-th receive RF chain, let $\bv_{i,t}^{(m)} \triangleq \left[\begin{array}{ccc} v_{i}^{(m)}[k_0[t]] & \ldots & v_{i}^{(m)}[k_0[t]+K-1] \\ \end{array}\right]$ contain the time-domain noise samples impairing the $t$-th \ac{OFDM} symbol, let $\bm \theta_t^{(m)} = \left[\begin{array}{ccc} \theta^{(m)}[k_0[t]] & \ldots & \theta^{(m)}[k_0[t] + K-1] \\ \end{array}\right]^T$ be the \ac{PN} vector corresponding to the $t$-th \ac{OFDM} symbol, and let $\bP\left(\bm \theta_t^{(m)}\right) \triangleq \bigoplus_{n=0}^{K-1} e^{\sfj \theta^{(m)}[k_0[t] + n]}$ be the diagonal \ac{PN} matrix impairing the received $t$-th \ac{OFDM} symbol. Letting $\bF \in \mathbb{C}^{K \times K}$ denote the $K$-point DFT matrix, the received time synchronized signal $\br^{(m)}[n]$ can be vectorized as
\begin{equation}
\begin{split}
	\underbrace{\left[\begin{array}{c}
	r_i^{(m)}[k_0[t]] \\ 
	\vdots \\
	r_i^{(m)}[k_0[t] + K-1] \\ 
	\end{array}\right]}_{\br_{i,t}^{(m)}} &= \bm \Omega_t\left(\Delta f^{(m)}\right) \bP_t\left(\bm \theta_t^{(m)}\right) \underbrace{\frac{1}{\sqrt{K}} \left[\begin{array}{ccc}
	e^{\sfj \frac{2\pi 0}{K}} & \ldots & e^{\sfj \frac{2 \pi 0 (K-1)}{K}} \\
	\vdots & \ddots & \vdots \\
	e^{\sfj \frac{2\pi 0 (K-1) }{K}} & \ldots & e^{\sfj \frac{2\pi (K-1)(K-1)}{K}} \\ \end{array}\right]}_{\bF^*} \\ &\times \underbrace{\left(\bigoplus_{k=0}^{K-1} \sfs_t^{(m)}[k] \right)}_{\bsfS_t^{(m)}} \underbrace{\left[\begin{array}{c}
	\sfg_i^{(m)}[0] \\
	\vdots \\
	\sfg_i^{(m)}[K-1] \\ \end{array}\right]}_{\bsfg_i^{(m)}} + \underbrace{\left[\begin{array}{c} v_i^{(m)}[k_0[t]] \\
	\vdots \\
	v_i^{(m)}[k_0[t] + K-1] \\ \end{array}\right]}_{\bv_{i,t}^{(m)}},
	\end{split}
	\label{equation:rx_signal_frame_OFDM_symbol_RF_chain_full}
\end{equation}
such that the vectorized received signal can be simplified as
\begin{equation}
	\br_{i,t}^{(m)} = \bm \Omega_t\left(\Delta f^{(m)}\right) \bP_t\left(\bm \theta_t^{(m)}\right) \bF^* \bsfS_t^{(m)} \bsfg_i^{(m)} + \bv_{i,t}^{(m)}.
	\label{equation:rx_signal_frame_OFDM_symbol_RF_chain}
\end{equation}
Now, the $K \times 1$ random vector $\br_{i,t}^{(m)}$ can be stacked for the different received \ac{OFDM} symbols $1 \leq t \leq \Ntr$ and \ac{RF} chains $1 \leq i \leq \Lr$ as
\begin{equation}
\begin{split}
	\underbrace{\left[\begin{array}{c}
	\br_{i,1}^{(m)} \\
	\vdots \\
	\br_{i,\Ntr}^{(m)} \\ \end{array}\right]}_{\br_i^{(m)}} &= \underbrace{\left(\bigoplus_{t=1}^{\Ntr} \bm \Omega_t \left(\Delta f^{(m)}\right)\right)}_{\bm \Omega\left(\Delta f^{(m)}\right)}\underbrace{\left( \bigoplus_{t=1}^{\Ntr} \bP_t\left(\bm \theta_t^{(m)}\right)\right)}_{\bP\left(\bm \theta^{(m)}\right)} \underbrace{\left(\bI_{\Ntr} \otimes \bF^*\right)}_{\bF_\otimes^*} \\ & \times \underbrace{\left[\begin{array}{c}
	\bsfS_1^{(m)} \\
	\vdots \\
	\bsfS_{\Ntr}^{(m)} \\ \end{array}\right]}_{\bsfS^{(m)}} \bsfg_i^{(m)} + \underbrace{\left[\begin{array}{c}
	\bv_{i,1}^{(m)} \\
	\vdots \\
	\bv_{i,\Ntr}^{(m)} \\ \end{array}\right]}_{\bv_i^{(m)}}. 
	\end{split}
	\label{equation:rx_signal_frame_RF_chain}
\end{equation}
Therefore, the received signal $\br_i^{(m)}$ is distributed according to ${\cal CN}\left(\bm \mu_i^{(m)}\left(\bm \xi^{(m)}\right), \sigma^2 \bI_{K \Ntr}\right)$, where $\bm \mu_i^{(m)}\left(\bm \xi^{(m)}\right) = \bm \Omega\left(\Delta f^{(m)}\right) \bP\left(\bm \theta^{(m)}\right) \bF_\otimes^* \bsfS^{(m)} \bsfg_i^{(m)}$. Finally, stacking the received signals $\br_i^{(m)}$ for the different RF chains yields
\begin{equation}
\left[\begin{array}{c}
\br_1^{(m)} \\
\vdots \\
\br_{\Lr}^{(m)} \\ \end{array}\right] = \left(\bI_{\Lr} \otimes \bm \Omega \left(\Delta f^{(m)}\right) \bP\left(\bm \theta^{(m)}\right) \bF_\otimes^* \bsfS^{(m)} \right) \left[\begin{array}{c} \bsfg_1^{(m)} \\
\vdots \\
\bsfg_{\Lr}^{(m)} \\ \end{array}\right] + \left[\begin{array}{c} \bv_1^{(m)} \\
\vdots
\\
\bv_{\Lr}^{(m)} \\ \end{array}\right]. 
\label{equation:rx_signal_frame_all_RF_chains}
\end{equation}

For the purpose of theoretically analyzing the estimation problem of finding the unknown parameters, let $\sfg_i^{(m)}[k]$ be further expressed as $\sfg_i^{(m)}[k] = \alpha_i^{(m)}[k] e^{\sfj \beta_i^{(m)}[k]}$, and let $\tilde{\bsfg}_i^{(m)}[k] \in \mathbb{C}^2$ be defined as $\tilde{\bsfg}_i^{(m)}[k] \triangleq [\alpha_i^{(m)}[k], \beta_i^{(m)}[k]]^T$. Finally, let $\tilde{\bsfg}_i^{(m)} \in \mathbb{C}^{2K \times 1}$ be given by $\tilde{\bsfg}_i^{(m)} \triangleq [\tilde{\bsfg}_i^{(m)T}[0], \ldots, \tilde{\bsfg}_i^{(m)T}[K-1] ]^T$, and $\tilde{\bsfg}^{(m)} \triangleq [\tilde{\bsfg}_1^{(m)T}, \ldots, \tilde{\bsfg}_{\Lr}^{(m)T}]^T$. Likewise, let $\bm \theta^{(m)} \triangleq [\bm \theta_0^{(m)T}, \ldots, \bm \theta_{\Ntr - 1}^{(m)T}]T$. Now, the vector of parameters to be estimated is defined as $\bm \xi^{(m)} \in \mathbb{C}^{(K(\Lr + \Ntr) + 1) \times 1}$, given by $\bm \xi^{(m)} \triangleq [\Delta f^{(m)} , \tilde{\bsfg}^{(m)T} , \bm \theta^{(m)T}]^T$.

\subsection{Computation of the \ac{HIM}}

In this section, I derive the \ac{HIM} of the vector of parameters $\bm \xi^{(m)}$ and derive the hybrid \ac{CRLB} for any unbiased estimator of $\bm \xi^{(m)}$. Since there is prior knowledge on the \ac{PN} parameters in $\bm \theta_t^{(m)}$, $0 \leq t \leq \Ntr$, the \ac{HIM} $\bH\left(\bm \xi^{(m)}\right)$ can be defined as \cite{Bayesian_Bounds_Van_Trees}
\begin{equation}
	\bH\left(\bm \xi^{(m)}\right) = \bI_\text{D}\left(\bm \xi^{(m)}\right) + \bI_\text{P}\left(\bm \xi^{(m)}\right),
	\label{equation:HIM}
\end{equation}
where 
\begin{equation}
\bI_\text{D}\left(\bm \xi^{(m)}\right) \triangleq \mathbb{E}_{\bm \theta^{(m)}} \left\{ \bI \left(\bm \xi^{(m)}\right) \right\},
\end{equation}
with $\bI\left(\bm \xi^{(m)}\right)$ denoting the \ac{FIM} and 
\begin{equation}
\bI_\text{P}\left(\bm \xi^{(m)}\right) \triangleq - \mathbb{E}_{\bm \theta^{(m)}|\tilde{\bsfg}^{(m)}, \Delta f^{(m)} } \left\{ \frac{\partial^2 \log{p\left(\bm \theta^{(m)} | \tilde{\bsfg}^{(m)}, \Delta f^{(m)}\right)}}{\partial \bm \xi^{(m)} \bm \xi^{(m)T}}\right\}
\label{equation:PIM}
\end{equation}
is the prior information matrix with $p\left(\bm \theta^{(m)} | \tilde{\bsfg}^{(m)}, \Delta f^{(m)} \right)$ denoting the prior distribution of the \ac{PN} vector given the equivalent beamformed channels $\tilde{\bsfg}^{(m)}$ and the \ac{CFO} $\Delta f^{(m)}$.

The \ac{FIM} associated to $\bm \xi^{(m)}$, $\bI\left(\bm \xi^{(m)}\right) \in \mathbb{R}^{ (K(\Lr + \Ntr) + 1) \times (K(\Lr + \Ntr) + 1) }$, can be expressed as \cite{Kay:Fundamentals-of-Statistical-Signal:93}
\begin{equation}
	[\bI\left(\bm \xi^{(m)}\right)]_{r,c} = \frac{2}{\sigma^2} \sum_{i=1}^{\Lr} \Real \left\{ \frac{\partial \bm \mu_i^{(m)*}\left(\bm \xi^{(m)}\right)}{\partial \xi_r^{(m)} } \frac{\partial \bm \mu_i^{(m)}\left(\bm \xi^{(m)}\right)}{\partial \xi_c^{(m)}} \right\}, \quad 1 \leq r,c \leq K ( \Lr + \Ntr) + 1.
	\label{equation:FIM}
\end{equation}
Let $\bee_m^n \in \mathbb{R}^{n}$ denote the $m$-th canonical vector in $\mathbb{R}^n$, and let $p[t,\ell] = n_0 + \Lc + (t-1)(K + \Lc) + \ell$. Then, the terms $\frac{\partial \bm \mu_i^{(m)}\left(\bm \xi^{(m)}\right)}{\partial \xi_r^{(m)}}$ are given by
\begin{equation}
\begin{split}
\frac{\partial \bm \mu_i^{(m)}\left(\bm \xi^{(m)}\right)}{\partial \xi_r^{(m)}} = \left\{ \begin{array}{cc}
\sfj \bM \bm \Omega\left(\Delta f^{(m)} \right) \bP\left(\bm \theta^{(m)}\right) \bF_\otimes^* \bsfS^{(m)} \bsfg_i^{(m)} & \xi_r^{(m)} = \Delta f^{(m)} \\
\sfj \diag\left\{ \bm \Omega \left(\Delta f^{(m)}\right) \bF_\otimes^* \bsfS^{(m)} \bsfg_i^{(m)} \right\} e^{\sfj \theta_t^{(m)}[\ell]} \bee_{p[t,\ell]}^{K\Ntr} & \xi_r^{(m)} = \theta_t^{(m)}[\ell] \\
e^{\sfj \beta_i^{(m)}[k]} \bm \Omega \left( \Delta f^{(m)} \right) \bP\left(\bm \theta^{(m)} \right) \bF_\otimes^* \bsfS^{(m)} \bee_k^K & \xi_r^{(m)} = \alpha_i^{(m)}[k] \\
\sfj \sfg_i^{(m)}[k] \bm \Omega \left( \Delta f^{(m)} \right) \bP\left( \bm \theta^{(m)} \right) \bF_\otimes^* \bsfS^{(m)} \bee_k^K & \xi_r^{(m)} = \beta_i^{(m)}[k]. \\ \end{array}\right.
\end{split}
\label{equation:partial_derivatives_FIM}
\end{equation}
where $\bM \in \mathbb{C}^{K \times K}$ is given by $\bM \triangleq \bigoplus_{t=0}^{\Ntr - 1} \bM[t]$, with $\bM[t]$ given by $\bM[t] = 2\pi \bigoplus_{n=0}^{K-1}(k_0[t] + n)$.
The \ac{FIM} can be structured as
\begin{equation}
	\bI\left(\bm \xi^{(m)}\right) = \frac{2}{\sigma^2} \Real \left\{  \left[\begin{array}{ccccccc}
	i_{\Delta f^{(m)},\Delta f^{(m)}}\left(\bm \xi^{(m)}\right) & \bi_{\Delta f^{(m)},\tilde{\bsfg}^{(m)}}\left(\bm \xi^{(m)}\right) & \bi_{\Delta f^{(m)},\bm \theta^{(m)}}\left(\bm \xi^{(m)}\right) \\
	\bi_{\tilde{\bsfg}^{(m)},\Delta f^{(m)}} \left(\bm \xi^{(m)}\right) & \bI_{\tilde{\bsfg}^{(m)},\tilde{\bsfg}^{(m)}}\left(\bm \xi^{(m)}\right) & \bI_{\tilde{\bsfg}^{(m)}, \bm \theta^{(m)}}\left(\bm \xi^{(m)}\right) \\
	\bi_{\bm \theta^{(m)},\Delta f^{(m)}}\left(\bm \xi^{(m)}\right) & \bI_{\bm \theta^{(m)},\tilde{\bsfg}^{(m)}}\left(\bm \xi^{(m)}\right) & \bI_{\bm \theta^{(m)},\bm \theta^{(m)}}\left(\bm \xi^{(m)}\right) \\	  
	   \end{array}\right] \right\}.
	   \label{equation:FIM_structured}
\end{equation}
The element $i_{\Delta f^{(m)},\Delta f^{(m)}}\left(\bm \xi^{(m)}\right)$ is given by
\begin{equation}
\begin{split}
i_{\Delta f^{(m)},\Delta f^{(m)}} \left(\bm \xi^{(m)}\right) &=  \sum_{i=1}^{\Lr} \bsfg_i^{(m)*} \bsfS^{(m)*} \bF_\otimes \bM^* \bM \bF_\otimes^* \bsfS^{(m)} \bsfg_i^{(m)} \\
&= \sum_{i=1}^{\Lr}\sum_{t=0}^{\Ntr-1}\bsfg_i^{(m)*} \bsfS_t^{(m)*} \bF \bM^*[t]\bM[t] \bF^* \bsfS_t^{(m)} \bsfg_i^{(m)}.
\end{split}
\label{equation:FIM_CFO_final_form}
\end{equation}
The vector $\bi_{\Delta f^{(m)}, \tilde{\bsfg}^{(m)}}\left(\bm \xi^{(m)}\right)$ can be expressed as
\begin{equation}
\bi_{\Delta f^{(m)},\tilde{\bsfg}^{(m)}} \left(\bm \xi^{(m)}\right) = \left[\begin{array}{ccc}
\bi_{\Delta f^{(m)}, \tilde{\bsfg}_1^{(m)}}\left(\bm \xi^{(m)}\right) & \ldots & \bi_{\Delta f^{(m)}, \tilde{\bsfg}_{\Lr}^{(m)}}\left(\bm \xi^{(m)}\right) \\ \end{array}\right],
\end{equation}
with $\bi_{\Delta f^{(m)}, \tilde{\bsfg}_i^{(m)}}\left(\bm \xi^{(m)}\right)$, $1 \leq i \leq \Lr$, being given by
\begin{equation}
	\bi_{\Delta f^{(m)},\tilde{\bsfg}_i^{(m)}}\left(\bm \xi^{(m)}\right) = \left[\begin{array}{ccc}
	\bi_{\Delta f^{(m)},\tilde{\bsfg}_i^{(m)}[0]}\left(\bm \xi^{(m)}\right) & \ldots & \bi_{\Delta f^{(m)}, \tilde{\bsfg}_i^{(m)}[K-1]}\left(\bm \xi^{(m)}\right) \\ \end{array}\right],
\end{equation}
and the vector $\bi_{\Delta f^{(m)},\tilde{\bsfg}_i^{(m)}[k]}\left(\bm \xi^{(m)}\right)$, $0 \leq k \leq K-1$, is given by
\begin{equation}
\begin{split}
	\bi_{\Delta f^{(m)},\tilde{\bsfg}_i^{(m)}[k]}\left(\bm \xi^{(m)}\right) &= \left[\begin{array}{ccc}
	i_{\Delta f^{(m)}, \alpha_i^{(m)}[k]}\left(\bm \xi^{(m)}\right) & i_{\Delta f^{(m)}, \beta_i^{(m)}[k]}\left(\bm \xi^{(m)}\right) \\ \end{array}\right] \\
	&= \sum_{t=0}^{\Ntr-1} \left[\begin{array}{cc}
 	-\sfj e^{\sfj \beta_i^{(m)}[k]} & \sfg_i^{(m)}[k] \\ \end{array}\right] \bsfg_i^{(m)*} \bsfS_t^{(m)*} \bF \bM[t] \bF^* \bsfS_t^{(m)} \bee_k^K.
\end{split}
\label{equation:FIM_CFO_gains_subcarrier_final_form}
\end{equation}
Furthermore, the vector $\bi_{\Delta f^{(m)},\bm\theta^{(m)}}\left(\bm \xi^{(m)}\right)$ can be written as
\begin{equation}
	\bI_{\Delta f^{(m)},\bm \theta^{(m)}}\left(\bm \xi^{(m)}\right) = \left[\begin{array}{ccc}
	\bI_{\Delta f^{(m)},\bm \theta_0^{(m)}}\left(\bm \xi^{(m)}\right) & \ldots & \bI_{\Delta f^{(m)},\bm \theta_{\Ntr-1}^{(m)}}\left(\bm \xi^{(m)}\right) \\ \end{array}\right],
\end{equation}
where $\bi_{\Delta f^{(m)},\bm \theta_t^{(m)}}\left(\bm \xi^{(m)}\right)$, $0 \leq t \leq \Ntr-1$, is given by
\begin{equation}
	\bi_{\Delta f^{(m)},\bm \theta_t^{(m)}}\left(\bm \xi^{(m)}\right) = \left[\begin{array}{ccc}
	i_{\Delta f^{(m)},\theta_t^{(m)}[0]}\left(\bm \xi^{(m)}\right) & \ldots & i_{\Delta f^{(m)},\theta_t^{(m)}[K-1]}\left(\bm \xi^{(m)}\right) \\ \end{array}\right].
\label{equation:FIM_CFO_PN_OFDM_symbol_t_almost_final_form}
\end{equation}
Each of the terms in \eqref{equation:FIM_CFO_PN_OFDM_symbol_t_almost_final_form} can be expressed as
\begin{equation}
i_{\Delta f^{(m)},\theta_t^{(m)}[\ell]}\left(\bm \xi^{(m)}\right) = \sum_{i=1}^{\Lr} \bsfg_i^{(m)*} \bsfS_t^{(m)*} \bF \bM[t] \diag\left\{ \bF^* \bsfS_t^{(m)} \bsfg_i^{(m)} \right\} \bee_\ell^K.
\label{equation:FIM_CFO_PN_OFDM_symbol_t_final_form}
\end{equation}
The submatrix $\bI_{\tilde{\bsfg}^{(m)},\tilde{\bsfg}^{(m)}}\left(\bm \xi^{(m)}\right)$  can be expressed as
\begin{equation}
\bI_{\tilde{\bsfg}^{(m)},\tilde{\bsfg}^{(m)}}\left(\bm \xi^{(m)}\right) = \left[\begin{array}{ccc}
\bI_{\tilde{\bsfg}_1^{(m)},\tilde{\bsfg}_{\Lr}^{(m)}}\left(\bm \xi^{(m)}\right) & \ldots & \bI_{\tilde{\bsfg}_1^{(m)},\tilde{\bsfg}_{\Lr}^{(m)}}\left(\bm \xi^{(m)}\right) \\
\vdots & \ddots & \vdots \\
\bI_{\tilde{\bsfg}_{\Lr}^{(m)},\tilde{\bsfg}_1^{(m)}}\left(\bm \xi^{(m)}\right) & \ldots & \bI_{\tilde{\bsfg}_{\Lr}^{(m)},\tilde{\bsfg}_{\Lr}^{(m)}}\left(\bm \xi^{(m)}\right) \\ \end{array}\right],
\label{equation:FIM_gains_matrix}
\end{equation}
which has a particularly interesting structure from an estimation theoretic perspective. By observing the Kronecker structure in \eqref{equation:rx_signal_frame_all_RF_chains}, as well as the derivatives in \eqref{equation:partial_derivatives_FIM}, it is clear that $\bI_{\tilde{\bsfg}_i^{(m)},\tilde{\bsfg}_j^{(m)}}\left(\bm \xi^{(m)}\right) = \bm 0$ for $i \neq j$. Therefore, only the terms $\bI_{\tilde{\bsfg}_i^{(m)},\tilde{\bsfg}_i^{(m)}}\left(\bm \xi^{(m)}\right)$ are non-zero valued, which can be computed as
\begin{equation}
\bI_{\tilde{\bsfg}_i^{(m)},\tilde{\bsfg}_i^{(m)}}\left(\bm \xi^{(m)}\right) = \left[\begin{array}{ccc}
\bI_{\tilde{\bsfg}_i^{(m)}[0],\tilde{\bsfg}_i^{(m)}[0]}\left(\bm \xi^{(m)}\right) & \ldots & \bI_{\tilde{\bsfg}i^{(m)}[0],\tilde{\bsfg}_i^{(m)}[K-1]}\left(\bm \xi^{(m)}\right) \\
\vdots & \ddots & \vdots \\
\bI_{\tilde{\bsfg}_i^{(m)}[K-1],\tilde{\bsfg}_i^{(m)}[0]}\left(\bm \xi^{(m)}\right) & \ldots & \bI_{\tilde{\bsfg}_i^{(m)}[K-1],\tilde{\bsfg}^{(m)}[K-1]}\left(\bm \xi^{(m)}\right) \\ \end{array}\right].
\label{equation:FIM_gains_matrix_subcarrier}
\end{equation}
The matrix in \eqref{equation:FIM_gains_matrix_subcarrier} has a similar structure to that of \eqref{equation:FIM_gains_matrix}. From \eqref{equation:partial_derivatives_FIM} and \eqref{equation:rx_signal_frame_RF_chain}, it is observed that $\bI_{\tilde{\bsfg}_i^{(m)}[k_1],\tilde{\bsfg}_i^{(m)}[k_2]}\left(\bm \xi^{(m)}\right) = \bm 0$ if $k_1 \neq k_2$. The non-zero matrices $\bI_{\tilde{\bsfg}_i^{(m)}[k],\tilde{\bsfg}_i^{(m)}[k]}\left(\bm \xi^{(m)}\right)$ can be written as
\begin{equation}
	\bI_{\tilde{\bsfg}_i^{(m)}[k],\tilde{\bsfg}_i^{(m)}[k]}\left(\bm \xi^{(m)}\right) = \left[\begin{array}{cc}
	\bI_{\alpha_i^{(m)}[k],\alpha_i^{(m)}[k]}\left(\bm \xi^{(m)}\right) & \bI_{\alpha_i^{(m)}[k],\beta_i^{(m)}[k]}\left(\bm \xi^{(m)}\right) \\
	\bI_{\beta_i^{(m)}[k],\alpha_i^{(m)}[k]}\left(\bm \xi^{(m)}\right) & \bI_{\beta_i^{(m)}[k],\beta_i^{(m)}[k]}\left(\bm \xi^{(m)}\right) \\ \end{array}\right].
\label{equation:FIM_gains_matrix_subcarrier_gain_phase}
\end{equation}
By plugging the corresponding partial derivatives in \eqref{equation:partial_derivatives_FIM} into \eqref{equation:FIM}, the matrix in \eqref{equation:FIM_gains_matrix_subcarrier_gain_phase} is given by
\begin{equation}
	\bI_{\tilde{\bsfg}_i^{(m)}[k],\tilde{\bsfg}_i^{(m)}[k]}\left(\bm \xi^{(m)}\right) = \sum_{t=0}^{\Ntr-1}\left|[\bsfS_t^{(m)}]_{k,k}\right|^2 \underbrace{\left[\begin{array}{cc}
	1 & 0 \\
	0 & \alpha_i^{(m)2}[k] \\ \end{array}\right]}_{\bm \Lambda_i^{(m)}[k]},
\end{equation}
which shows that estimation of both the amplitude and phase of a given subchannel $\sfg_i^{(m)}[k]$ do not interfere with each other, a result that has been shown in \cite{RodGon:CFO:TWC:2019}. Let $\bsfs_k^{(m)} \in \mathbb{C}^{\Ntr \times 1}$ be the vector containing the training pilots for a given subcarrier and all the transmitted \ac{OFDM} symbols, $\bsfs_k^{(m)} = [\sfs_0^{(m)}[k], \ldots, \sfs_{\Ntr-1}^{(m)}[k]]^T$. Then, the matrix in \eqref{equation:FIM_gains_matrix_subcarrier} can be expressed as
\begin{equation}
\bI_{\tilde{\bsfg}_i^{(m)},\tilde{\bsfg}_i^{(m)}}\left(\bm \xi^{(m)}\right) =  \bigoplus_{k=0}^{K-1} \left\| \bsfs_k^{(m)} \right\|_2^2 \bm \Lambda_i^{(m)}[k],
\end{equation}
and \eqref{equation:FIM_gains_matrix} can be written as
\begin{equation}
	\bI_{\tilde{\bsfg}^{(m)},\tilde{\bsfg}^{(m)}}\left(\bm \xi^{(m)}\right) =  \bigoplus_{i=1}^{\Lr}  \bigoplus_{k=0}^{K-1} \left\| \bsfs_k^{(m)} \right\|_2^2 \bm \Lambda_i^{(m)}[k].
	\label{equation:FIM_gains_matrix_final_form}
\end{equation}
Now, the block $\bI_{\tilde{\bsfg}^{(m)},\bm \theta^{(m)}}\left(\bm \xi^{(m)}\right)$ can be expressed as
\begin{equation}
	\bI_{\tilde{\bsfg}^{(m)},\bm \theta^{(m)}}\left(\bm \xi^{(m)}\right) = \left[\begin{array}{ccc}
	\bI_{\tilde{\bsfg}_1^{(m)},\bm \theta_0^{(m)}}\left(\bm \xi^{(m)}\right) & \ldots & \bI_{\tilde{\bsfg}_1^{(m)},\bm \theta_{\Ntr-1}^{(m)}}\left(\bm \xi^{(m)}\right) \\
	\vdots & \ddots & \vdots \\
	\bI_{\tilde{\bsfg}_{\Lr}^{(m)},\bm \theta_0^{(m)}}\left(\bm \xi^{(m)}\right) & \ldots & \bI_{\tilde{\bsfg}_{\Lr}^{(m)},\bm \theta_{\Ntr-1}^{(m)}}\left(\bm \xi^{(m)}\right) \\ \end{array}\right],
\end{equation}
wherein the blocks $\bI_{\tilde{\bsfg}_i^{(m)},\bm \theta_t^{(m)}}\left(\bm \xi^{(m)}\right)$ are of the form
\begin{equation}
	\bI_{\tilde{\bsfg}_i^{(m)},\bm \theta_t^{(m)}}\left(\bm \xi^{(m)}\right) = \left[\begin{array}{c}
	\bI_{\tilde{\bsfg}_i^{(m)}[0],\bm \theta_t^{(m)}}\left(\bm \xi^{(m)}\right) \\
	\vdots \\
	\bI_{\tilde{\bsfg}_i^{(m)}[K-1],\bm \theta_t^{(m)}}\left(\bm \xi^{(m)}\right) \\ \end{array}\right],
\end{equation}
with $\bI_{\tilde{\bsfg}_i^{(m)}[k],\bm \theta_t^{(m)}}\left(\bm \xi^{(m)}\right)$ given by
\begin{equation}
	\bI_{\tilde{\bsfg}_i^{(m)}[k],\bm \theta_t^{(m)}}\left(\bm \xi^{(m)}\right) = \left[\begin{array}{ccc}
	\bi_{\tilde{\bsfg}_i^{(m)}[k],\theta_t^{(m)}[0]}\left(\bm \xi^{(m)}\right) & \ldots & \bi_{\tilde{\bsfg}_i^{(m)}[k],\theta_t^{(m)}[K-1]}\left(\bm \xi^{(m)}\right) \\ \end{array}\right].
	\label{equation:FIM_gains_subcarrier_mag_phase_PN}
\end{equation}
Let $\bsfs_t^{(m)} \in \mathbb{C}^{K \times K}$ be the column vector containing the training pilots for the $t$-th transmitted \ac{OFDM} symbol. This vector is given by $\bsfs_t^{(m)} = \vect\{\diag\{\bsfS_t^{(m)}\}\}$. Furthermore, let $\bff_{\ell} \in \mathbb{C}^{K\times 1}$ be the $\ell$-th column in the \ac{DFT} matrix $\bF$. Then, plugging the corresponding derivatives from \eqref{equation:partial_derivatives_FIM} into \eqref{equation:FIM} allows expressing each column in \eqref{equation:FIM_gains_subcarrier_mag_phase_PN} as
\begin{equation}
	\bI_{\tilde{\bsfg}_i^{(m)}[k],\theta_t^{(m)}[\ell]}\left(\bm \xi^{(m)}\right) =  \Real \left\{ \left[\begin{array}{c}
	\sfj e^{-\sfj \beta_i^{(m)}[k]} \\
	\sfg_i^{(m)\text{C}}[k] \\ \end{array}\right] \left(\bee_k^K \right)^T \left(\bsfs_t^{(m)\text{C}} \circ \bff_{\ell}\right) \left(\bsfs_t^{(m)T} \circ \bff_{\ell}^*\right)\bsfg_i^{(m)} \right\}.
	\label{equation:FIM_gains_subcarrier_mag_phase_PN_col_final_form}
\end{equation}
Finally, the block matrix $\bI_{\bm \theta^{(m)},\bm \theta^{(m)}}\left(\bm \xi^{(m)}\right)$ can be expressed in the form
\begin{equation}
	\bI_{\bm \theta^{(m)},\bm \theta^{(m)}}\left(\bm \xi^{(m)}\right) = \left[\begin{array}{ccc}
	\bI_{\bm \theta_0^{(m)},\bm \theta_0^{(m)}}\left(\bm \xi^{(m)}\right) & \ldots & \bI_{\bm \theta_0^{(m)},\bm \theta_{\Ntr-1}^{(m)}}\left(\bm \xi^{(m)}\right) \\
	\vdots & \ddots & \vdots \\
	\bI_{\bm \theta_{\Ntr-1}^{(m)},\bm \theta_0^{(m)}}\left(\bm \xi^{(m)}\right) & \ldots & \bI_{\bm \theta_{\Ntr-1}^{(m)},\bm \theta_{\Ntr-1}^{(m)}}\left(\bm \xi^{(m)}\right) \\ \end{array}\right].
	\label{equation:FIM_PN}
\end{equation}
Owing to the structure of the partial derivative of $\bm \mu_i^{(m)}\left(\bm \xi^{(m)}\right)$ with respect to $\theta_t^{(m)}[\ell]$ in \eqref{equation:partial_derivatives_FIM}, the different matrices $\bI_{\bm \theta_t^{(m)},\bm \theta_u^{(m)}}\left(\bm \xi^{(m)}\right)$ can be checked to be zero-valued for $t \neq u$. Further, the matrices in the main block diagonal of $\bI_{\bm \theta^{(m)},\bm \theta^{(m)}}\left(\bm \xi^{(m)}\right)$ can be expressed as
\begin{equation}
	\bI_{\bm \theta_t^{(m)},\bm \theta_t^{(m)}}\left(\bm \xi^{(m)}\right) = \left[ \begin{array}{ccc}
	i_{\theta_t^{(m)}[0],\theta_t^{(m)}[0]}\left(\bm \xi^{(m)}\right) & \ldots & i_{\theta_t^{(m)}[0],\theta_t^{(m)}[K-1]}\left(\bm \xi^{(m)}\right) \\
	\vdots & \ddots & \vdots \\
	i_{\theta_t^{(m)}[K-1],\theta_t^{(m)}[0]}\left(\bm \xi^{(m)}\right) & \ldots & i_{\theta_t^{(m)}[K-1],\theta_t^{(m)}[K-1]}\left(\bm \xi^{(m)}\right) \\ \end{array}\right],
	\label{equation:FIM_PN_OFDM_symbols}
\end{equation}
which again, due to the structure of the partial derivative of $\bm \mu_i^{(m)}\left(\bm \xi^{(m)}\right)$ with respect to $\theta_t^{(m)}[\ell]$ in \eqref{equation:partial_derivatives_FIM}, is a diagonal matrix given by
\begin{equation}
	\bI_{\bm \theta_t^{(m)},\bm \theta_t^{(m)}}\left(\bm \xi^{(m)}\right) = \bigoplus_{k=0}^{K-1}\left(\sum_{i=1}^{\Lr} \left\| \diag\left\{ \bF^* \bsfS_t^{(m)}\bsfg_i^{(m)} \right\} \bee_{p[t,k]}^{K\Ntr} \right\|_2^2 \right).
	\label{equation:FIM_PN_OFDM_symbol_t}
\end{equation}
Using \eqref{equation:FIM_PN_OFDM_symbol_t}, the matrix in \eqref{equation:FIM_PN} can be expressed as
\begin{equation}
	\bI_{\bm \theta^{(m)},\bm \theta^{(m)}}\left(\bm \xi^{(m)}\right) = \bigoplus_{t=0}^{\Ntr-1} \bigoplus_{k=0}^{K-1} \left(\sum_{i=1}^{\Lr} \left\| \diag\left\{ \bF^* \bsfS_t^{(m)}\bsfg_i^{(m)} \right\} \bee_{p[t,k]}^{K\Ntr} \right\|_2^2\right).
	\label{equation:FIM_PN_final_form}
\end{equation}
Due to the structure of the \ac{FIM}, the matrices below the main block diagonal in \eqref{equation:FIM_structured} are given by $\bI_{\tilde{\bsfg}^{(m)},\Delta f^{(m)}}\left(\bm \xi^{(m)}\right) = \bI_{\Delta f^{(m)},\tilde{\bsfg}^{(m)}}^T\left(\bm \xi^{(m)}\right)$, $\bI_{\bm \theta^{(m)},\Delta f^{(m)}}\left(\bm \xi^{(m)}\right) = \bI_{\Delta f^{(m)},\bm \theta^{(m)}}^T\left(\bm \xi^{(m)}\right)$, and $\bI_{\bm \theta^{(m)},\tilde{\bsfg}^{(m)}}\left(\bm \xi^{(m)}\right) = \bI_{\tilde{\bsfg}^{(m)},\bm \theta^{(m)}}^T\left(\bm \xi^{(m)}\right)$.

Finally, to obtain $\bI_\text{D}\left(\bm \xi^{(m)}\right)$, notice that the terms in \eqref{equation:FIM_CFO_final_form}, \eqref{equation:FIM_CFO_gains_subcarrier_final_form}, \eqref{equation:FIM_CFO_PN_OFDM_symbol_t_final_form}, \eqref{equation:FIM_gains_matrix_final_form}, \eqref{equation:FIM_gains_subcarrier_mag_phase_PN_col_final_form} and \eqref{equation:FIM_PN_final_form} do not depend on $\bm \theta^{(m)}$ since the \ac{PN} exponentials get canceled by their conjugates. Hence, there is no need to calculate the explicit expectation of $\bI\left(\bm \xi^{(m)}\right)$ over $\bm \theta^{(m)}$, and $\bI_\text{D}\left(\bm \xi^{(m)}\right) = \bI\left(\bm \xi^{(m)}\right)$.

Now, the only matrix left to compute in order to find the \ac{HIM} $\bH\left(\bm \xi^{(m)}\right)$ is $\bI_\text{P}\left(\bm \xi^{(m)}\right)$ in \eqref{equation:HIM}. From the expression in \eqref{equation:PIM}, since no prior knowledge on either $\tilde{\bsfg}^{(m)}$ or $\Delta f^{(m)}$ is assumed, $\bI_\text{P}\left(\bm \xi^{(m)}\right)$ is structured as
\begin{equation}
\begin{split}
	\bI_\text{P}\left(\bm \xi^{(m)}\right) &\triangleq -\left[\begin{array}{ccc}
	\mathbb{E}_{\bm \theta^{(m)}}\left\{ \frac{\partial^2 \log{p\left(\bm \theta^{(m)}\right)}}{\partial \Delta f^{(m)2}} \right\} & \mathbb{E}_{\bm \theta^{(m)}}\left\{ \frac{\partial^2 \log{p\left(\bm \theta^{(m)}\right)}}{\partial \Delta f^{(m)} \partial \tilde{\bsfg}^{(m)T}} \right\} & \mathbb{E}_{\bm \theta^{(m)}}\left\{ \frac{\partial^2 \log{p\left(\bm \theta^{(m)}\right)}}{\partial \Delta f^{(m)} \bm \theta^{(m)T}} \right\} \\ 
	\mathbb{E}_{\bm \theta^{(m)}}\left\{ \frac{\partial^2 \log{p\left(\bm \theta^{(m)}\right)}}{\partial \tilde{\bsfg}^{(m)} \partial \Delta f^{(m)}} \right\} & \mathbb{E}_{\bm \theta^{(m)}}\left\{ \frac{\partial^2 \log{p\left(\bm \theta^{(m)}\right)}}{\partial \tilde{\bsfg}^{(m)} \partial \tilde{\bsfg}^{(m)T} } \right\} & \mathbb{E}_{\bm \theta^{(m)}}\left\{ \frac{\partial^2 \log{p\left(\bm \theta^{(m)}\right)}}{\partial \tilde{\bsfg}^{(m)} \bm \theta^{(m)T}} \right\} \\
	\mathbb{E}_{\bm \theta^{(m)}}\left\{ \frac{\partial^2 \log{p\left(\bm \theta^{(m)}\right)}}{\partial \bm \theta^{(m)} \partial \Delta f^{(m)}} \right\} & \mathbb{E}_{\bm \theta^{(m)}}\left\{ \frac{\partial^2 \log{p\left(\bm \theta^{(m)}\right)}}{\partial \bm \theta^{(m)} \partial \tilde{\bsfg}^{(m)T} } \right\} & \mathbb{E}_{\bm \theta^{(m)}}\left\{ \frac{\partial^2 \log{p\left(\bm \theta^{(m)}\right)}}{\partial \bm \theta^{(m)} \bm \theta^{(m)T}} \right\} \\ \end{array}\right] \\
	&= -\left[\begin{array}{ccc}
	\bm 0 & \bm 0 & \bm 0 \\
	\bm 0 & \bm 0 & \bm 0 \\
	\bm 0 & \bm 0 & \mathbb{E}_{\bm \theta^{(m)}}\left\{ \frac{\partial^2 \log{p\left(\bm \theta^{(m)}\right)}}{\partial \bm \theta^{(m)} \bm \theta^{(m)T}} \right\} \\ \end{array}\right],
\end{split}
\label{equation:PIM_full}
\end{equation}
where the last equality comes from the \ac{PDF} of the \ac{PN} being independent of the \ac{CFO} and equivalent channel gains. The \ac{LLF} of the \ac{PN} is given by
\begin{equation}
	\log{p\left(\bm \theta^{(m)}\right)} = -\frac{K\Ntr}{2}\log{(2\pi)} - \frac{\Ntr}{2}\log{\det\{\bC_{\bm \theta^{(m)},\bm \theta^{(m)}}\}} - \frac{1}{2} \bm \theta^{(m)T} \bC_{\bm \theta^{(m)},\bm \theta^{(m)}}^{-1} \bm \theta^{(m)},
\end{equation}
and its Hessian reads
\begin{equation}
\frac{\partial^2 \log{p\left(\bm \theta^{(m)}\right)}}{\partial \bm \theta^{(m)}\bm \theta^{(m)T}} = -\bC_{\bm \theta^{(m)},\bm \theta^{(m)}}^{-1}.
\end{equation}
Therefore, by combining $\bI_\text{D}\left(\bm \xi^{(m)}\right)$ and $\bI_\text{P}\left(\bm \xi^{(m)}\right)$, the \ac{HIM} $\bH\left(\bm \xi^{(m)}\right)$ is obtained as
\begin{equation}
\begin{split}
	\bH \left(\bm \xi^{(m)}\right) &= \left[\begin{array}{ccc}
	i_{\text{H},1,1}\left(\bm \xi^{(m)}\right) & \bi_{\text{H},1,2}^T\left(\bm \xi^{(m)}\right) & \bi_{\text{H},1,3}\left(\bm \xi^{(m)}\right) \\
	\bi_{\text{H},2,1}\left(\bm \xi^{(m)}\right) & \bI_{\text{H},2,2}\left(\bm \xi^{(m)}\right) & \bI_{\text{H},2,3}\left(\bm \xi^{(m)}\right) \\
	\bi_{\text{H},3,1}\left(\bm \xi^{(m)}\right) & \bI_{\text{H},3,2}\left(\bm \xi^{(m)}\right) & \bI_{\text{H},3,3}\left(\bm \xi^{(m)}\right) \\ \end{array}\right] \\
	&= \left[\begin{array}{ccc}
	i_{\Delta f^{(m)},\Delta f^{(m)}}\left(\bm \xi^{(m)}\right) & \bi_{\Delta f^{(m)},\tilde{\bsfg}^{(m)}}\left(\bm \xi^{(m)}\right) & \bi_{\Delta f^{(m)},\bm \theta^{(m)}}\left(\bm \xi^{(m)}\right) \\
	\bi_{\tilde{\bsfg}^{(m)},\Delta f^{(m)}} \left(\bm \xi^{(m)}\right) & \bI_{\tilde{\bsfg}^{(m)},\tilde{\bsfg}^{(m)}}\left(\bm \xi^{(m)}\right) & \bI_{\tilde{\bsfg}^{(m)}, \bm \theta^{(m)}}\left(\bm \xi^{(m)}\right) \\
	\bi_{\bm \theta^{(m)},\Delta f^{(m)}}\left(\bm \xi^{(m)}\right) & \bI_{\bm \theta^{(m)},\tilde{\bsfg}^{(m)}}\left(\bm \xi^{(m)}\right) & \bI_{\bm \theta^{(m)},\bm \theta^{(m)}}\left(\bm \xi^{(m)}\right) + \bC_{\bm \theta^{(m)},\bm \theta^{(m)}}^{-1} \\	  
	   \end{array}\right].
	   \end{split}
\end{equation}
Finally, the hybrid \ac{CRLB} is given by the inverse of the \ac{HIM}, $\bH^{-1}\left(\bm \xi^{(m)}\right)$. In particular, using the formula for the inverse of block matrices \cite{Kay:Fundamentals-of-Statistical-Signal:93}, the hybrid \ac{CRLB} for the \ac{CFO} can be found as follows. Let $\tilde{i}_{{\Delta f}^{(m)},{\Delta f}^{(m)}}\left(\bm \xi^{(m)}\right) \in \mathbb{R}$, $\tilde{\bI}_{\tilde{\bsfg}^{(m)},\tilde{\bsfg}^{(m)}} \left(\bm \xi^{(m)}\right) \in \mathbb{R}^{2K\Lr \times 2K\Lr}$ and $\bx\left(\bm \xi^{(m)}\right) \in \mathbb{R}^{2K\Lr \times 1}$ denote the \ac{HIM} for the \ac{CFO} parameter when the channel $\tilde{\bsfg}$ is known, the \ac{HIM} for the channels when the \ac{CFO} is known, and a vector accounting for the coupling between the \ac{PN}, channel, and \ac{CFO} parameters. These parameters are given by 
\begin{equation}
	\tilde{i}_{{\Delta f}^{(m)},{\Delta f}^{(m)}}\left(\bm \xi^{(m)}\right) = i_{\text{H},1,1}\left(\bm \xi^{(m)}\right) - \bi_{\text{H},1,3}\left(\bm \xi^{(m)}\right) \bI_{\text{H},3,3}^{-1}\left(\bm \xi^{(m)}\right) \bi_{\text{H},3,1}\left(\bm \xi^{(m)}\right)
\end{equation}
\begin{equation}
	\tilde{\bI}_{\tilde{\bsfg}^{(m)},\tilde{\bsfg}^{(m)}}\left(\bm \xi^{(m)}\right) = \bI_{\text{H},2,2}\left(\bm \xi^{(m)}\right) - \bI_{\text{H},2,3}\left(\bm \xi^{(m)}\right) \bI_{\text{H},3,3}\left(\bm \xi^{(m)}\right) \bI_{\text{H},3,2}\left(\bm \xi^{(m)}\right).
\end{equation}
\begin{equation}
	\bx \left(\bm \xi^{(m)}\right) = \bi_{\text{H},1,2}\left(\bm \xi^{(m)}\right) - \bi_{\text{H},2,3}\left(\bm \xi^{(m)}\right) \bI_{\text{H},3,3}^{-1}\left(\bm \xi^{(m)}\right)\bI_{\text{H},2,3}\left(\bm \xi^{(m)}\right).
\end{equation}
Then, the hybrid \ac{CRLB} for any unbiased estimator of $\Delta f^{(m)}$, $\tilde{\bsfg}^{(m)}$ are given by
\begin{equation}
	\var\left\{ \widehat{\Delta f}^{(m)} \right\} \geq \frac{1}{\tilde{i}_{\Delta f}^{(m)}\left(\bm \xi^{(m)}\right) - \bx^T \left(\bm \xi^{(m)}\right) \bI_{\tilde{\bsfg}^{(m)},\tilde{\bsfg}^{(m)}}^{-1}\left(\bm \xi^{(m)}\right) \bx \left(\bm \xi^{(m)}\right) },
\label{equation:HCRLB_CFO}
\end{equation}
\begin{equation}
	\covar\left\{ \hat{\tilde{\bsfg}}^{(m)},\hat{\tilde{\bsfg}}^{(m)} \right\} \geq \tilde{\bI}_{\tilde{\bsfg}^{(m)},\tilde{\bsfg}^{(m)}}^{-1}\left(\bm \xi^{(m)}\right) + \frac{  \bI_{\tilde{\bsfg}^{(m)},\tilde{\bsfg}^{(m)}}^{-1}\left(\bm \xi^{(m)}\right) \bx \left(\bm \xi^{(m)}\right)  \bx^T \left(\bm \xi^{(m)}\right) \bI_{\tilde{\bsfg}^{(m)},\tilde{\bsfg}^{(m)}}^{-1}\left(\bm \xi^{(m)}\right)}{\tilde{i}_{\Delta f}^{(m)}\left(\bm \xi^{(m)}\right) - \bx^T \left(\bm \xi^{(m)}\right) \bI_{\tilde{\bsfg}^{(m)},\tilde{\bsfg}^{(m)}}^{-1}\left(\bm \xi^{(m)}\right) \bx \left(\bm \xi^{(m)}\right) }.
	\label{equation:HCRLB_channels}
\end{equation}

\section{Estimation of beamformed channels and high-dimensional MIMO channel}
\label{subsec:ML_estimation_broadband}

In this section, I formulate and present novel solutions to the problem of estimating both the \ac{CFO}, the equivalent frequency-selective beamformed channels, and the \ac{PN} vector for the signal model in Section \ref{sec:system_model_broadband}. Then, I formulate the problem of estimating the high-dimensional frequency-selective \ac{mmWave} \ac{MIMO} channel $\{\bsfH[k]\}_{k=0}^{K-1}$ from the estimates of the equivalent channel accounting for both the estimates for these parameters and their hybrid \ac{CRLB}. Since prior statistical information on the \ac{PN} vector is available, it is well-known that the optimum estimator for the \ac{PN} is the \ac{MMSE} estimator, which is well-known to be unbiased and attain the hybrid \ac{CRLB}. The main problem concerning applying the \ac{MMSE} estimator is that it requires knowledge of the \ac{CFO} and the equivalent channels, which is not available a priori. Another strategy to find the \ac{PN} vector relies on using the \ac{MAP} estimator, which is attractive due to its simplicity, but it present the drawback of being, in general, biased. Due to this, the application of the \ac{MAP} estimator may well lead to the different estimates $\hat{\bsfg}_i^{(m)}$, $1 \leq i \leq \Lr$, $1 \leq m \leq M$ having random phase errors that could destroy incoherence in the measurements, thereby invalidating the application of \ac{CS}-based algorithms to retrieve the frequency-selective channel $\{ \bsfH[k]\}_{k=0}^{K-1}$. For this reason, it is crucial to consider an unbiased estimator for the different parameters. Owing to the difficulty in finding a closed-form solution for the estimation of $\Delta f^{(m)}$, $\{\bsfg_i^{(m)}\}_{i=1}^{\Lr}$, and $\{\bm \theta_t^{(m)}\}_{t=0}^{\Ntr-1}$, I propose to use the \ac{EM} approach \cite{Kay:Fundamentals-of-Statistical-Signal:93} to find these estimators. I will show that this leads to finding the \ac{MMSE} estimator for the \ac{PN} impairment, parameterized by the current estimates of the unknown \ac{CFO} and equivalent channels, which can be computed as it will soon become apparent. The \ac{EM} method is a well-known iterative approach to find the \ac{ML} estimators for unknown parameters when the \ac{LLF} is unknown, and hence impossible to optimize directly. The first proposed algorithm aims at finding the \ac{LMMSE} estimator for the \ac{PN} by batch processing the $\Lr K \Ntr$ received measurements at once, thereby providing very good performance. The second proposed algorithm also aims at finding the \ac{LMMSE} estimator for the \ac{PN} but, unlike the first proposed algorithm, it processes the received measurements in sets of $\Lr$ samples to reduce computational complexity.

\subsection{LMMSE-EM Algorithm}

In this subsection, I present the first proposed algorithm to find the \ac{ML} estimates for the \ac{CFO} and the equivalent beamformed channels using the \ac{EM} iterative estimation approach. At each iteration, this algorithm processes all the $\Lr K \Ntr$ received measurements using single-shot estimation to find a closed-form solution to the problem of estimating the \ac{PN} vector. The \ac{EM} algorithm consists of two steps:
\begin{itemize}
\item \textbf{E-step}: in the first step of the \ac{EM} algorithm, the posterior expected value of the joint \ac{LLF} of $\br^{(m)}$ and $\bm \xi_\text{R}^{(m)}$ is computed. Let us consider a partition of the vector of parameters to be estimated, $\bm \xi^{(m)}$, into a vector of deterministic parameters $\bm \xi_\text{D}^{(m)} = [\Delta f^{(m)}, \bsfg_1^{(m)T}, \ldots, \bsfg_{\Lr}^{(m)T}]^T$, and a vector of random parameters $\bm \xi_\text{R}^{(m)} = [\bm \theta_0^{(m)T}, \ldots, \bm \theta_{\Ntr-1}^{(m)T}]^T$. Then, the expectation step for the $n$-th step can be formalized as
\begin{equation}
	Q\left(\bm \xi_\text{D}^{(m)}, \hat{\bm \xi}_\text{D}^{(m,n-1)}\right) \triangleq \mathbb{E}_{\bm \xi_\text{R}^{(m)} | \br^{(m)}, \hat{\bm \xi}_\text{D}^{(m,n-1)}} \left\{ \log{ p\left(\br^{(m)}, \bm \xi_\text{R}^{(m)}; \bm \xi_\text{D}^{(m)}\right)} \right\},
	\label{equation:Q_function_EM}
\end{equation}
where $\hat{\bm \xi}_\text{D}^{(m,n)}$ is the estimate of $\bm \xi_\text{D}^{(m)}$ found at the $n$-th iteration of the algorithm.
\item \textbf{M-step}: this step consists of finding $\hat{\bm \xi}_\text{D}^{(m,n)}$, which is defined as the maximizer of the function found during the E-step. The maximization step is formalized as
\begin{equation}
	\hat{\bm \xi}_\text{D}^{(m,n)} = \underset{\bm \xi_\text{D}^{(m)}}{\arg\,\max\,}Q\left(\bm \xi_\text{D}^{(m)}, \hat{\bm \xi}_\text{D}^{(m,n-1)}\right).
	\label{equation:M_step}
\end{equation}
Due to the independence of the \ac{PN} sequence on the deterministic parameters in $\bm \xi_\text{D}^{(m)}$, the $Q$ function in \eqref{equation:Q_function_EM} can be expressed as
\begin{equation}
	Q\left(\bm \xi_\text{D}^{(m)}, \hat{\bm \xi}_\text{D}^{(m,n)}\right) = -\frac{1}{\sigma^2}\sum_{i=1}^{\Lr}\left\| \br_i^{(m)} - \bm \Omega\left(\Delta f^{(m)}\right) \bP\left(\hat{\bm \theta}_\text{MMSE}^{(m,n)}\right) \bF_\otimes^* \bsfS^{(m)} \bsfg_i^{(m)}\right\|_2^2,
	\label{equation:Q_function_final}
\end{equation}
where $\hat{\bm \theta}_\text{MMSE}^{(m,n)} \triangleq \mathbb{E}_{\bm \xi_\text{R}^{(m)} | \br^{(m)}, \hat{\bm \xi}_\text{D}^{(m,n-1)}}\{\bm \theta^{(m)}\}$ is the \ac{MMSE} estimator of the \ac{PN} sequence found during the $n$-th E-step.
\end{itemize}

Finding the \ac{MMSE} estimator of the \ac{PN} sequence requires finding the posterior \ac{PDF} of the \ac{PN} sequence, given the received measurements $\br_i^{(m)}$, $1 \leq i \leq \Lr$. Finding this \ac{PDF}, however, requires multi-dimensional integration over the joint \ac{PDF} of the received measurements and the \ac{PN} sequence, which is difficult to find, in general. For this reason, I propose another approach to estimate the \ac{PN} as follows. Exploiting the fact that the \ac{PN} sequence typically has small amplitude \cite{Cabric:CompressiveIA:JSTSP:2019}, I use a first-order Taylor series approximation to linearize the received measurement with respect to the \ac{PN} sequence around the expected value of $\bm \theta^{(m)}$, given by $\bm \mu_{\bm \theta^{(m)}}$, as
\begin{equation}
	\br^{(m)} \approx \underbrace{\left( \bI_{\Lr} \otimes \bm \Omega\left(\Delta f^{(m)}\right) \bP\left(\bm \theta^{(m)}\right) \bF_\otimes^* \bsfS^{(m)} \right) \bsfg^{(m)}}_{\bh\left(\bm \theta^{(m)},\bm \xi_\text{D}^{(m)}\right)} + \bm \nabla_\bff\left(\bm \theta^{(m)}\right){\bigg{|}}_{\bm \theta^{(m)} = \bm \mu_{\bm \theta^{(m)}}} \bm \theta^{(m)} + \bv^{(m)},
	\label{equation:linear_received_measurements}
\end{equation}
where $\bm \nabla_\bff\left(\bm \theta^{(m)},\bm \xi_\text{D}^{(m)}\right)$ is the Jacobian matrix of $\bff\left(\bm \theta^{(m)},\bm \xi_\text{D}^{(m)}\right):\mathbb{R}^{K\Lr\Ntr}\to \mathbb{C}^{K\Lr\Ntr}$, $\bff\left(\bm \theta^{(m)},\bm \xi_\text{D}^{(m)}\right) = \left(\bI_{\Lr} \otimes \bm \Omega\left(\Delta f^{(m)}\right)\right) \bP\left(\bm \theta^{(m)}\right) \left(\bI_{\Lr} \otimes \bF_\otimes^* \bsfS^{(m)}\right) \bsfg^{(m)}$, which is given by
\begin{equation}
	\bm \nabla_\bff\left(\bm \theta^{(m)}\right) = \left[\begin{array}{ccc}
	\frac{\partial \bff\left(\bm \theta^{(m)},\bm \xi_\text{D}^{(m)}\right)}{\partial \bm \theta_0^{(m)T}} & \ldots & \frac{\partial \bff\left(\bm \theta^{(m)},\bm \xi_\text{D}^{(m)}\right)}{\partial \bm \theta_{\Ntr-1}^{(m)T}} \\ \end{array}\right].
	\label{equation:jacobian_F_PN}
\end{equation}
Each of the submatrices $\frac{\partial \bff\left(\bm \theta^{(m)},\bm \xi_\text{D}^{(m)}\right)}{\partial \bm \theta_t^{(m)T}} \in \mathbb{C}^{K\Lr \Ntr \times K}$ is given by
\begin{equation}
	\frac{\partial \bff\left(\bm \theta^{(m)},\bm \xi_\text{D}^{(m)}\right)}{\partial \bm \theta_t^{(m)T}} = \left[\begin{array}{ccc}
	\frac{\partial \bff\left(\bm \theta^{(m)},\bm \xi_\text{D}^{(m)}\right)}{\partial \theta_{t,0}^{(m)}} & \ldots & \frac{\partial \bff\left(\bm \theta^{(m)},\bm \xi_\text{D}^{(m)}\right)}{\partial \theta_{t,K-1}^{(m)}} \\ \end{array}\right],
	\label{equation:derivative_PN_total} 
\end{equation}
wherein $\frac{\partial \bff\left(\bm \theta^{(m)}\right)}{\partial \theta_{t,k}^{(m)}}$ is given by
\begin{equation}
\begin{split}
\frac{\partial \bff\left(\bm \theta^{(m)},\bm \xi_\text{D}^{(m)}\right)}{\partial \theta_{t,k}^{(m)}} &= \sfj e^{\sfj \theta_{t,k}^{(m)}} \left(\bI_{\Lr} \otimes \bee_{p[t,\ell]}^{K\Ntr} \left(\bee_{p[t,\ell]}^{K\Ntr}\right)^T \bm \Omega\left(\Delta f^{(m)}\right) \bF_\otimes^* \bsfS^{(m)}\right) \bsfg^{(m)}\\
&= \sfj e^{\sfj \theta_{t,k}^{(m)}} \left[\begin{array}{c}
 \diag\left\{ \bm \Omega\left(\Delta f^{(m)}\right) \bF_\otimes^* \bsfS^{(m)} \bsfg_1^{(m)} \right\} \bee_{p[t,\ell]}^{K\Ntr} \\
 \vdots \\
  \diag\left\{ \bm \Omega\left(\Delta f^{(m)}\right) \bF_\otimes^* \bsfS^{(m)} \bsfg_{\Lr}^{(m)} \right\} \bee_{p[t,\ell]}^{K\Ntr} \\ \end{array}\right]
\end{split} 
\label{equation:derivative_PN_each_symbol}
\end{equation}
Using \eqref{equation:jacobian_F_PN}-\eqref{equation:derivative_PN_each_symbol}, the time and measurement update equations for the estimation of $\bm \xi_\text{R}^{(m)} = \bm \theta^{(m)}$ at the $n$-th E-step are given by
\begin{itemize}
\item \textbf{Time update}
\begin{equation}
\begin{split}
	\hat{\bm \theta}_\text{TU}^{(m,n)} &= \bm \mu_{\bm \theta}^{(m)} \\
	\hat{\bC}_{\hat{\bm \theta}_\text{TU}^{(m)},\hat{\bm \theta}_\text{TU}^{(m)}}^{(n)} &= \bC_{\bm \theta^{(m)},\bm \theta^{(m)}}^{(m)}.
\end{split}
\label{equation:TU_LMMSE}
\end{equation}
\item \textbf{Measurement update}
\begin{equation}
\begin{split}
	\hat{\bC}_{\bm \theta^{(m)},\br^{(m)}}^{(n)} &= \hat{\bC}_{\hat{\bm \theta}_\text{TU}^{(m)},\hat{\bm \theta}_\text{TU}{(m)}}^{(n)} \bm \nabla_\bff^*\left(\hat{\bm \theta}_\text{TU}^{(m,n)},\hat{\bm \xi}_\text{D}^{(m,n-1)}\right) \\
	\hat{\bC}_{\br^{(m)},\br^{(m)}}^{(n)} &= \bm \nabla_\bff\left(\hat{\bm \theta}_\text{TU}^{(m,n)},\hat{\bm \xi}_\text{D}^{(m,n-1)} \right) \hat{\bC}_{\hat{\bm \theta}_\text{TU}^{(m)},\hat{\bm \theta}_\text{TU}^{(m)}}^{(n)} \bm \nabla_\bff^*\left(\hat{\bm \theta}_\text{TU}^{(m,n)}, \hat{\bm \xi}_\text{D}^{(m,n-1)} \right)  + \sigma^2 \bI_{K \Ntr \Lr} \\ 
	\hat{\bm \theta}_\text{MU}^{(m,n)} &= \hat{\bm \theta}_\text{TU}^{(m,n)} + \hat{\bC}_{\bm \theta^{(m)},\br^{(m)}}^{(n)} \left(\hat{\bC}_{\br^{(m)},\br^{(m)}}^{(n)}\right)^{-1} \left(\br^{(m)} - \bh\left(\hat{\bm \theta}_\text{TU}^{(m)},\hat{\bm \xi}_\text{D}^{(m,n-1)}\right) \right) \\
	\hat{\bC}_{\hat{\bm \theta}_\text{MU}^{(m)},\hat{\bm \theta}_\text{MU}^{(m)}}^{(n)} &= \hat{\bC}_{\hat{\bm \theta}_\text{TU}^{(m)},\hat{\bm \theta}_\text{TU}^{(m)}}^{(n)} - \hat{\bC}_{\bm \theta^{(m)},\br^{(m)}}^{(n)} \left(\hat{\bC}_{\br^{(m)},\br^{(m)}}^{(n)}\right)^{-1} \hat{\bC}_{\br^{(m)},\bm \theta^{(m)}}.
\end{split}
	\label{equation:MU_LMMSE}
\end{equation}
\end{itemize}
Finally, motivated by the linearization in \eqref{equation:linear_received_measurements} and the assumption that the \ac{PN} sequence is Gaussian \cite{TSP_PN,Channel_PN_HCRLB}, the \ac{MMSE} estimator for the \ac{PN} sequence at the $n$-th E-step is substituted by the approximate \ac{LMMSE} estimate obtained by the \ac{EKF} recursions in \eqref{equation:TU_LMMSE}-\eqref{equation:MU_LMMSE}. 

Then, the optimum \ac{ML} estimator found during the $n$-th M-step is found by maximizing \eqref{equation:Q_function_EM}. Optimizing \eqref{equation:Q_function_EM} directly is, however, computationally complex because of the lack of closed-form solutions for the estimation of $\Delta f^{(m)}$ \cite{Asilomar_2018}. Therefore, to circumvent this issue, I propose to reduce the complexity associated with the M-step by carrying out the optimization in \eqref{equation:M_step} with respect to one of the parameters while keeping the remaining parameters at their most recently updated values. First, by using the equivalent channel estimates at the $(n-1)$-th E-step, $\hat{\bsfg}^{(m,n-1)}$, and the \ac{PN} vector estimate from the E-step, $\hat{\bm \theta}_\text{MU}^{(m,n)}$, the function in \eqref{equation:Q_function_EM} is maximized with respect to $\Delta f^{(m)}$ to obtain the estimate for the $n$-th iteration, $\widehat{\Delta f}^{(m,n)}$ as
\begin{equation}
\widehat{\Delta f}_\text{ML}^{(m,n)} = \underset{\Delta f^{(m)}}{\arg\,\min\,}\sum_{i=1}^{\Lr}\left\| \br_i^{(m)} - \bm \Omega\left(\Delta f^{(m)}\right) \bP\left(\hat{\bm \theta}_\text{MU}^{(m,n)}\right) \bF_\otimes^* \bsfS^{(m)} \hat{\bsfg}_i^{(m,n-1)}\right\|_2^2.
\label{equation:optimize_CFO_M_step}
\end{equation}
After simplifying \eqref{equation:optimize_CFO_M_step}, it is obtained that
\begin{equation}
	\widehat{\Delta f}_\text{ML}^{(m,n)} = \underset{\Delta f^{(m)}}{\arg\,\max\,} \sum_{i=1}^{\Lr} \sum_{t=0}^{\Ntr-1} \Real \left\{ \br_{i,t}^{(m)*} \bm \Omega_t\left(\Delta f^{(m)}\right) \bP_t\left( \hat{\bm \theta}_{t,\text{MU}}^{(m,n)}\right) \bF^* \bsfS_t^{(m)} \hat{\bsfg}_{i,\text{ML}}^{(m,n-1)} \right\}.
\label{equation:optimize_CFO_simplified}
\end{equation}
To resolve the nonlinearity in \eqref{equation:optimize_CFO_simplified}, I resort to a second-order Taylor series expansion of the function in \eqref{equation:optimize_CFO_simplified} around the previous \ac{CFO} estimate, $\widehat{\Delta f}^{(m,n-1)}$. For this purpose, let $\bh_{i,t}^{(m,n)} \triangleq \bm \Omega_t\left(\widehat{\Delta f}_\text{ML}^{(m,n-1)}\right) \bP_t\left(\hat{\bm \theta}_{t,\text{MU}}^{(m,n)}\right) \bF^* \bsfS_t^{(m)} \hat{\bsfg}_{i,\text{ML}}^{(m,n-1)}$. Then, \eqref{equation:optimize_CFO_simplified} can be approximated as
\begin{equation}
\begin{split}
	\widehat{\Delta f}_\text{ML}^{(m,n)} =& \underset{\Delta f^{(m)}}{\arg\,\max\,}\sum_{i=1}^{\Lr}\sum_{t=0}^{\Ntr-1}\Real\left\{ \br_{i,t}^{(m)*} \bh_{i,t}^{(m,n)}\right\} \\
	&+ \left(\Delta f^{(m)} - \widehat{\Delta f}^{(m,n-1)}\right) \sum_{i=1}^{\Lr}\sum_{t=0}^{\Ntr-1}\Real\left\{ \br_{i,t}^{(m)*} \sfj \bM \bh_{i,t}^{(m,n)}\right\} \\
	&+ \frac{1}{2}\left(\Delta f^{(m)} - \widehat{\Delta f}^{(m,n-1)}\right)^2 \sum_{i=1}^{\Lr}\sum_{t=0}^{\Ntr-1}\Real\left\{ \br_{i,t}^{(m)*} \sfj^2 \bM^2 \bh_{i,t}^{(m,n)}\right\}.
	\end{split}
	\label{equation:optimize_CFO_Taylor}
\end{equation}
Setting the partial derivative of \eqref{equation:optimize_CFO_Taylor} to zero allows finding the estimate of $\Delta f^{(m)}$ at the $n$-th iteration as
\begin{equation}
	\widehat{\Delta f}_\text{ML}^{(m,n)} = \widehat{\Delta f}_\text{ML}^{(m,n-1)} - \frac{\sum_{i=1}^{\Lr}\sum_{t=0}^{\Ntr}\Imag\left\{ \br_{i,t}^{(m)*} \bM \bh_{i,t}^{(m,n)} \right\}}{\sum_{i=1}^{\Lr}\sum_{t=0}^{\Ntr}\Real\left\{ \br_{i,t}^{(m)*} \bM^2 \bh_{i,t}^{(m,n)} \right\}}.
\label{equation:optimum_CFO_M_step}
\end{equation}
Finally, using \eqref{equation:optimize_CFO_M_step}, we can find the estimator of $\bsfg_i^{(m)}$ at the $n$-th M-step as
\begin{equation}
	\hat{\bsfg}_{i,\text{ML}}^{(m,n)} = \left(\bsfS^{(m)*}\bsfS^{(m)}\right)^{-1} \bsfS^{(m)*} \bF_\otimes \bP^*\left(\hat{\bm \theta}_{\text{MU}}^{(m,n)}\right) \bm \Omega^*\left(\widehat{\Delta f}_\text{ML}^{(m,n)}\right) \br_i^{(m)}.
\label{equation:optimum_channel_M_step}
\end{equation} 
Therefore, using \eqref{equation:TU_LMMSE}, \eqref{equation:MU_LMMSE}, \eqref{equation:optimum_CFO_M_step}, and \eqref{equation:optimum_channel_M_step}, the proposed algorithm iteratively updates the \ac{PN}, \ac{CFO}, and equivalent channel gains respectively. The algorithm is terminated when the difference between the \ac{LF} at two iterations is smaller than a threshold $\eta$, i.e.,
\begin{equation}
\begin{split}
	&{\bigg{|}} \sum_{i=1}^{\Lr}\left\| \br_i^{(m)} - \bm \Omega\left(\widehat{\Delta f}^{(m,n)}\right) \bP\left(\hat{\bm \theta}_\text{MU}^{(m,n)}\right) \bF_\otimes^* \bsfS^{(m)} \hat{\bsfg}_i^{(m,n)}\right\|_2^2 \\
	&\qquad - \left\| \br_i^{(m)} - \bm \Omega\left(\widehat{\Delta f}^{(m,n-1)}\right) \bP\left(\hat{\bm \theta}_\text{MU}^{(m,n-1)}\right) \bF_\otimes^* \bsfS^{(m)} \hat{\bsfg}_i^{(m,n-1)}\right\|_2^2 {\bigg{|}} \leq \eta.
\end{split}
\label{equation:termination_criterion_EM}
\end{equation}
The overall \ac{LMMSE}-\ac{EM} estimation algorithm is summarized in Algorithm \ref{alg:LMMSE_EM}.

\begin{algorithm}
\caption{\ac{LMMSE}-\ac{EM} algorithm}\label{alg:LMMSE_EM}
\begin{algorithmic}[1]
\State \textbf{Initialize \ac{CFO}, beamformed channel estimates, and initial difference in \ac{LLF}} \\ 
\qquad \qquad $\widehat{\Delta f}^{(0)} = \underset{\Delta f}{\arg\,\max}\, \sum_{i=1}^{\Lr} \| \bm \Omega(\Delta f) \bF_\otimes^* \bsfS^{(m)} \bsfS^{(m)*} \bF_\otimes \bm \Omega^*(\Delta f) \br_i^{(m)} \|_2^2 $ \\ 
\qquad \qquad $\hat{\bsfg}_i^{(m,0)} = \left(\bsfS^{(m)*}\bsfS^{(m)}\right)^{-1} \bsfS^{(m)*} \bF_\otimes \bm \Omega\left(\widehat{\Delta f}^{(0)}\right) \br_i^{(m)}$, $i = 1,\ldots,\Lr$\\
\qquad \qquad $\hat{\eta}^{(m,n)} = \infty$, $n = 1$
\While {$\hat{\eta}^{(m,n)} > \eta$}
\qquad \State \textbf{Update \ac{PN} estimate} \\ 
\qquad \qquad \qquad \eqref{equation:jacobian_F_PN}-\eqref{equation:MU_LMMSE} 
\qquad \State \textbf{Update \ac{CFO} estimate} \\ 
\qquad \qquad \qquad $\bh_{i,t}^{(m,n)} \triangleq \bm \Omega_t\left(\widehat{\Delta f}_\text{ML}^{(m,n-1)}\right) \bP_t\left(\hat{\bm \theta}_{t,\text{MU}}^{(m,n)}\right) \bF^* \bsfS_t^{(m)} \hat{\bsfg}_{i,\text{ML}}^{(m,n-1)}$ \\
\qquad \qquad \qquad $\widehat{\Delta f}_\text{ML}^{(m,n)} = \widehat{\Delta f}_\text{ML}^{(m,n-1)} - \frac{\sum_{i=1}^{\Lr}\sum_{t=0}^{\Ntr}\Imag\left\{ \br_{i,t}^{(m)*} \bM \bh_{i,t}^{(m,n)} \right\}}{\sum_{i=1}^{\Lr}\sum_{t=0}^{\Ntr}\Real\left\{ \br_{i,t}^{(m)*} \bM^2 \bh_{i,t}^{(m,n)} \right\}}$ 
\qquad \State \textbf{Update the beamformed channel estimates} \\ 
\qquad \qquad \qquad $\hat{\bsfg}_{i,\text{ML}}^{(m,n)} = \left(\bsfS^{(m)*}\bsfS^{(m)}\right)^{-1} \bsfS^{(m)*} \bF \bP^*\left(\hat{\bm \theta}_{\text{MU}}^{(m,n)}\right) \bm \Omega^*\left(\widehat{\Delta f}_\text{ML}^{(m,n)}\right) \br_i^{(m)}$, $i = 1,\ldots,\Lr$ 
\qquad \State \textbf{Iteration update} \\ 
\qquad \qquad \qquad $n = n + 1$
\qquad \State \textbf{Update difference in likelihood function}\\ 
\qquad \qquad \qquad	$\hat{\eta}^{(m,n)} = {\bigg{|}} \sum_{i=1}^{\Lr}\left\| \br_i^{(m)} - \bm \Omega\left(\widehat{\Delta f}^{(m,n)}\right) \bP\left(\hat{\bm \theta}_\text{MU}^{(m,n)}\right) \bF_\otimes^* \bsfS^{(m)} \hat{\bsfg}_i^{(m,n)}\right\|_2^2 - \left\| \br_i^{(m)} - \bm \Omega\left(\widehat{\Delta f}^{(m,n-1)}\right) \bP\left(\hat{\bm \theta}_\text{MU}^{(m,n-1)}\right) \bF_\otimes^* \bsfS^{(m)} \hat{\bsfg}_i^{(m,n-1)}\right\|_2^2$ 
\EndWhile
\end{algorithmic}
\captionof{figure}{Detailed steps of the first proposed LMMSE-EM algorithm.}
\end{algorithm}

\subsection{EKF-RTS-EM Algorithm}

In this subsection, I present an alternative strategy to using our first proposed \ac{LMMSE}-\ac{EM} algorithm. Despite the simplicity of \eqref{equation:optimize_CFO_M_step} and \eqref{equation:optimum_channel_M_step}, the algorithm introduced in the previous subsection exhibits high computational complexity.  The main computational bottleneck of the \ac{LMMSE}-\ac{EM} algorithm is the inversion of $\hat{\bC}_{\br^{(m)},\br^{(m)}}^{(n)}$ in \eqref{equation:MU_LMMSE}, which has complexity $\mathcal{O}(\left(K\Ntr\Lr)^3\right)$ in the worst case. This high complexity comes at the cost of batch processing the $\Lr K \Ntr$ measurements at once to find the \ac{LMMSE} estimator for the \ac{PN}, which exhibits very good performance but it may not be computationally feasible if the number of subcarriers is in the order of a few thousands. However, a trade-off between estimation performance and computational complexity can be achieved if the size of the matrix inversion in \eqref{equation:MU_LMMSE} is reduced. To reduce computational complexity, I propose to sequentially process every set of $\Lr$ received measurements to reduce complexity to be ${\cal O}\left(\Lr^3\right)$ at most, which is computationally affordable since $\Lr$ is usually a small number \cite{ElAyach:SSP:TWC:2014}. The \ac{PN} estimate can be found using a combination of the \ac{EKF} and the \ac{RTS} smoother \cite{Bayesian_Filtering_Smoothing}, which exploits a first-order linearization of the received measurement vector and uses the \ac{RTS} smoother on the linearized vector as follows.

Using \eqref{equation:rx_signal_frame_OFDM_symbol_RF_chain}, the $\Lr$-dimensional time-domain received measurement can be expressed as
\begin{equation}
	\underbrace{\left[\begin{array}{c}
	r_{1,t}^{(m)}[k_0[t] + \ell] \\
	\vdots \\
	r_{\Lr,t}^{(m)}[k_0[t]+\ell] \\ \end{array}\right]}_{\br_t^{(m)}[k_0[t]+\ell]} = \underbrace{e^{\sfj 2\pi \Delta f^{(m)} (k_0[t] + \ell)} e^{\sfj \theta_t^{(m)}[k_0[t]+\ell]} \left[\begin{array}{c}
	\bff_\ell^* \bsfS_t^{(m)} \bsfg_1^{(m)} \\
	\vdots \\
		\bff_\ell^* \bsfS_t^{(m)} \bsfg_{\Lr}^{(m)} \\ \end{array}\right]}_{\bh_{\ell,t}\left(\bm \xi_\text{D}^{(m)},\theta_t^{(m)}[k_0[t]+\ell]\right)} + \left[\begin{array}{c}
		v_{1,t}^{(m)}[k_0[t] + \ell] \\
		\vdots \\
		v_{\Lr,t}^{(m)}[k_0[t] + \ell] \\ \end{array}\right].
		\label{equation:rx_signal_RTS}
\end{equation}
The \ac{RTS} smoother consists of a backward filter that follows the \ac{EKF} recursion given by the following:
\begin{itemize}
\item \textbf{Forward recursion}: Time Update Equations:
\begin{equation}
\begin{split}
	\hat{\theta}_{t,\text{TU}}^{(m)}[k_0[t]+\ell] &= \left\{\begin{array}{cc}
	0 & t = 0, \ell = 0 \\
	 \hat{\theta}_{t,\text{MU}}^{(m)}[k_0[t] + \ell-1] & \ell > 0, \\
	 \hat{\theta}_{t-1,\text{MU}}^{(m)}[k_0[t-1] + K-1] & \ell = 0, t > 0,  \\ \end{array}\right. \\
	\left(\hat{\sigma}_{\hat \theta_{t,\text{TU}}}^{(\ell)}\right)^2 &= \left\{\begin{array}{cc}
	\left\|\left[\bC_{\bm \theta^{(m)},\bm \theta^{(m)}}\right]_{1,:}\right\|_2^2 & \ell = 0, t = 0, \\
	\left(\hat{\sigma}_{\hat \theta_{t,\text{MU}}}^{(\ell-1)}\right)^2 + \left\| \bm \Delta \bd_{t,k-1}^{t,k}\right\|_2^2 & \ell > 0 \\
	\left(\hat{\sigma}_{t-1,\text{MU}}^{(K-1)}\right)^2 & \ell = 0, t > 0, \\ \end{array}\right.
\end{split}
\label{equation:TU_EKF}
\end{equation}
Measurement Update Equations:
\begin{equation}
\begin{split}
\hat{\theta}_{t,\text{MU}}^{(m)}[k_0[t] + \ell] &= \hat{\theta}_{t,\text{TU}}^{(m)}[k_0[t]+\ell] +\Real{\bigg{\{}} \hat{\bc}_{\theta_t^{(m)},\br_t^{(m)}}^{(\ell)} \left(\hat{\bC}_{\br_t^{(m)},\br_t^{(m)}}^{(\ell)}\right)^{-1} \\
&\times \left(\br_t^{(m)}[k_0[t] + \ell] - \bh_{\ell,t}\left(\hat{\bm \xi}_\text{D}^{(m,n-1)},\hat{\theta}_{t,\text{TU}}^{(m)}[k_0[t] + \ell]\right)\right){\bigg{\}}} \\
\left(\hat{\sigma}_{\hat \theta_{t,\text{MU}}^{(m)}}^{(\ell,t)}\right)^2 &= \Real\left\{\left(\hat{\sigma}_{\hat \theta_{t,\text{TU}}}^{(\ell)}\right)^2 - \left(\hat{\bc}_{\br_t^{(m)},\theta_t^{(m)}}^{(\ell)}\right)^* \left(\hat{\bC}_{\br_t^{(m)},\br_t^{(m)}}^{(\ell)}\right)^{-1} \hat{\bc}_{\br_t^{(m)},\theta_t^{(m)}}^{(\ell)}\right\},
\end{split}
\label{equation:MU_EKF}
\end{equation}
where $\hat{\bc}_{\br_t^{(m)},\theta_t^{(m)}}^{(\ell)} \in \mathbb{C}^{\Lr \times 1}$, $\hat{\bC}_{\br_t^{(m)},\br_t^{(m)}}^{(\ell)} \in \mathbb{C}^{\Lr \times \Lr}$ are the covariance matrix of $\br_t^{(m)}[k_0[t] + \ell]$ and $\theta_t^{(m)}[k_0[t]+\ell]$, and the autocovariance matrix of $\br_t^{(m)}[k_0[t]+\ell]$, which are given by
\begin{equation}
\begin{split}
	\hat{\bc}_{\br_t^{(m)},\theta_t^{(m)}}^{(\ell)} &= \sfj \bh_{\ell,t}\left(\hat{\bm \xi}_\text{D}^{(m,n-1)},\hat{\theta}_{t,\text{TU}}^{(m)}[k_0[t]+\ell]\right) \left(\hat{\sigma}_{\hat \theta_\text{MU}}^{(\ell,t)}\right)^2 \\
	\hat{\bC}_{\br_t^{(m)},\br_t^{(m)}}^{(\ell)} &= \bh_{\ell,t}\left(\hat{\bm \xi}_\text{D}^{(m,n-1)},\hat{\theta}_{t,\text{TU}}^{(m)}[k_0[t]+\ell]\right) \left(\hat{\sigma}_{\hat \theta_\text{MU}}^{(\ell,t)}\right)^2 \bh_{\ell,t}^*\left(\hat{\bm \xi}_\text{D}^{(m,n-1)},\hat{\theta}_{t,\text{TU}}^{(m)}[k_0[t]+\ell]\right) + \sigma^2 \bI_{\Lr}.
	\end{split}
	\label{equation:covariances_EKF}
\end{equation}
\item \textbf{Backward recursion}: 
\begin{equation}
\begin{split}
	G_t^{(\ell)} &= \frac{\left(\hat{\sigma}_{\hat \theta_{t,\text{MU}}^{(m)}}^{(\ell,t)}\right)^2}{\left(\hat{\sigma}_{\hat \theta_{t,\text{MU}}^{(m)}}^{(\ell,t)}\right)^2 + \left\|\bm \Delta \bd_{t,k}^{t,k+1}\right\|_2^2} \\
	\hat{\theta}_{t,\text{RTS}}^{(m)}[k_0[t]+\ell] &= \hat{\theta}_{t,\text{MU}}^{(m)}[k_0[t]+\ell] + G_t^{(\ell)} \left(\hat{\theta}_{t,\text{RTS}}^{(m)}[k_0[t] + \ell+1] - \hat{\theta}_{t,\text{MU}}[k_0[t]+\ell+1]\right) \\
\left(\hat{\sigma}_{\hat \theta_{t,\text{RTS}}^{(m)}}^{(\ell,t)}\right)^2 &= \left(\hat{\sigma}_{\hat \theta_{t,\text{MU}}^{(m)}}^{(\ell,t)}\right)^2 + \left(G_t^{(\ell)}\right)^2 \left( \left(\hat{\sigma}_{\hat \theta_{t,\text{RTS}}^{(m)}}^{(\ell+1,t)}\right)^2 - \left(\hat{\sigma}_{\hat \theta_{t,\text{MU}}^{(m)}}^{(\ell+1,t)}\right)^2 - \left\|\bm \Delta \bd_{t,k}^{t,k+1}\right\|_2^2 \right).
\end{split}
\label{equation:MU_ERTS}
\end{equation}
\end{itemize}
In \eqref{equation:TU_EKF}, the proposed algorithm is initialized as $\hat{\theta}_{t,\text{TU}}[k_0[0]] = 0$ since the \ac{PN} vector is assumed to have zero mean, and the predicted variance is initialized as $\left(\hat{\sigma}_{\hat{\theta}_{0,\text{TU}}^{(m)}}^{(0)}\right)^2 = \left\| \left[\bC_{\bm \theta^{(m)},\bm \theta^{(m)}}\right]_{1,:} \right\|_2^2$. Also, notice that the \ac{CP} is removed after timing offset synchronization, which requires properly updating the \ac{PN} predicted statistics from the last sample of the $t$-th \ac{OFDM} symbol to the first sample of the $(t+1)$-th \ac{OFDM} symbol, as reflected in \eqref{equation:TU_EKF}.

Thereby, using \eqref{equation:TU_EKF}-\eqref{equation:MU_ERTS}, and then \eqref{equation:optimum_CFO_M_step} and \eqref{equation:optimum_channel_M_step}, the second proposed algorithm can iteratively update the \ac{PN} sample estimates, \ac{CFO}, and equivalent channel gains, respectively. The termination criterion for the proposed algorithm is analogous to the termination criterion for the first proposed \ac{LMMSE}-\ac{EM} algorithm, given in \eqref{equation:termination_criterion_EM}. The detailed steps the proposed \ac{EKF}-\ac{RTS} algorithm follows are summarized in Algorithm \ref{alg:EKF_RTS_EM}.

\begin{algorithm}
\caption{\ac{EKF}-\ac{RTS}-\ac{EM} algorithm}\label{alg:EKF_RTS_EM}
\begin{algorithmic}[1]
\State \textbf{Initialize \ac{CFO}, beamformed channel estimates, and initial difference in \ac{LF}} \\ 
\qquad \qquad $\widehat{\Delta f}^{(0)} = \underset{\Delta f}{\arg\,\max}\, \sum_{i=1}^{\Lr} \| \bm \Omega(\Delta f) \bF_\otimes^* \bsfS^{(m)} \bsfS^{(m)*} \bF_\otimes \bm \Omega^*(\Delta f) \br_i^{(m)} \|_2^2 $ \\ 
\qquad \qquad $\hat{\bsfg}_i^{(m,0)} = \left(\bsfS^{(m)*}\bsfS^{(m)}\right)^{-1} \bsfS^{(m)*} \bF_\otimes \bm \Omega\left(\widehat{\Delta f}^{(0)}\right) \br_i^{(m)}$, $i = 1,\ldots,\Lr$\\
\qquad \qquad $\hat{\eta}^{(m,n)} = \infty$, $n = 1$
\While {$\hat{\eta}^{(m,n)} > \eta$}
\qquad \State \textbf{Update \ac{PN} estimate} \\ 
\qquad \qquad \qquad \eqref{equation:TU_EKF}-\eqref{equation:MU_ERTS} 
\qquad \State \textbf{Update \ac{CFO} estimate} \\ 
\qquad \qquad \qquad $\bh_{i,t}^{(m,n)} \triangleq \bm \Omega_t\left(\widehat{\Delta f}_\text{ML}^{(m,n-1)}\right) \bP_t\left(\hat{\bm \theta}_{t,\text{RTS}}^{(m,n)}\right) \bF^* \bsfS_t^{(m)} \hat{\bsfg}_{i,\text{ML}}^{(m,n-1)}$ \\
\qquad \qquad \qquad $\widehat{\Delta f}_\text{ML}^{(m,n)} = \widehat{\Delta f}_\text{ML}^{(m,n-1)} - \frac{\sum_{i=1}^{\Lr}\sum_{t=0}^{\Ntr}\Imag\left\{ \br_{i,t}^{(m)*} \bM \bh_{i,t}^{(m,n)} \right\}}{\sum_{i=1}^{\Lr}\sum_{t=0}^{\Ntr}\Real\left\{ \br_{i,t}^{(m)*} \bM^2 \bh_{i,t}^{(m,n)} \right\}}$ 
\qquad \State \textbf{Update the beamformed channel estimates} \\ 
\qquad \qquad \qquad $\hat{\bsfg}_{i,\text{ML}}^{(m,n)} = \left(\bsfS^{(m)*}\bsfS^{(m)}\right)^{-1} \bsfS^{(m)*} \bF \bP^*\left(\hat{\bm \theta}_{\text{RTS}}^{(m,n)}\right) \bm \Omega^*\left(\widehat{\Delta f}_\text{ML}^{(m,n)}\right) \br_i^{(m)}$, $i = 1,\ldots,\Lr$ 
\qquad \State \textbf{Iteration update} \\ 
\qquad \qquad \qquad $n = n + 1$
\qquad \State \textbf{Update difference in likelihood function}\\ 
\qquad \qquad \qquad	$\hat{\eta}^{(m,n)} = {\bigg{|}} \sum_{i=1}^{\Lr}\left\| \br_i^{(m)} - \bm \Omega\left(\widehat{\Delta f}^{(m,n)}\right) \bP\left(\hat{\bm \theta}_\text{RTS}^{(m,n)}\right) \bF_\otimes^* \bsfS^{(m)} \hat{\bsfg}_i^{(m,n)}\right\|_2^2 - \left\| \br_i^{(m)} - \bm \Omega\left(\widehat{\Delta f}^{(m,n-1)}\right) \bP\left(\hat{\bm \theta}_\text{RTS}^{(m,n-1)}\right) \bF_\otimes^* \bsfS^{(m)} \hat{\bsfg}_i^{(m,n-1)}\right\|_2^2$ 
\EndWhile
\end{algorithmic}
\captionof{figure}{Detailed steps of the second proposed EKF-RTS-EM algorithm.}
\end{algorithm}

\subsection{Initialization and Convergence}

Appropriate initialization of the \ac{CFO}, $\Delta f^{(m)}$, and equivalent beamformed channels, $\{\bsfg_i^{(m)}\}_{i=1}^{\Lr}$, is essential to ensure global convergence of the proposed algorithms. The initialization process can be summarized as follows:
\begin{itemize}
	\item Similar to \cite{Channel_PN_HCRLB}, an initial \ac{CFO} estimate $\widehat{\Delta f}^{(m,0)}$ is obtained by applying an exhaustive search for the value of $\Delta f^{(m)}$ that minimizes the cost function in the absence of \ac{PN}. This cost function is given in (\cite{Asilomar_2018}, equation (17)). Simulation results in Section \ref{subsec:results_broadband} show that an exhaustive search with a coarse step size of $0.02$ is sufficient to initialize the proposed algorithms.
	\item Using $\widehat{\Delta f}^{(m,0)}$, the initial channel estimates $\{\hat{\bsfg}_i^{(m,0)}\}_{i=1}^{\Lr}$ are obtained by applying $\hat{\bsfg}_i^{(m,0)} = \left(\bsfS^{(m)*}\bsfS^{(m)}\right)^{-1} \bsfS^{(m)*} \bF \bm \Omega^*\left(\widehat{\Delta f}^{(m,0)}\right) \br_i^{(m)}$.
\end{itemize}
Based on the equivalent system model in \eqref{equation:rx_signal_frame_OFDM_symbol_RF_chain} and the simulation results in Section \ref{subsec:results_broadband}, it can be concluded that the proposed \ac{LMMSE}-\ac{EM} and \ac{EKF}-\ac{RTS}-\ac{EM} algorithms converge globally when the \ac{PN} vector is initialized as $\hat{\bm \theta}^{(m,0)} = \bm 0_{K\Ntr \times 1}$.

\subsection{Dictionary-Constrained Channel Estimation}

In this section, I formulate the problem of estimating the frequency-selective \ac{mmWave} \ac{MIMO} channel using the \ac{ML} statistics already estimated using the proposed \ac{LMMSE}-\ac{EM} and \ac{EKF}-\ac{RTS}-\ac{EM} algorithms. Once $M$ training frames are processed, each comprising of $\Ntr$ \ac{OFDM} symbols, the estimated equivalent beamformed channels can be stacked to form the signal model
\begin{equation}
	\underbrace{\left[\begin{array}{c}
	\hat{\bsfg}_\text{ML}^{(1,N)}[k] \\
	\vdots \\
	\hat{\bsfg}_\text{ML}^{(M,N)}[k] \\ \end{array}\right]}_{\hat{\bsfg}_\text{ML}^{(N)}} = \underbrace{\left[\begin{array}{c}
	\bsfq^{(1)T} \bsfF_\text{tr}^{(1)T}\otimes \bsfD_\text{w}^{(1)-*} \bsfW_\text{tr}^{(1)*} \\
	\vdots \\
	\bsfq^{(M)T} \bsfF_\text{tr}^{(M)T}\otimes \bsfD_\text{w}^{(M)-*} \bsfW_\text{tr}^{(M)*} \\ \end{array}\right]}_{\bm \Phi_\text{w}} \vect\{\bsfH[k]\} + \underbrace{\left[\begin{array}{c}
	\tilde{\bsfv}^{(1,N)}[k] \\
	\vdots \\
	\tilde{\bsfv}^{(M,N)}[k] \\ \end{array}\right]}_{\tilde{\bsfv}^{(N)}[k]},
	\label{equation:signal_model_MIMO_estimation}
\end{equation}
where $\tilde{\bsfv}^{(m,N)} \in \mathbb{C}^{\Lr \times 1}$ is the estimation error of $\hat{\bsfg}_\text{ML}^{(m,N)}[k]$, $1\leq m \leq M$, and $\bm \Phi_\text{w} \in \mathbb{C}^{M\Lr \times \Nt\Nr}$ is the post-whitened measurement matrix. Now, the channel matrix in \eqref{equation:channel_extended_fd} can be vectorized and plugged into \eqref{equation:signal_model_MIMO_estimation} to obtain
\begin{equation}
\hat{\bsfg}_\text{ML}^{(N)}[k] \approx \bm \Phi_\text{w} \underbrace{\left(\tilde{\bsfA}_\text{T}^\text{C} \otimes \tilde{\bsfA}_\text{R}\right)}_{\bm \Psi} \underbrace{\vect\{\bsfG^\text{v}[k]\}}_{\bsfg^\text{v}[k]} + \tilde{\bsfv}^{(N)}[k],
\label{equation:signal_model_MIMO_estimation_vec}
\end{equation}
where $\bm \Psi \in \mathbb{C}^{\Nt\Nr \times \Gt\Gr}$ is the angular dictionary matrix, and $\bsfg^\text{v}[k] \in \mathbb{C}^{\Gr\Gt \times 1}$ is the sparse vector containing the complex channel path gains in its non-zero coefficients \cite{RodGonVenHea:TWC:2018}. To estimate the frequency-selective sparse vectors $\{\bsfg^\text{v}[k]\}_{k=0}^{K-1}$, the design of the measurement matrix $\bm \Phi_\text{w}$ in \eqref{equation:signal_model_MIMO_estimation_vec} needs to be such that this matrix has as small correlation between columns as possible, which is a result proven in the \ac{CS} literature to ensure that the estimation of the channel's support will be robust, and this depends on the design of the precoding and combining matrices $\bsfF_\text{tr}^{(m)}$, $\bsfq^{(m)}$, and $\bsfW_\text{tr}^{(m)}$. As discussed in \cite{Asilomar_2018}, the precoders and combiners should be designed accounting for the lack of timing synchronization, such that the equivalent measurement matrix design is suitable for compressive estimation and the estimated timing offset $\hat{n}_0$ matches the actual timing offset. For this reason, I adopt the design method in \cite{Asilomar_2018} to generate hybrid precoders and combiners, which has been shown to offer excellent performance at the low \ac{SNR} regime. 

Another issue to overcome when estimating the sparse channel vectors is how to obtain prior information on either the sparsity level of the channel or the variance of the noise in \eqref{equation:signal_model_MIMO_estimation_vec}. As discussed in \cite{RodGon:CFO:TWC:2019}, knowing the sparsity level is unrealistic in practice, and even if it were known, there is no guarantee that the best sparse approximation of $\{\bsfg^\text{v}[k]\}_{k=0}^{K-1}$ has as many non-zero components as the actual number of multipath components in the frequency-selective channel. This mismatch is even more severe in the frequency-selective scenario, in which transmit and receive pulse-shaping bandlimit the channel, thereby limiting the resolution to detect multipath components at baseband level \cite{Heath:IntroWirelessComm}. For this reason, I will focus on finding the variance of the estimation error in \eqref{equation:signal_model_MIMO_estimation_vec}. 

From the property of asymptotic efficiency of \ac{ML} estimators it is known that, if the \ac{SNR} is not too low, and the number of samples used to estimate the different parameters is large enough, the estimation errors $\tilde{\bsfv}^{(m,N)}[k]$ are Gaussian, with zero mean and covariance given by the hybrid \ac{CRLB} matrix for the estimation of the complex path gains $\hat{\bsfg}_\text{ML}^{(m,N)}[k]$. Since the received noise vectors $\bv^{(m)}[n]$ in \eqref{equation:rx_frame} are independent and identically distributed, it is clear that estimation errors for $\hat{\bsfg}_\text{ML}^{(m,N)}[k]$ are independent as well, although not identically distributed. For this reason, it is necessary to compute the covariance matrix for each of the estimation error vectors corresponding to the $M$ different training frames. Let $\bC_{\hat{\bg}_i^{(m)},\hat{\bg}_i^{(m)}} \in \mathbb{C}^{\Lc \times \Lc}$ denote the hybrid \ac{CRLB} matrix for the estimation of $\bg_i^{(m)}$. Using that $g_i^{(m)}[d] = [\bg_i^{(m)}]_d = \alpha_i^{(m)}[d] e^{\sfj \beta_i^{(m)}[d]}$, it follows that
\begin{equation}
\underbrace{\left[\begin{array}{c}
g_1^{(m)}[0] \\
\vdots \\
g_1^{(m)}[\Lc -1] \\
\vdots \\
g_{\Lr}^{(m)}[0] \\
\vdots \\
g_{\Lr}^{(m)}[\Lc-1] \\ \end{array}\right]}_{\bg^{(m)}} = \underbrace{\left[\begin{array}{c}
\alpha_1^{(m)}[0] e^{\sfj \beta_1^{(m)}[0]} \\
\vdots \\
\alpha_1^{(m)}[\Lc-1] e^{\sfj \beta_1^{(m)}[\Lc-1]} \\
\vdots \\
\alpha_{\Lr}^{(m)}[0] e^{\sfj \beta_{\Lr}^{(m)}[0]} \\
\vdots \\
\alpha_{\Lr}^{(m)}[\Lc-1] e^{\sfj \beta_{\Lr}^{(m)}[\Lc-1]} \\ \end{array}\right]}_{\bff\left(\bm \alpha^{(m)}, \bm \beta^{(m)}\right)}.
\label{equation:complex_td_channels}
\end{equation}
Using \eqref{equation:complex_td_channels}, the covariance matrix of any unbiased estimator $\hat{\bg}^{(m)}$ of $\bg^{(m)}$ is lower bounded by the hybrid \ac{CRLB} as
\begin{equation}
	\bC_{\hat{\bg}^{(m)},\hat{\bg}^{(m)}} \geq \bJ_{\bff}\left(\bm \alpha^{(m)},\bm \beta^{(m)}\right) \bC_{\hat{\tilde{\bg}}^{(m)},\hat{\tilde{\bg}}^{(m)}} \bJ_{\bff}\left(\bm \alpha^{(m)}, \bm \beta^{(m)} \right)^*,
	\label{equation:HCRLB_time_domain}
\end{equation}
where $\bJ_{\bff}\left(\bm \alpha^{(m)}, \bm \beta^{(m)}\right) \in \mathbb{C}^{\Lr \Lc \times 2 \Lr \Lc}$ is the Jacobian matrix of $\bff\left(\bm \alpha^{(m)}, \bm \beta^{(m)}\right)$
\begin{equation}
	\bJ_{\bff}\left(\bm \alpha^{(m)}, \bm \beta^{(m)}\right) = \bigoplus_{i=1}^{\Lr}\bigoplus_{d=0}^{D-1} \left[\begin{array}{cc} e^{\sfj \beta_i^{(m)}[d]} & \sfj g_i^{(m)}[d] \\ \end{array}\right].
	\label{equation:Jacobian_complex_td}
\end{equation}
Next, the hybrid \ac{CRLB} for the estimation of $\bsfg^{(m)}$ is computed as follows. Notice that the different frequency-domain channel vectors $\bsfg_i^{(m)}$ are related to their time-domain counterparts through a Fourier transform, mathematically represented using $\bF_1 \in \mathbb{C}^{K \times \Lc}$, which comprises of the first $\Lc$ in $\bF$. Thereby, using \eqref{equation:HCRLB_time_domain} the covariance for any unbiased estimator $\hat{\bsfg}^{(m)}$ is simply given by
\begin{equation}
	\bC_{\hat{\bsfg}^{(m)}, \hat{\bsfg}^{(m)}} \geq \left(\bI_{\Lr} \otimes \bF_1\right) \bC_{\hat{\bg}^{(m)},\hat{\bg}^{(m)}}  \left(\bI_{\Lr} \otimes \bF_1\right)^*.
	\label{equation:HCRLB_freq_domain}
\end{equation}
Now, the hybrid \ac{CRLB} for the estimation of $\bg^{(m)}[k] = \left[\begin{array}{ccc} \sfg_1^{(m)}[k] & \ldots & \sfg_{\Lr}^{(m)}[k] \\ \end{array}\right]^T$ is related to the hybrid \ac{CRLB} in \eqref{equation:HCRLB_freq_domain} through a selection matrix as
\begin{equation}
\begin{split}
	\bsfg^{(m)}[k] &= \left[\begin{array}{c}
	\bee_k^T \bsfg_1^{(m)} \\
	\vdots \\
	\bee_k^T \bsfg_{\Lr}^{(m)} \\ \end{array}\right] \\
	&= \left(\bI_{\Lr} \otimes \bee_k^T\right) \bsfg^{(m)}, 
\end{split}
\label{equation:freq_domain_for_CS}
\end{equation}
whereby the hybrid \ac{CRLB} for any unbiased estimator $\hat{\bsfg}^{(m)}[k]$ of $\bsfg^{(m)}[k]$ is given by
\begin{equation}
	\bC_{\hat{\bsfg}^{(m)}[k], \hat{\bsfg}^{(m)}[k]} \geq \left(\bI_{\Lr} \otimes \bee_k^T\right) \bC_{\hat{\bsfg}^{(m)}, \hat{\bsfg}^{(m)}} \left(\bI_{\Lr} \otimes \bee_k\right).
	\label{equation:HCRLB_for CS}
\end{equation}
Finally, the overall covariance matrix for the estimation error vector $\tilde{\bsfv}^{(N)}[k]$ in \eqref{equation:signal_model_MIMO_estimation_vec} needs to be found. Using the fact that the received noise at the antenna level is temporally white, the covariance matrix of $\tilde{\bsfv}^{(N)}[k]$ is given by the hybrid \ac{CRLB} for any unbiased estimator of $\bsfg[k] = \left[\begin{array}{ccc} \bsfg^{(1)T}[k] & \ldots & \bsfg^{(M)T}[k] \\ \end{array}\right]^T$. The final hybrid \ac{CRLB} is given by
\begin{equation}
	\bC_{\hat{\bsfg}[k], \hat{\bsfg}[k]} \geq \bigoplus_{m=1}^{M} \bC_{\hat{\bsfg}^{(m)}[k], \hat{\bsfg}^{(m)}[k]}.
	\label{equation:HCRLB_final_CS}
\end{equation}
Then, the estimation error is distributed as $\tilde{\bsfv}^{(N)}[k] \sim {\cal CN}\left(\bm 0,\bsfC_{\tilde{\bsfv}^{(N)}[k],\tilde{\bsfv}^{(N)}[k]}\right)$. Let $\bsfD_{\tilde{\bsfv}^{(N)}[k]} \in \mathbb{C}^{M\Lr \times M\Lr}$ be the Cholesky factor of $\bsfC_{\tilde{\bsfv}^{(N)}[k],\tilde{\bsfv}^{(N)}[k]}$, i.e., $\bsfC_{\tilde{\bsfv}^{(N)}[k],\tilde{\bsfv}^{(N)}[k]} = \bsfD_{\tilde{\bsfv}^{(N)}[k]}^* \bsfD_{\tilde{\bsfv}^{(N)}[k]}$.  Thereby, the problem of estimating $\{\bsfg^\text{v}[k]\}_{k=0}^{K-1}$ can be formulated as
\begin{equation}
	\hat{\bsfg}^\text{v}[k] = \underset{\{\bsfg^\text{v}[\ell]\}_{\ell=0}^{K-1}}{\arg\,\min\,}\sum_{k=0}^{K-1}\left\|\bsfg^\text{v}[k]\right\|^1, \qquad \text{subject to } \frac{1}{K}\sum_{k=0}^{K-1}\left\|\bsfD_{\tilde{\bsfv}^{(N)}[k]}^{-*}\left(\hat{\bsfg}_\text{ML}^{(N)}[k] - \bm \Phi_\text{w} \bm \Psi \bsfg^\text{v}[k] \right) \right\|_2^2 \leq \epsilon,
	\label{equation:channel_estimation_sparse}
\end{equation}
where $\epsilon \in \mathbb{R}$ is a design parameter defining the maximum allowable reconstruction error for the sparse vectors $\{\bsfg^\text{v}[k]\}_{k=0}^{K-1}$. From a computational complexity standpoint, the main difficulty in \eqref{equation:channel_estimation_sparse} comes from the fact that post-whitening the proxy estimates $\hat{\bsfg}_\text{ML}^{(n)}[k]$ results in frequency-dependent measurement matrices $\bm \Upsilon[k] = \bsfD_{\bsfv^{(N)}[k]}^{-*} \bm \Phi_\text{w} \bm \Psi$, which increases the complexity of sparse recovery algorithms by a factor of $K$. Since $K$ can be in the order of hundreds or thousands of subcarriers, using frequency-dependent measurement matrices results in high-complexity channel estimation algorithms. To circumvent this issue, I propose to find a covariance matrix $\bsfC_{\bsfv^{(N)},\bsfv^{(N)}}$ that accurately represents the covariance matrix of the estimation error for every subcarrier in the \ac{MMSE} sense. Let $\hat{\tilde{\bsfg}}_\text{ML}^{(N)}[k] \in \mathbb{C}^{M \Lr \times 1}$ denote an approximate estimate of $\hat{\bsfg}_\text{ML}^{(N)}[k]$ given by
\begin{equation}
	\hat{\tilde{\bsfg}}_\text{ML}^{(N)}[k] \approx \bsfg[k] + \tilde{\bsfv}^{(N)}[k],
	\label{equation:approximate_ML_beamformed_channel}
\end{equation}
in which $\tilde{\bsfv}^{(N)}[k] \sim {\cal CN}\left(\bm 0, \bsfC_{\bsfv^{(N)},\bsfv^{(N)}}\right)$. Then, the problem of finding the covariance matrix $\bsfC_{\bsfv^{(N)},\bsfv^{(N)}}$ can be stated as
\begin{equation}
\bsfC_{\bsfv^{(N)},\bsfv^{(N)}} = \underset{\bsfC}{\arg\,\min}\, \sum_{k=0}^{K-1}\mathbb{E}\left\{ \left\| \hat{\tilde{\bsfg}}_\text{ML}^{(N)}[k] - \hat{\bsfg}_\text{ML}^{(N)} \right\|_2^2 \right\}.
\label{equation:optimization_problem_covariance}
\end{equation}
Upon developing the cost function in \eqref{equation:optimization_problem_covariance}, and letting $\bsfD \in \mathbb{C}^{M \Lr \times M \Lr}$ be the Cholesky factor of $\bsfC$, i.e., $\bsfC = \bsfD^* \bsfD$, the optimal covariance matrix can be found as the solution to the problem
\begin{equation}
	\bsfC_{\bsfv^{(N)},\bsfv^{(N)}} = \underset{\bsfC}{\arg\,\min}\, \sum_{k=0}^{K-1}\mathbb{E}\left\{ \left\| \bsfD - \bsfD_{\bsfv^{(N)}[k],\bsfv^{(N)}[k]} \right\|_F^2 \right\},
\end{equation}
which is a \ac{LS} problem with solution given by
\begin{equation}
	\bsfC_{\bsfv^{(N)},\bsfv^{(N)}} = \left(\frac{1}{K}\sum_{k=0}^{K-1} \bsfD_{\bsfv^{(N)}[k],\bsfv^{(N)}[k]}\right)^* \left(\frac{1}{K}\sum_{k=0}^{K-1} \bsfD_{\bsfv^{(N)}[k],\bsfv^{(N)}[k]}\right).
	\label{equation:optimal_covariance}
\end{equation}
The result in \eqref{equation:optimal_covariance} indicates that the covariance matrix that best represents the covariance of every estimation error vector in the \ac{MMSE} sense has a Cholesky decomposition with Cholesky factor given by the average of the Cholesky factors for the covariance matrices of the estimation error at the different subcarriers.

The last step to close the estimation problem in \eqref{equation:channel_estimation_sparse} is the definition of $\epsilon$. Since the training precoders and combiners might lead, in general, to different covariance matrices $\bsfC_{\tilde{\bsfv}^{(N)}[k],\tilde{\bsfv}^{(N)}[k]}$, an overall representative for the noise variance of the entire vector $\tilde{\bsfv}^{(N)} = [\tilde{\bsfv}^{(N)T}[0], \ldots, \tilde{\bsfv}^{(N)T}[K-1]]^T$ is needed. To overcome this issue, similarly to \cite{RodGon:CFO:TWC:2019}, I propose to design $\epsilon$ as a convex combination of the hybrid \ac{CRLB} for the different $\sfg_i^{(m)}[k]$, $1 \leq \Lr$, $0\leq k \leq K-1$, $1 \leq m \leq M$, using estimates of the \ac{SNR} per \ac{RF} chain. Using the property of asymptotic invariance of \ac{ML} estimators, the \ac{ML} estimate of the \ac{SNR} per \ac{RF} chain can be written as $\hat{\gamma}_{i,\text{ML}}^{(m)}[k] = \hat{\alpha}_{i,\text{ML}}^{(m)2}[k]/\sigma^2$. Letting $\hat{\bm \Gamma}^{(m)}[k] = \bigoplus_{i=1}^{\Lr}\hat{\gamma}_{i,\text{ML}}^{(m)}[k]$, the parameter $\epsilon$ can be set as
\begin{equation}
\epsilon = \sum_{m=1}^{M}\sum_{k=0}^{K-1} \frac{\trace\left\{\hat{\bm \Gamma}^{(m)}[k] \bC_{\hat{\bsfg}^{(m)}[k],\hat{\bsfg}^{(m)}[k]} \right\}}{\sum_{m=1}^{M}\sum_{k=0}^{K-1}\trace\left\{\hat{\bm \Gamma}^{(m)}[k]\right\}}.
\label{equation:epsilon_design}
\end{equation}
Then, the \ac{SW-OMP} or \ac{SS-SW-OMP+Th} algorithms in \cite{RodGonVenHea:TWC:2018} can be used to solve the problem in \eqref{equation:channel_estimation_sparse}. These algorithms have been shown to offer very good performance even when the \ac{mmWave} \ac{MIMO} channel has several clusters with non-negligible \ac{AS}. It is important to highlight that the hybrid \ac{CRLB} for the estimation of the channel matrices $\{\bsfH[k]\}_{k=0}^{K-1}$ is not computed. The reason is that, for realistic \ac{mmWave} \ac{MIMO} channel models such as NYUSIM \cite{NYUSIM}, \ac{QuaDRiGa} \cite{QuaDRiGa_IEEE}, \cite{QuaDRiGa_Tech_Rep}, and the 5G \ac{NR} channel model \cite{5G_channel_model}, the finite antenna resolution, bandlimitedness of the baseband equivalent channel, and lack of knowledge of the number of multipath components, make it impossible to assume that an unbiased estimator for the channel can be found. For this reason, the estimates $\{\hat{\bsfg}^\text{v}[k]\}_{k=0}^{K-1}$ will, in general, have a different number of entries than the number of multipath components the channel actually comprises of. Consequently, the theory of \ac{CRLB} cannot be directly applied to this problem.

\section{Numerical Results}
\label{subsec:results_broadband}

This section includes preliminar numerical results obtained with the proposed synchronization algorithms. These results are obtained after performing Monte Carlo simulations averaged over $100$ trials to evaluate the \ac{NMSE}, ergodic spectral efficiency, and \ac{BER}.

Unless otherwise stated, the typical parameters for the system configuration are as follows. Both the transmitter and the receiver are assumed to use a \ac{ULA} with half-wavelength separation. Such a \ac{ULA} has array response vectors given by $\left[\ba_\text{T}(\theta_\ell)\right]_n = \sqrt{\frac{1}{N_\text{t}}} e^{\jj n \pi\cos{(\theta_\ell)}}, \quad n = 0,\ldots,N_\text{t}-1$ and $\left[\ba_\text{R}(\phi_\ell)\right]_m = \sqrt{\frac{1}{N_\text{r}}} e^{\jj m \pi\cos{(\phi_\ell)}}, \quad m= 0,\ldots,N_\text{r}-1$, for both transmitter and receiver, respectively. The
I take $N_\text{t} = 64$ and $N_\text{r}=32$ for illustration, and $G_\text{t}=G_\text{r}=128$. The phase-shifters used in both the transmitter and the receiver are assumed to have $N_\text{Q}$ quantization bits, so that the entries of the analog training precoders and combiners $\bsfF_\text{tr}^{(m)}$, $\bsfW_\text{tr}^{(m)}$, $m = 1,2,\ldots,M$ are drawn from a set ${\cal A} = \left\{0,\frac{2\pi}{2^{N_\text{Q}}},\ldots,\frac{2\pi (2^{N_\text{Q}}-1)}{2^{N_\text{Q}}} \right\}$. The number of quantization bits is set to $N_\text{Q} = 6$. The number of RF chains is set to $L_\text{t} = 8$ at the transmitter and $L_\text{r} = 4$ at the receiver. The number of \ac{OFDM} subcarriers is set to $K = 256$, and the carrier frequency is set to $60$ GHz.

The frequency-selective \ac{mmWave} \ac{MIMO} channel is generated using \eqref{eqn:channel_model} with small-scale parameters taken from the \ac{QuaDRiGa} channel simulator \cite{QuaDRiGa_IEEE}, \cite{QuaDRiGa_Tech_Rep}, which implements the 3GPP 38.901 \ac{UMi} channel model in \cite{5G_channel_model}. The channel samples are generated with an average Rician factor of $-10$ dB, and the distance between the transmitter and the receiver is set to $d = 30$ meters for illustration.

\begin{figure}[t!]
\centering
\begin{tabular}{cc}
\includegraphics[width=0.5\textwidth]{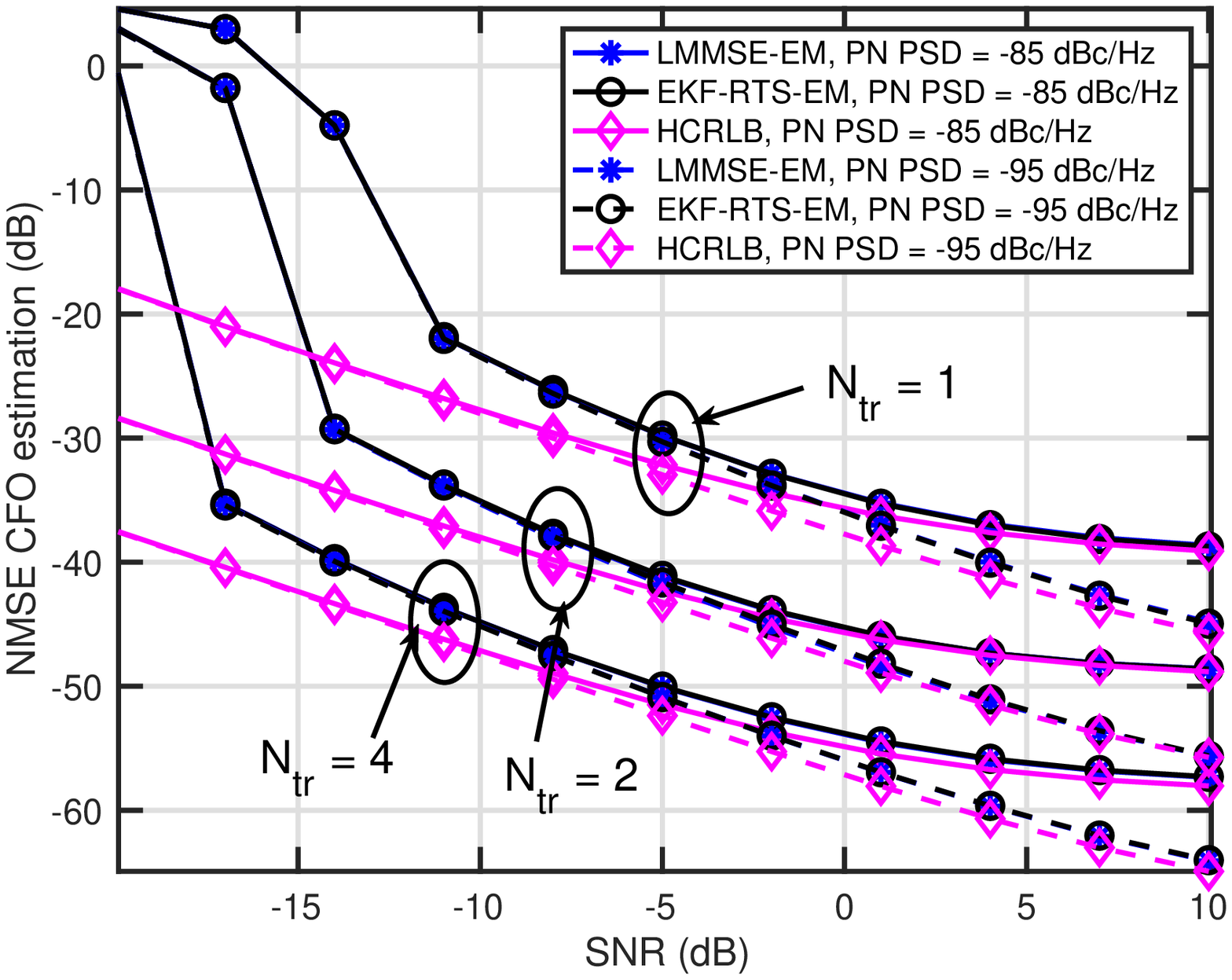} & \includegraphics[width=0.5\textwidth]{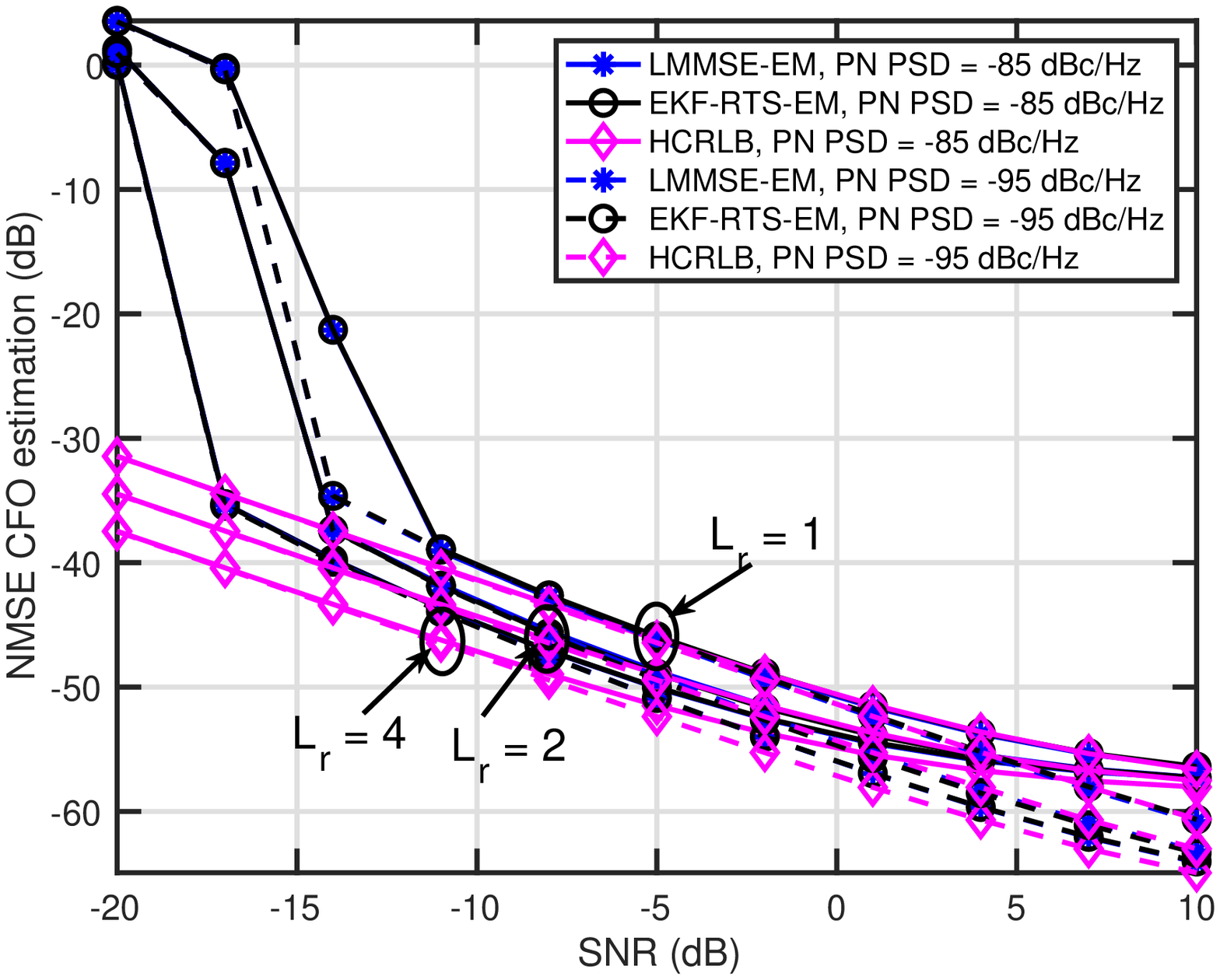} \\ (a) & (b) 
\end{tabular}
\caption{Evolution of the \ac{NMSE} of the CFO estimates obtained using the proposed algorithms versus $\SNR$. The hybrid \ac{CRLB} is also provided as a performance bound.}
\label{fig:NMSE_vs_CFO_several_Ntr}
\end{figure}

In Fig. \ref{fig:NMSE_vs_CFO_several_Ntr}, I show the evolution of the \ac{NMSE} of the \ac{CFO} estimates versus $\SNR$ for the proposed \ac{LMMSE}-\ac{EM} and \ac{EKF}-\ac{RTS}-\ac{EM} algorithms. The hybrid \ac{CRLB} is also provided as an estimation performance bound. I evaluate both algorithms using two different values for the \ac{PN} variance, which are $G_\theta = -85$ dBc/Hz and $G_\theta = -95$ dBc/Hz. In Fig. \ref{fig:NMSE_vs_CFO_several_Ntr} (a), I set the number of receive \ac{RF} chains to $\Lr = 4$ and sweep $\Ntr$ within the range $\{1,2,4\}$ \ac{OFDM} training symbols. Conversely, in Fig. \ref{fig:NMSE_vs_CFO_several_Ntr} (b), the number of \ac{OFDM} training symbols is set to $\Ntr = 4$, and the number of receive \ac{RF} chains is swept within the range $\{1,2,4\}$. 

\begin{figure}[t!]
\centering
\includegraphics[width=0.5\textwidth]{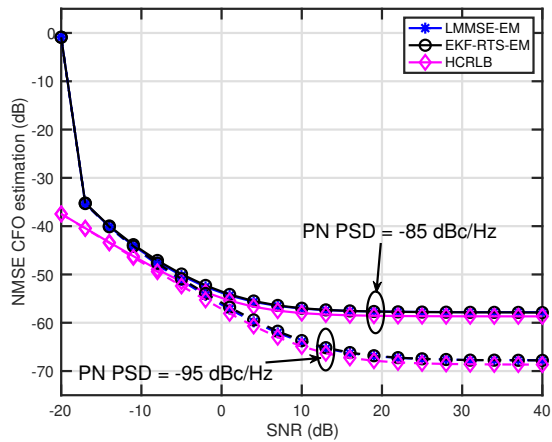}
\caption{Asymptotic evolution of the \ac{NMSE} of the CFO estimates obtained using the proposed algorithms versus $\SNR$. The hybrid \ac{CRLB} is also provided as a performance bound.}
\label{fig:NMSE_vs_CFO_asymptotics}
\end{figure}

Several observations can be made from Fig. \ref{fig:NMSE_vs_CFO_several_Ntr}:
 
\begin{itemize}
	\item The proposed \ac{LMMSE}-\ac{EM} and \ac{EKF}-\ac{RTS}-\ac{EM} algorithms exhibit very similar estimation performance, which suggests that the proposed \ac{EKF}-\ac{RTS}-\ac{EM} algorithm does not compromise estimation performance while dramatically reducing computational complexity during the measurement update in \ac{PN} estimation.
	
	\item The estimation performance of the proposed algorithms exhibits a small gap with respect to the hybrid \ac{CRLB}. At low \ac{SNR}, the gap between the \ac{NMSE} and the hybrid \ac{CRLB} is more noticeable, but it shrinks as $\SNR \to \infty$. It is also observed that the \ac{NMSE} and the hybrid \ac{CRLB} are monotonically decreasing proportionally to the $\SNR$. There is, however, a certain $\SNR$ value beyond which both the performance of the proposed algorithms and the hybrid \ac{CRLB} saturate and exhibit a plateau effect. This behavior sets the distinction between the noise-limited regime and the \ac{PN}-limited regime, whereby estimation performance cannot longer improve even if $\SNR \to \infty$. This behavior is shown in Fig. \ref{fig:NMSE_vs_CFO_asymptotics} for $\Ntr = 4$ and $\Lr = 4$.

	\item The estimation performance in the low \ac{SNR} regime does not depend on the \ac{PSD} of the \ac{PN}, which indicates that the synchronization performance is limited by the \ac{AWGN}. Notice, however, that as $\SNR \to \infty$, 	\item The estimation performance in the low \ac{SNR} regime does not depend on the \ac{PSD} of the \ac{PN}, which indicates that the synchronization performance is limited by the \ac{AWGN}. Notice, however, that as $\SNR \to \infty$, both the estimation performance of the proposed algorithms and the hybrid \ac{CRLB} are different for the two values of the \ac{PSD} $G_\theta$ of the \ac{PN}. Intuitively, as $G_\theta$ increases, the information coupling between the \ac{PN} and the \ac{CFO} impairments increases, thereby reducing both the achievable \ac{CFO} estimation performance of the proposed algorithms and the hybrid \ac{CRLB}. Conversely, reducing $G_\theta$ reduces this information coupling, which results in better \ac{CFO} estimates and lower hybrid \ac{CRLB}.
		
	\item When the $\SNR$ is very low, the \ac{NMSE} of the proposed algorithms is high. However, when the \ac{SNR} increases, a waterfall effect is observed, and the $\SNR$ at which this effect happens depends on both the number of \ac{RF} chains $\Lr$ and the number of \ac{OFDM} training symbols $\Ntr$. More especifically, increasing $\Ntr$ or $\Lr$ shifts the minimum $\SNR$ at which this waterfall effect is observed. Thereby, increasing $\Ntr$ and $\Lr$ results in more accurate estimates of the \ac{CFO} parameter, even for $\SNR < -10$ dB.
	
	\item Last, increasing the number of \ac{OFDM} training symbols $\Ntr$ and the number of receive \ac{RF} chains $\Lr$ have a different impact on both the \ac{CFO} estimation performance and the hybrid \ac{CRLB}. More especifically, doubling $\Ntr$ results in a performance gain of approximately $9$ dB, which indicates that the estimation performance depends on $\Ntr^3$, which is a similar result to the \ac{CFO} estimation performance and \ac{CRLB} in \cite{RodGon:CFO:TWC:2019}. Regarding the number of receive \ac{RF} chains, doubling $\Lr$ enhances estimation performance by a factor of $3$ dB, which indicates that the estimation approach averages the receive noise across multiple receive \ac{RF} chains, thereby exhibiting an \ac{NMSE} estimation performance proportional to $\Lr^{-1}$.
	
\end{itemize}

\begin{figure}[t!]
\centering
\begin{tabular}{cc}
\includegraphics[width=0.5\textwidth]{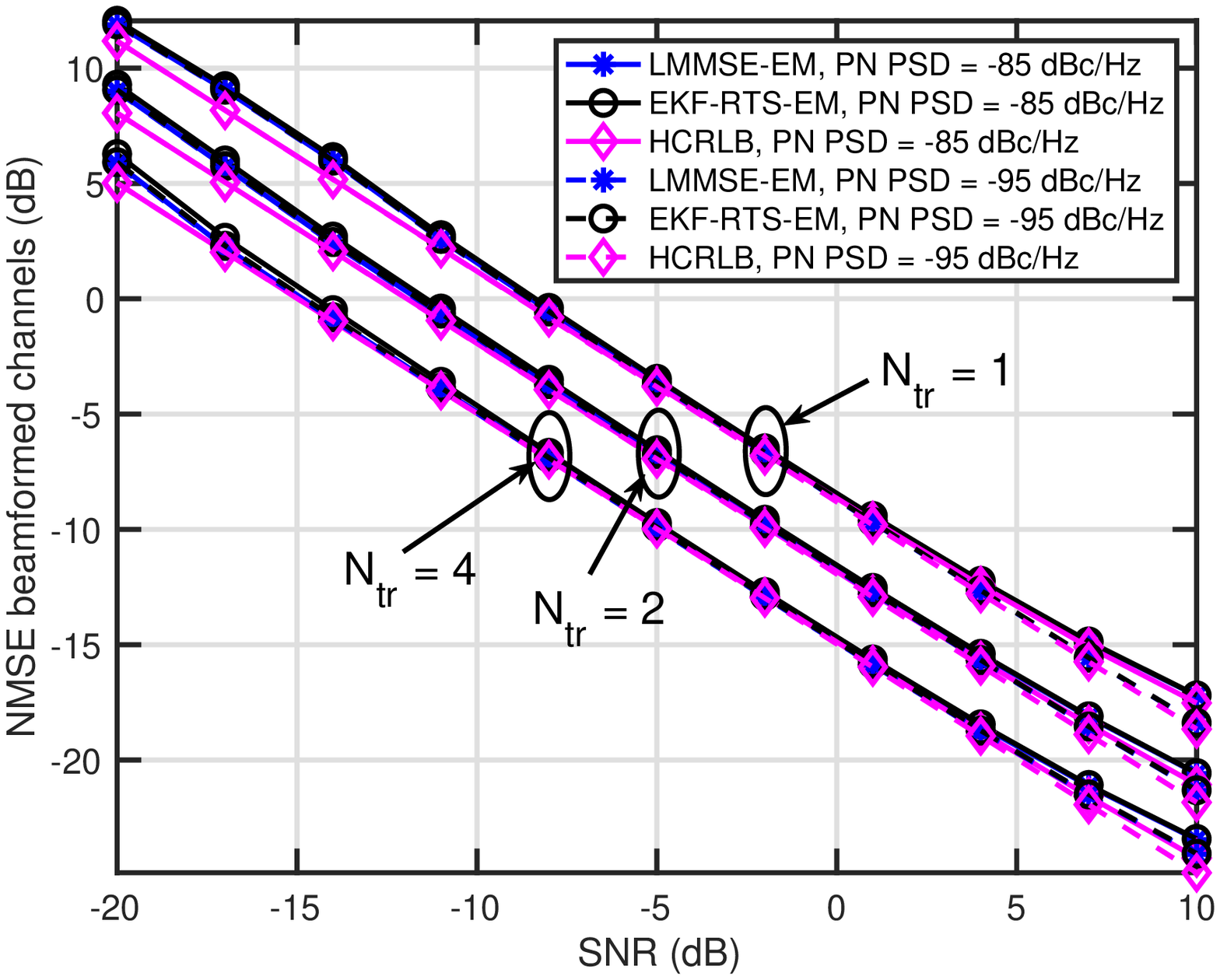} & \includegraphics[width=0.5\textwidth]{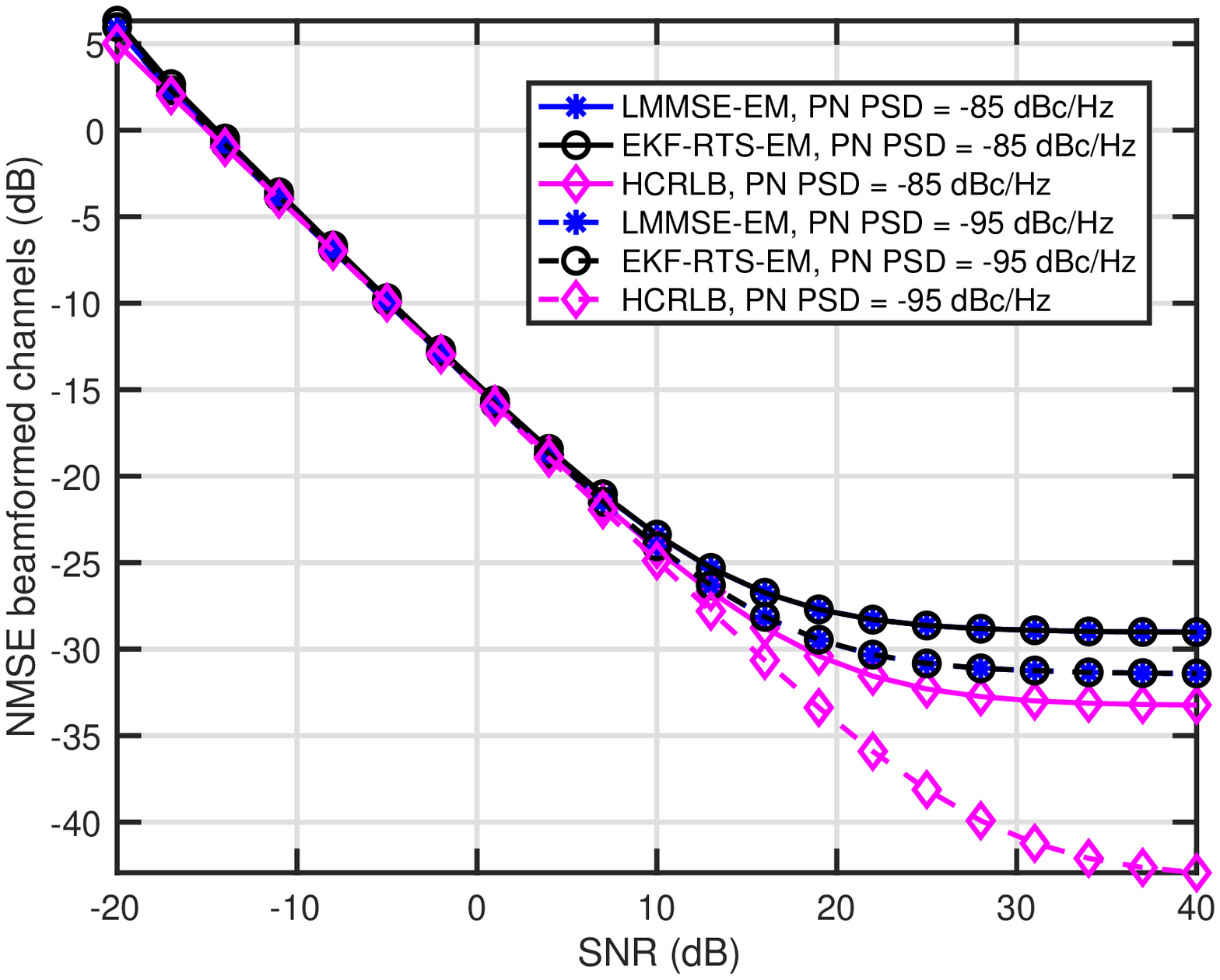} \\ (a) & (b)\end{tabular}
\caption{Evolution of the \ac{NMSE} of the beamformed channel estimates obtained using the proposed algorithms versus $\SNR$. The hybrid \ac{CRLB} is also provided as a performance bound.}
\label{fig:NMSE_vs_channel_several_Ntr}
\end{figure}

In Fig. \ref{fig:NMSE_vs_channel_several_Ntr} (a), I show the \ac{NMSE} evolution of the equivalent beamformed channels versus $\SNR$, for both the proposed \ac{LMMSE}-\ac{EM} and the \ac{EKF}-\ac{RTS}-\ac{EM} algorithms. The number of receive \ac{RF} chains is set to $\Lr = 4$, and the number of \ac{OFDM} training symbols $\Ntr$ is swept within $\{1,2,4\}$. A similar behavior to that in Fig. \ref{fig:NMSE_vs_CFO_several_Ntr} is observed. For both proposed algorithms, the estimation performance is very close to the hybrid \ac{CRLB}, although there is a more noticeable performance gap for $\SNR <10$ dB. Similar to Fig. \ref{fig:NMSE_vs_CFO_several_Ntr}, it is observed that the reduction in computational complexity of the second proposed \ac{EKF}-\ac{RTS}-\ac{EM} algorithm does not compromise estimation performance, thereby showing that synchronization in the low \ac{SNR} regime can be successfully accomplished with reduced computational complexity. It is also observed that doubling the number of \ac{OFDM} training symbols $\Ntr$ results in enhanced estimation performance by a factor of $3$ dB, which is expected since it was observed that the Fisher information of the channel coefficients increases linearly with the number of training samples.

Furthermore, the performance of the proposed algorithms, as well as the hybrid \ac{CRLB}, do not depend on the \ac{PSD} of the \ac{PN} in both the mid and low \ac{SNR} regimes. Notice, however, that for $\SNR > 0$ dB, the estimation performance and hybrid \ac{CRLB} depend on the \ac{PSD} of the \ac{PN}, similar to Fig. \ref{fig:NMSE_vs_CFO_several_Ntr}. This behavior sets the beginning of the \ac{PN}-limited regime, which is more pronounced as $\SNR \to \infty$, as shown in Fig. \ref{fig:NMSE_vs_channel_several_Ntr} (b). In Fig. \ref{fig:NMSE_vs_channel_several_Ntr}, the number of \ac{OFDM} training symbols is set to $\Ntr = 4$. It is observed that the estimation performance gap between the proposed algorithms and the hybrid \ac{CRLB} increases as $\SNR \to \infty$, and it is more pronounced for smaller values of the \ac{PSD} of the \ac{PN}. This behavior is due to two different factors: i) instead of using the \ac{MMSE} estimator for the \ac{PN}, the proposed algorithms attempt to approximate this estimator using statistical linearization (\ac{LMMSE}), such that non-linearities are not dealt with, and ii) for smaller values of the \ac{PSD} of the \ac{PN}, the covariance matrix of the \ac{PN} has smaller eigenvalues, thereby reducing the amount of prior information on this parameter. This second fact makes the covariance after propagation update be significantly larger than the \ac{AWGN} impairment, thereby making the Kalman gain for the \ac{PN} estimator rely more heavily on the measurement and less on the \ac{AWGN}. Consequently, the \ac{PN} impairment is more difficult to estimate, which affects the estimation of the equivalent beamformed channels.

\begin{figure}[t!]
\centering
\includegraphics[width=0.5\textwidth]{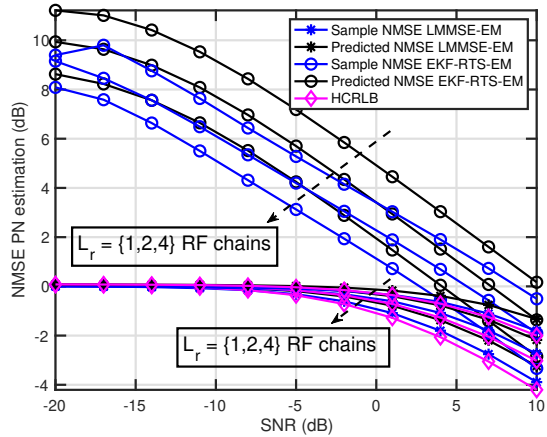} 
\caption{Evolution of the \ac{NMSE} of the \ac{PN} estimates obtained using the proposed algorithms versus $\SNR$. The hybrid \ac{CRLB} is also provided as a performance bound.}
\label{fig:NMSE_vs_PN_several_Ntr}
\end{figure}

\begin{figure}[t!]
\centering
\begin{tabular}{cccc}
\includegraphics[width=0.5\textwidth]{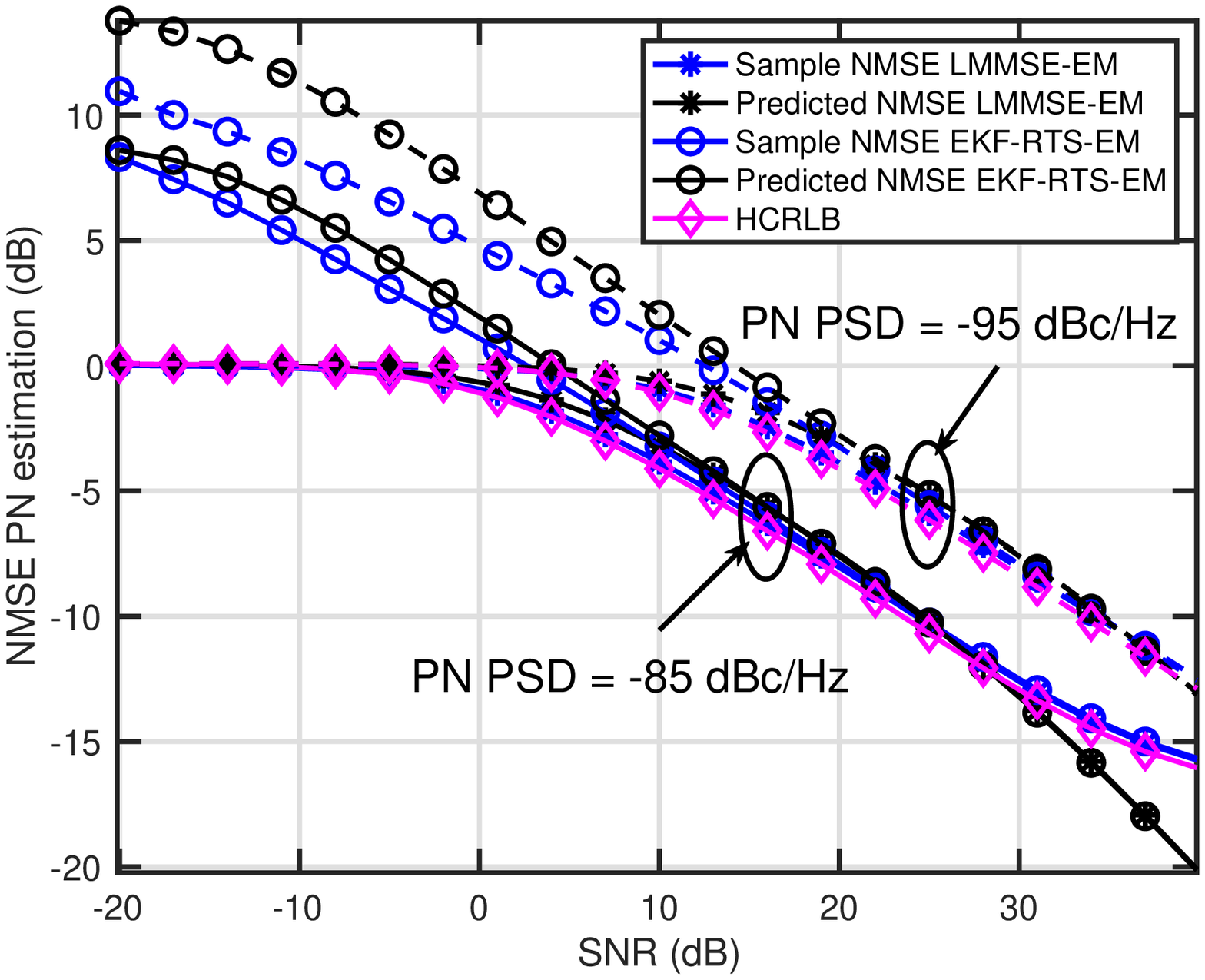} & \includegraphics[width=0.5\textwidth]{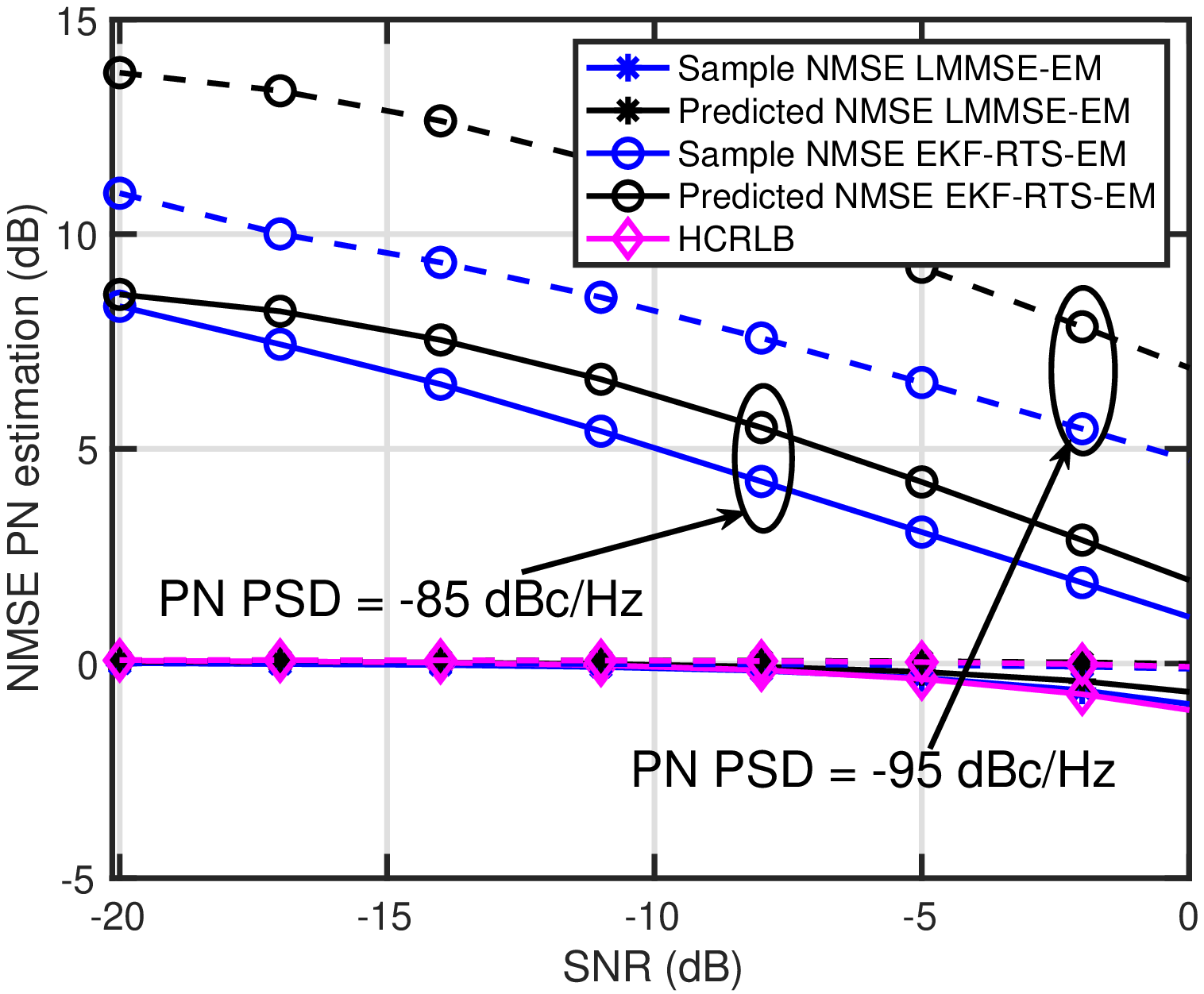} \\
(a) & (b) \\
\includegraphics[width=0.5\textwidth]{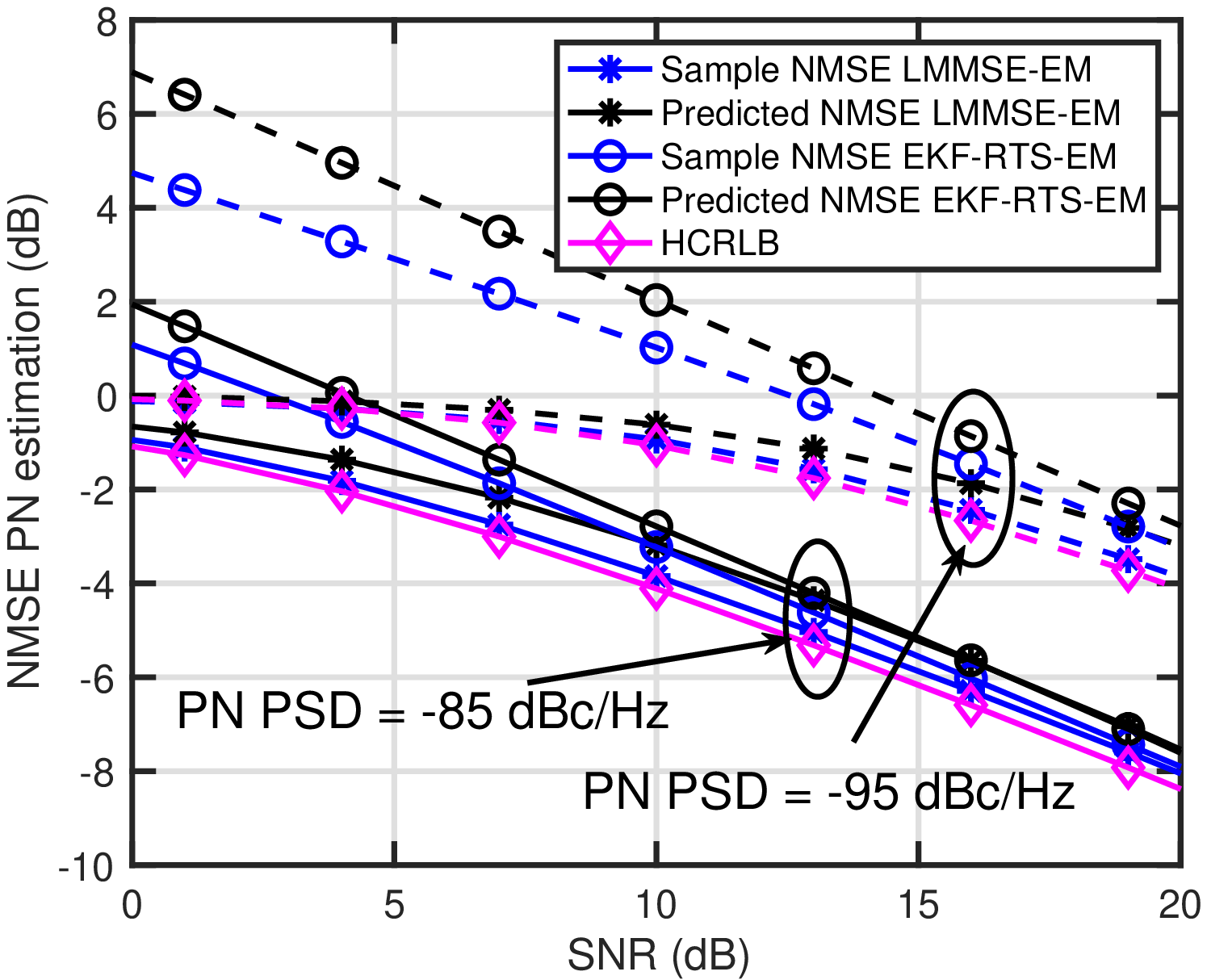} & \includegraphics[width=0.5\textwidth]{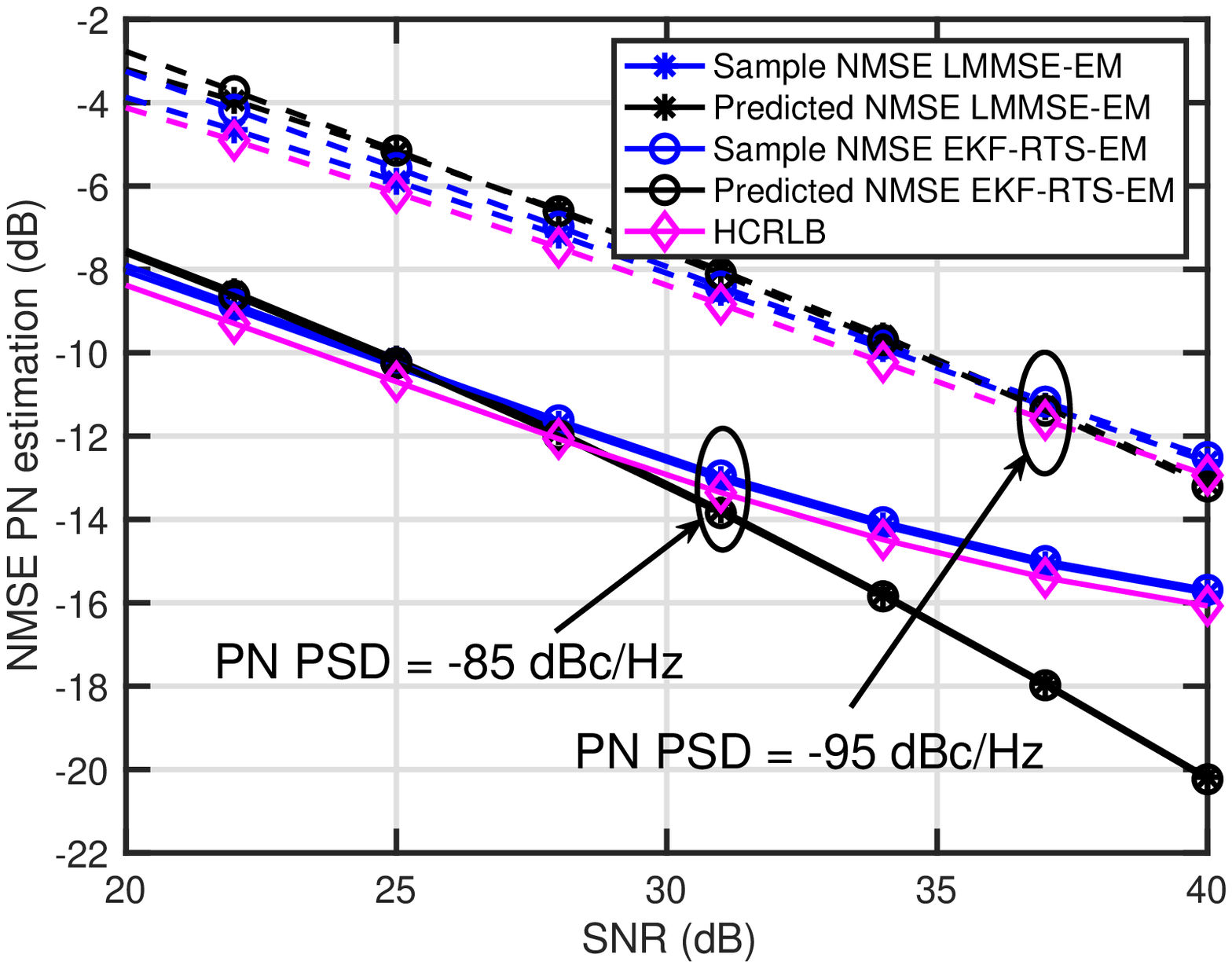}  \\
(c) & (d) \\
\end{tabular}
\caption{Asymptotic evolution of the \ac{NMSE} of the \ac{PN} estimates obtained using the proposed algorithms versus $\SNR$. The curves in (a) show this asymptotic evolution for $\SNR$ values between $-20$ and $40$ dB. Magnified curves of the asymptotic evolution are also shown for $\SNR \in [-20,0]$ dB in (b), $\SNR \in [0,20]$ dB in (c), and $\SNR \in [20,40]$ dB in (d). The hybrid \ac{CRLB} is also provided as a performance bound.}
\label{fig:NMSE_vs_PN_asymptotics}
\end{figure}

Last, I show the estimation performance of the proposed \ac{LMMSE}-\ac{EM} and \ac{EKF}-\ac{RTS}-\ac{EM} algorithms versus $\SNR$ in Fig. \ref{fig:NMSE_vs_PN_several_Ntr} and Fi.g \ref{fig:NMSE_vs_PN_asymptotics}. In Fig. \ref{fig:NMSE_vs_PN_several_Ntr}, the number of receive \ac{RF} chains is swept within $\{1,2,4\}$, and it is set to $\Lr = 4$ in Fig. \ref{fig:NMSE_vs_PN_asymptotics}. The \ac{PSD} of the \ac{PN} is set to $G_\theta = -85$ dB in Fig. \ref{fig:NMSE_vs_PN_several_Ntr}, which corresponds to a stronger \ac{PN} process. The number of \ac{OFDM} training symbols is set to $\Ntr = 4$ in both Fig. \ref{fig:NMSE_vs_PN_several_Ntr} and Fig. \ref{fig:NMSE_vs_PN_asymptotics}. 

The first observation from both Fig. \ref{fig:NMSE_vs_PN_several_Ntr} and Fig. \ref{fig:NMSE_vs_PN_asymptotics} is that the \ac{PN} estimation performance of the first proposed \ac{LMMSE}-\ac{EM} algorithm significantly outperforms that of the second proposed \ac{EKF}-\ac{RTS}-\ac{EM} algorithm. This is not surprising, since the estimation approach in both cases comes from a linearization of the measurement signal through its Jacobian matrix, which is very sensitive to \ac{AWGN} if a small number of measurements are processed. If a single $\Lr$-dimensional measurement is processed, as in the \ac{EKF}-\ac{RTS}-\ac{EM} algorithm, the Jacobian matrix of the measurement varies significantly depending on the \ac{AWGN} variance and the \ac{PSD} of the \ac{PN} process, thereby making it harder for the second proposed algorithm to track the \ac{PN} variations. If every measurement is processed in a larger-dimensional batch, as in the first proposed \ac{LMMSE}-\ac{EM} algorithm, the Kalman measurement update is performed taking the multiple measurements into account, thereby contributing to stabilizing the Jacobian matrix and making it easier to track the \ac{PN} variations. This is a significant effect only in the low \ac{SNR} regime, and the performances of both proposed algorithms exhibit convergence as $\SNR$ increases, as shown in Fig. \ref{fig:NMSE_vs_PN_asymptotics}. 

It is also observed that increasing $\Lr$ results in a noise averaging effect, which results in an estimation performance improvement of approximately $3$ dB. This is a similar effect to that in Fig. \ref{fig:NMSE_vs_channel_several_Ntr} as a function of $\Ntr$, which indicates that the \ac{AWGN} can be more effectively filtered out as $\Lr$ increases.

It is also observed that, for a wide range of $\SNR$ values, the first proposed \ac{LMMSE}-\ac{EM} algorithm exhibits estimation performance lying very close to the hybrid \ac{CRLB}, and divergence from the bound is observed as $\SNR \to \infty$, for similar reasons as with Fig. \ref{fig:NMSE_vs_channel_several_Ntr} (b). It is also observed that the \ac{NMSE} predicted from the first proposed \ac{LMMSE}-\ac{EM} algorithm is very close to the actual \ac{NMSE} performance, while for the second proposed \ac{EKF}-\ac{RTS}-\ac{EM} algorithm it is more difficult to predict the \ac{NMSE} performance in the low \ac{SNR} regime, which is expected due to the varying nature of the Jacobian matrix of the measurement when the received samples are sequentially processed, instead of performing simultaneous batch-processing, as in the first proposed \ac{LMMSE}-\ac{EM} algorithm.

\begin{figure}[t!]
\centering
\begin{tabular}{cc}
\includegraphics[width=0.5\textwidth]{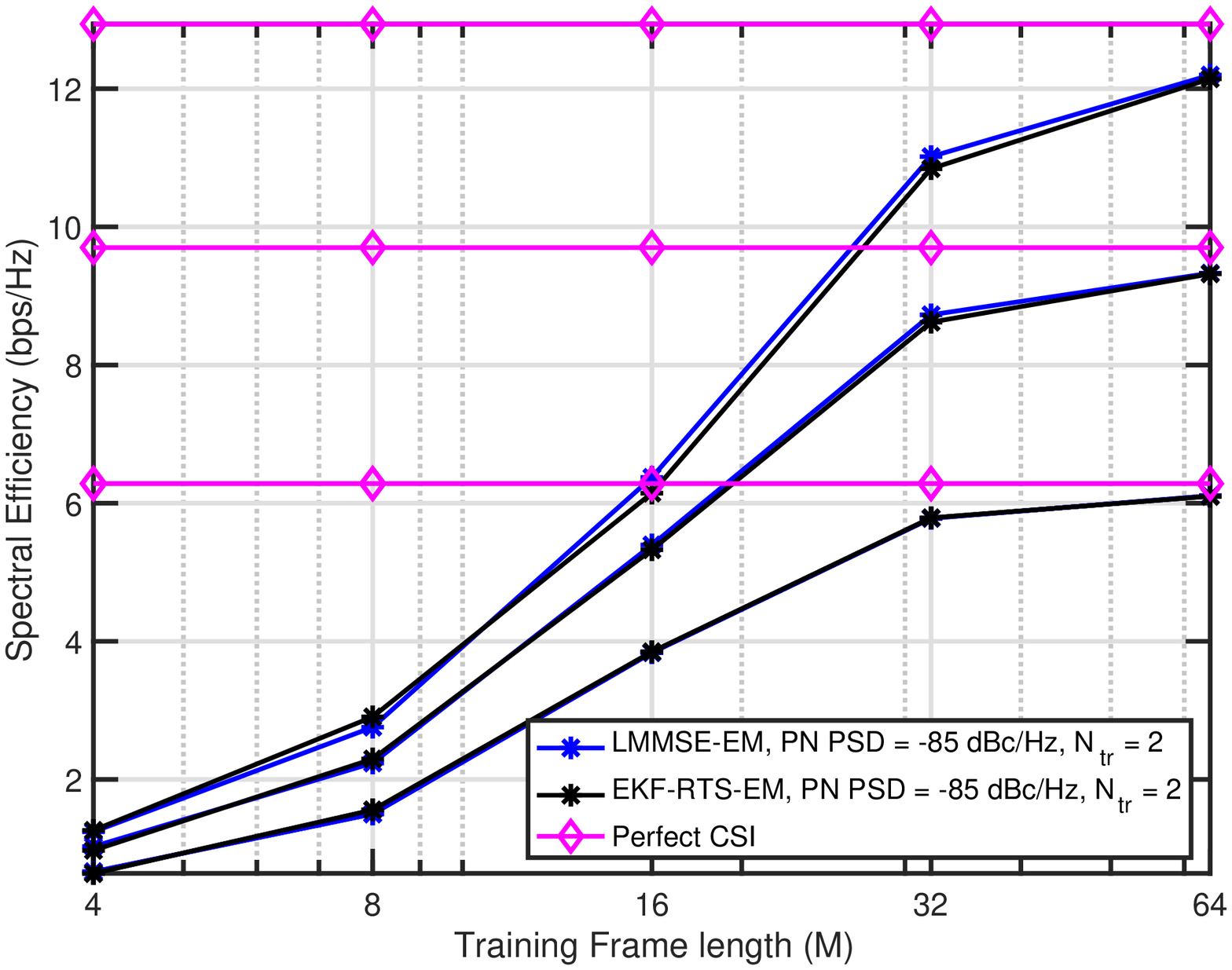} & \includegraphics[width=0.5\textwidth]{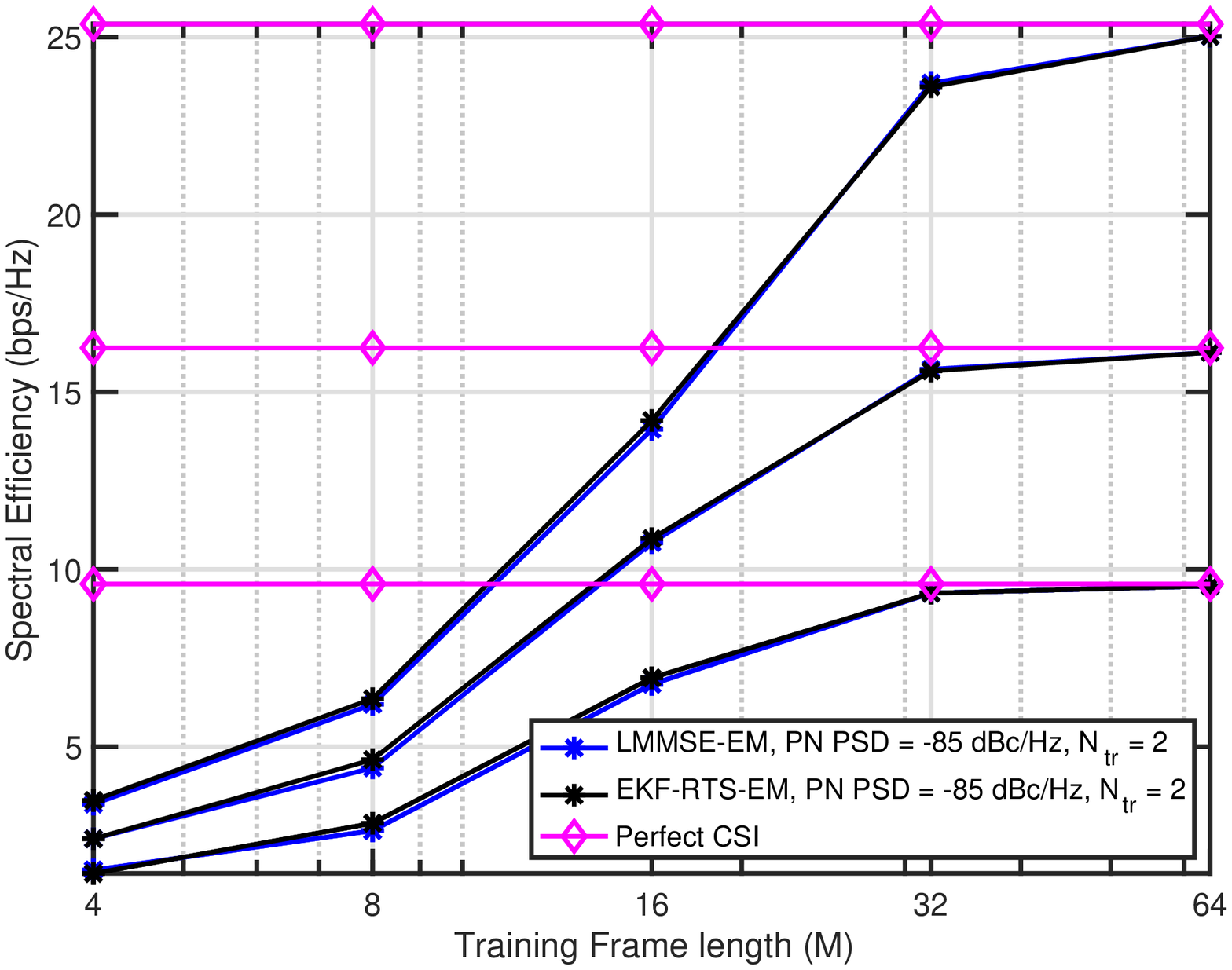} \\ (a) & (b)\end{tabular}
\caption{Evolution of the spectral efficiency versus number of training frames $M$, obtained using the proposed algorithms for $\SNR = -10$ dB (a) and $\SNR = 0$ dB (b). The number of \ac{OFDM} training symbols is set to $\Ntr = 2$.}
\label{fig:SE_vs_M}
\end{figure}

Finally, in Fig. \ref{fig:SE_vs_M} I show the spectral efficiency evolution as a function of $M$, for $\SNR = \{-10,0\}$ dB, and for \ac{PN} \ac{PSD} $G_\theta = -85$ dBc/Hz. It is observed that the proposed algorithms cannot accurately estimate the dominant components of the column and row spaces of the broadband channels for $M < 32$, and there is slight performance improvement for $M > 32$. By comparing $M = 32$ with $M = 64$, it is observed that the proposed algorithms are able to unlock higher spectral efficiency as the training length increases, but the marginal increase in spectral efficiency does not compensate for doubling the training overhead and computational complexity.
This behavior is more pronounced at $\SNR = -10$ dB in Fig. \ref{fig:SE_vs_M} (a), but the same trend can be observed in Fig. \ref{fig:SE_vs_M} (b), for $\SNR = 0$ dB.

\section{Conclusions}
\label{sec:conclusions}

In this paper, I proposed a joint synchronization and compressive channel estimation strategy suitable for \ac{mmWave} \ac{MIMO} systems using hybrid architectures. Leveraging the available information on the received data, I analyzed and provided closed-form expressions for the hybrid \ac{CRLB} associated to the synchronization problem, and designed two low-complexity \ac{EM}-based algorithms to iteratively estimate the \ac{TO}, \ac{CFO}, \ac{PN} and channel impairments. Furthermore, fully leveraging information on the estimates of the unknown parameters, I showed how the obtained channel estimates can be exploited to estimate the high-dimensional frequency-selective \ac{mmWave} \ac{MIMO} channel. The numerical results show that the proposed algorithms can be used to estimate the communication channel under the 5G \ac{NR} wireless channel model, and that near-optimum data rates can be achieved while keeping overhead low and regardless of lack of synchronization.



\bibliographystyle{IEEEtran}

\end{document}